\documentclass[twocolumn,tighten]{aastex63}

\usepackage{newtxtext,newtxmath}
\usepackage[T1]{fontenc}

\usepackage{graphicx}
\usepackage{amssymb}
\usepackage{amsmath}
\usepackage{natbib}

\newcommand{\mb}{m_{\rm b}}

\newcommand{\cm}{\,{\rm cm}}
\newcommand{\m}{\,{\rm m}}
    
\newcommand{\kpc}{\,{\rm kpc}}
\newcommand{\Mpc}{\,{\rm Mpc}}

\newcommand{\s}{\,{\rm s}}

\newcommand{\Myr}{\,{\rm Myr}}
\newcommand{\Gyr}{\,{\rm Gyr}}
\newcommand{\erg}{\,\rm erg}

\newcommand{\g}{\,{\rm g}}
\newcommand{\K}{\,{\rm K}}

\newcommand{\msun}{\,{\rm M_{\odot}}}
\newcommand{\zsun}{\,{\rm Z_{\odot}}}

\newcommand{\cs}{c_{\rm s}}
\newcommand{\vc}{v_{\rm c}}

\newcommand{\Mstar}{M_*}
\newcommand{\Mdot}{{\dot M}}
\newcommand{\nH}{n_{\rm H}}

\newcommand{\kms}{\,\rm km\ s^{-1}}

\newcommand{\Mvir}{M_{\rm vir}}

\newcommand{\Rvir}{R_{\rm vir}}
\newcommand{\Tvir}{T_{\rm vir}}

\newcommand{\tcool}{t_{\rm cool}}

\newcommand{\thubble}{t_{\rm H}}

\newcommand{\tff}{t_{\rm ff}}

\newcommand{\Mhalo}{M_{\rm halo}}

\newcommand{\Tc}{T^{({\rm s})}}
\newcommand{\rhoc}{\rho^{({\rm s})}}
\newcommand{\rhocrit}{\rho_{\rm crit}}
\newcommand{\nHc}{n_{\rm H}^{({\rm s})}}

\newcommand{\Deltac}{\Delta_{\rm c}}

\newcommand{\Omo}{\Omega_{\rm m,0}}

\newcommand{\Rcirc}{R_{\rm circ}}
\newcommand{\rvir}{R_{\rm vir}}
\newcommand{\fvc}{f_{\vc}}
\newcommand{\ft}{f_{t}}

\newcommand{\Mthres}{M_{\rm thresh}}

\newcommand{\dex}{\,{\rm dex}}

\newcommand{\Tmed}{\langle T\rangle}
\newcommand{\tcoolsh}{\tcool^{({\rm s})}}
\newcommand{\Vrot}{V_{\rm rot}}
\newcommand{\hi}{\ion{H}{1}}
\newcommand{\Ha}{\text{H$\alpha$}}
\newcommand{\fg}{f_{\rm gas}}

\received{}
\revised{}
\accepted{}

\submitjournal{ApJ}
\shorttitle{Inner CGM Virialization in FIRE}

\begin{document}

\title{Virialization of the inner CGM in the FIRE simulations and implications for galaxy discs, star formation and feedback}

\author{Jonathan Stern}
\altaffiliation{E-mail: jonathan.stern@northwestern.edu}
\altaffiliation{CIERA Fellow}
\affiliation{Department of Physics \& Astronomy and CIERA, Northwestern University, 1800 Sherman Ave, Evanston, IL 60201, USA}

\author{Claude-Andr{\'e} Faucher-Gigu{\`e}re}
\affiliation{Department of Physics \& Astronomy and CIERA, Northwestern University, 1800 Sherman Ave, Evanston, IL 60201, USA}

\author{Drummond Fielding}
\affiliation{Center for Computational Astrophysics, Flatiron Institute, 162 5th Ave, New York, NY 10010, USA}

\author{Eliot Quataert}
\affiliation{Astronomy Department and Theoretical Astrophysics Center, University of California Berkeley, Berkeley, CA 94720, USA}

\author{Zachary Hafen}
\affiliation{Department of Physics \& Astronomy and CIERA, Northwestern University, 1800 Sherman Ave, Evanston, IL 60201, USA}

\author{Alexander B. Gurvich}
\affiliation{Department of Physics \& Astronomy and CIERA, Northwestern University, 1800 Sherman Ave, Evanston, IL 60201, USA}

\author{Xiangcheng Ma}
\affiliation{Astronomy Department and Theoretical Astrophysics Center, University of California Berkeley, Berkeley, CA 94720, USA}

\author{Lindsey Byrne}
\affiliation{Department of Physics \& Astronomy and CIERA, Northwestern University, 1800 Sherman Ave, Evanston, IL 60201, USA}

\author{Kareem El-Badry}
\affiliation{Astronomy Department and Theoretical Astrophysics Center, University of California Berkeley, Berkeley, CA 94720, USA}

\author{Daniel Angl{\'e}s-Alc{\'a}zar}
\affiliation{Center for Computational Astrophysics, Flatiron Institute, 162 5th Ave, New York, NY 10010, USA}
\affiliation{Department of Physics, University of Connecticut, 196 Auditorium Road, U-3046, Storrs, CT 06269-3046, USA}

\author{T.K. Chan}
\affiliation{Department of Physics and Center for Astrophysics and Space Science, University of California at San Diego, 9500 Gilman Drive, La Jolla, CA 92093, USA}
\affiliation{Institute for Computational Cosmology, Durham University, South Road, Durham DH1 3LE, UK}

\author{Robert Feldmann}
\affiliation{Institute for Computational Science, University of Zurich, Zurich CH-8057, Switzerland}

\author{Du\v{s}an Kere\v{s}}
\affiliation{Department of Physics and Center for Astrophysics and Space Science, University of California at San Diego, 9500 Gilman Drive, La Jolla, CA 92093, USA}

\author{Andrew Wetzel}
\affiliation{Department of Physics, University of California, Davis, CA 95616, USA}

\author{Norman Murray}
\affiliation{Canadian Institute for Theoretical Astrophysics, 60 St. George Street, University of Toronto, ONM5S 3H8, Canada}

\author{Philip F. Hopkins}
\affiliation{TAPIR, Mailcode 350-17, California Institute of Technology, Pasadena, CA 91125, USA}





\begin{abstract}
We use the FIRE-2 cosmological simulations to study the formation of a quasi-static, virial-temperature gas phase in the circumgalactic medium (CGM) at redshifts $0<z<5$, and how the formation of this virialized phase affects the evolution of galactic discs. 
We demonstrate that when the halo mass crosses $\sim10^{12}\msun$, the cooling time of shocked gas in the inner CGM ($\sim0.1\Rvir$, where $\Rvir$ is the virial radius) exceeds the local free-fall time. 
The inner CGM then experiences a transition from on average sub-virial temperatures ($T\ll\Tvir$), large pressure fluctuations and supersonic inflow/outflow velocities, to virial temperatures ($T\sim\Tvir$), uniform pressures and subsonic velocities.  
This transition occurs when the outer CGM ($\sim0.5\Rvir$) is already subsonic and has a temperature $\sim\Tvir$, indicating that the longer cooling times at large radii allow the outer CGM to virialize at lower halo masses than the inner CGM. 
This outside-in CGM virialization scenario is in contrast with inside-out scenarios commonly envisioned
based on more idealized simulations. 
We demonstrate that inner CGM virialization coincides with abrupt changes in the central galaxy and its stellar feedback: 
the galaxy settles into a stable rotating disc, star formation transitions from `bursty' to `steady,' and stellar-driven galaxy-scale outflows are suppressed. 
Our results thus suggest that CGM virialization is initially associated with the formation of rotation-dominated thin galactic discs, rather than with the quenching of star formation as often assumed.
\end{abstract} 

\keywords{
cosmology: theory
-- galaxies: evolution
-- galaxies: formation 
-- galaxies: star formation
}

\section{Introduction}

\newcommand{\sigmag}{\sigma_{\rm g}}

While star forming $\sim$$L^*$  galaxies in the local Universe are typically thin discs with axial ratios of $\approx0.2$ \citep[e.g.,][]{Padilla08},
star forming dwarf galaxies with stellar masses $\Mstar\lesssim10^{9}\msun$ and circular velocities $\vc\lesssim50\kms$ tend to have irregular morphologies \citep{Roberts69,Hunter97, Kennicutt08, Dale09, Karachentsev13_catalog} and thick discs with axial ratios of $0.3-0.7$ \citep{StaveleySmith92, Hunter06, Roychowdhury13}. 
This transition from irregulars to thin discs with increasing mass coincides with a transition from dispersion-dominated to rotation-dominated kinematics.
\cite{simons15} showed that in irregular dwarfs the characteristic rotational velocity can be comparable to or even smaller than the random velocity component ($\Vrot\lesssim\sigmag$), while in higher mass discs typically $\Vrot>3\sigmag$. Dwarf irregulars also exhibit a larger dispersion than discs in \Ha\ equivalent width and \Ha\ to UV luminosity ratio (\citealt{Lee07, Lee09,Karachentsev13_SFR_vs_UV}), suggesting larger fluctuations in the recent SFR measured by H$\alpha$ ($\lesssim10\Myr$) relative to the average SFR on longer timescales measured by UV and optical emission \citep{Weisz12}. 

At earlier cosmic times, rotation appears to dominate the kinematics above a higher characteristic mass than in the local Universe, roughly $\sim5\cdot10^{10}\msun$ at redshift $z\sim1$ (\citealt{kassin12,simons17}). These observed mass and redshift trends suggest that galaxies have become more disk-dominated with cosmic time, a phenomenon known as `disc settling.'
Disc settling is also suggested by the observed relation between stellar age and vertical height in the Milky-Way and nearby blue galaxies, a trend often interpreted as a `thick disc' component with an old stellar population and a younger `thin disc' component \citep[e.g.,][]{gilmore83,Yoachim06}. 
\cite{Bovy12} quantified this trend based on a sample of stars within a few kpc from the Sun, and deduced that the oldest stars have a vertical to radial scale ratio of $\approx 0.5$, 
compared to a substantially smaller ratio of $\lesssim 0.1$ in the youngest stars. 
This relation also supports the notion that Milky-Way-mass discs become increasingly thin and rotation-dominated with decreasing redshift.

Why are rotating thin discs common only in massive star forming galaxies?
In standard disc formation theories (e.g.~\citealt{fall80,dalcanton97,mo98}), gas associated with the halo radiates away its gravitational energy and accretes towards the center of the halo. 
In the limit that energy loss to radiation is efficient, the gas will inflow down to the `circularization radius' $\Rcirc$ at which it is rotationally-supported against gravity (see \citealt{WetzelNagai15} for a demonstration of this process in a cosmological simulation).
The size of the galaxy is thus expected to be of order $\Rcirc$. 
One can estimate $\Rcirc$ by assuming gas inherits the specific angular momentum of the halo, which in turn is induced by tidal torques from the surrounding matter distribution (\citealt{peebles69,doroshkevich70,white84}). For a standard cosmology the expected $\Rcirc$ is $\sim \lambda\Rvir$ where $\rvir$ is the halo virial radius and $\lambda\approx0.02-0.06$ is the halo spin parameter \citep[][using the \citealt{bullock01} definition of $\lambda$]{rodriguezpuebla16}.
This prediction for the characteristic sizes of galaxies is borne out by comparing observed galaxies with abundance matching-based estimates of $\Rvir$
(\citealt{kravtsov13,shibuya15}).
However, $\lambda$ is predicted to be almost independent of halo mass and redshift, and thus the standard framework for disc formation does not explain why star-forming galaxies become thinner and more rotation-dominated with increasing mass and decreasing redshift. 


In this paper we argue that disc settling is linked to, and potentially a result of, the virialization of the inner circumgalactic medium (CGM), a predicted transition not previously incorporated in disc formation theory.  The basic physics of CGM virialization was originally discussed by \cite{rees77}, \cite{Silk77} and \cite{White78}, and has since been corroborated by more detailed (but still idealized) calculations \citep{Birnboim03, Dekel06, Fielding17}.
These studies showed that in halo masses below a threshold $\Mthres \sim 10^{12}\msun$ the cooling time $\tcoolsh$ of gas shocked to the halo virial temperature $T_{\rm vir}$ is shorter\footnote{The superscript $(s)$ denotes that this is the cooling time of shocked gas, as detailed below.}  than the free-fall time $\tff$, and thus shocked gas will immediately cool, lose pressure support and free-fall towards the galaxy.
In contrast when the halo mass is above the threshold $\tcoolsh$ exceeds $\tff$ and compressional heating from the inflow can compensate radiative losses, so the shocked gas remains hot 
and contracts quasi-statically.  In this paper we extend these previous results on the virialization of the CGM when the halo mass reaches $\Mthres$, by demonstrating that virialization proceeds from the outer CGM inwards, and by identifying the implications of virialization for galaxy evolution.

The idealized studies mentioned above did not model the central galaxy or the role of feedback, or did so in a highly idealized manner. In cosmological simulations, where these processes are modelled more realistically, early studies initially focused on identifying whether gas accreted onto galaxies was shocked to $\sim\Tvir$ prior to accretion \citep{Keres05, Keres09, brooks09, Oppenheimer10, vandeVoort11, Sales12, Nelson13}. However, one cannot directly infer from the thermal history of accreted gas if the volume-filling CGM phase has virialized, since even after CGM virialization gas may still accrete via cold filaments which do not shock \citep[e.g.,][]{Keres05}, and since prior to CGM virialization inflows may still shock to $\sim T_{\rm vir}$ against outflows from the galaxy \citep{Fielding17}. 
More recently, \cite{vandeVoort16} analyzed the X-ray emission from halos in the FIRE-1 cosmological zoom simulations \citep{Hopkins14}. They showed that the X-ray emission from halo masses below $\Mthres\sim10^{12}\msun$ is highly variable and correlated with the star formation rate (SFR), while above $\Mthres$ the X-ray emission is time-steady and uncorrelated with the SFR. This suggests that a quasi-static virialized CGM indeed forms at $\sim \Mthres$ also in a fully-cosmological setting, while in lower halo masses the hot X-ray emitting gas is more closely related to stellar-driven outflows. \cite{Correa18} reinforced this conclusion using the EAGLE cosmological simulations \citep{Schaye15}, by showing that essentially all CGM particles have $\tcool\lesssim\tff$ below $\Mthres$, in contrast with a large fraction of CGM particles with $\tcool>\tff$ above $\Mthres$. 
Existing studies of cosmological simulations thus support the idea that the CGM virializes at $\sim\Mthres$, though they leave open the question of how virialization occurs and what are its implications for galaxy evolution, which are the focus of this work.

 In \cite{Stern19,Stern20}, hereafter Paper I and Paper II, we addressed the question of CGM virialization using a new idealized approach. We modelled the volume-filling CGM phase as a spherical steady-state cooling flow, similar to the classic cooling flow solutions developed for the inner intracluster medium \citep[e.g.~][]{mathews78,Cowie80,Fabian84}. We showed that in these solutions the ratio of the cooling time to free-fall time increases with halo radius, while the overall normalization of the $\tcoolsh/\tff$ profile increases with halo mass. Thus, if these solutions are roughly valid also in a time-dependent scenario where the halo mass grows with time, then $\tcoolsh$ exceeds $\tff$ and a virialized CGM forms first at large radii and then at smaller radii. This outside-in virialization scenario is opposite to the inside-out scenario suggested by the idealized 1D simulations of \cite{Birnboim03}. 
In this paper we utilize the insights from steady-state solutions to study how the CGM virializes in the FIRE-2 cosmological simulations \citep{hopkins18}. 

While aspects of the analysis in this paper are motivated by the results of Papers I -- II on cooling flows, in this work we do not assume that halos consist of pure cooling flows but rather directly analyze the CGM simulated in FIRE, in which many complications in CGM physics are accounted for. These include time-dependent and non-spherical inflows, satellite galaxies, and outflows driven by stellar feedback (see discussion of the `baryon cycle' in FIRE in \citealt{anglesalcazar17a} and \citealt{hafen19, hafen20}).
Using FIRE also allows us to connect CGM virialization to transitions in the simulated central galaxy identifed by previous studies, including disc settling and the relation between rotation and galaxy mass \citep{Ma17,GarrisonKimmel18,elbadry18a,elbadry18b}, 
the transition from bursty to steady star formation \citep{muratov15,Sparre17,anglesalcazar17b, FaucherGiguere18,Emami19}, and the cessation of outflows in massive galaxies at low redshift \citep{muratov15,muratov17,anglesalcazar17a}. We show below that all these transitions coincide and potentially follow from the virialization of the CGM at small radii.

This paper is organized as follows. 
In \S\ref{s:methods} we review the results of the idealized calculations in Papers I -- II, review the FIRE cosmological simulations, and describe how we measure $\tcoolsh/\tff$ in FIRE. In \S\ref{s:virialization} we analyze how the CGM virializes in FIRE and how it relates to the ratio $\tcoolsh/\tff$, while in \S\ref{s:implications} we connect CGM virialization to several transitions in the simulated galaxies. 
We discuss our results in \S\ref{s:discussion} and summarize in \S\ref{s:summary}. 
Throughout the paper we assume a flat $\Lambda$CDM cosmology with Hubble constant $H_0=67\kms\Mpc^{-1}$ and $\Omo=0.32$ (\citealt{planck18}).\footnote{Some of our simulations
were evolved with slightly different cosmological parameters, but this does not significantly affect our results.}

\section{Methods}\label{s:methods}

In this section we review the condition for CGM virialization implied by spherical steady-state solutions, and then present how it is measured in the FIRE simulations. 

\subsection{Virialized CGM in steady-state solutions}\label{s:theoretical bkg}

Classic cooling flow solutions for the intracluster medium (e.g.~\citealt{mathews78,Bertschinger89}) highlight two characteristic radii of the flow, the `cooling radius' and the `sonic radius'. 
At the cooling radius, the cooling time of shocked gas $\tcoolsh$ (an exact definition is given below) equals the Hubble time $\thubble$ or age of the Universe, so only within this radius substantial cooling is expected. 
At the sonic radius, $\tcoolsh$ equals the free-fall time $\tff$, so within this radius cooling is so rapid that it cannot be balanced by heating due to gravitational compression of the inflow. 
The sonic radius thus separates between an outer subsonic region where $\tff \lesssim \tcoolsh \lesssim \thubble$ and the gas temperature is $\sim\Tvir$, and an inner supersonic region 
where $\tcoolsh \lesssim \tff$ and the flow is free-falling with a temperature $\ll \Tvir$. 
The outer subsonic region is expected to be smooth since in subsonic flows thermal instabilities develop on the same timescale as the flow time, while the inner supersonic region is expected to be clumpy since thermal instabilities develop faster than the flow time \citep[e.g.,][Paper I]{BalbusSoker89}. 

In Papers I and II we adapted the classic cooling flow solutions to gas in galaxy-scale halos, i.e.\ the CGM. 
We demonstrated that the cooling radius is typically outside the halo virial radius, while the sonic radius can either be at galaxy radii, at halo radii, or beyond the halo. 
Steady-state CGM solutions thus fall in one of three possible regimes: 
a fully subsonic CGM, where $\tcoolsh>\tff$ at all halo radii; 
a transonic CGM, where $\tcoolsh\sim\tff$ in the halo; 
and a fully supersonic CGM, where $\tcoolsh<\tff$ at all halo radii (see figure~3 in Paper II). Since the gas temperature is $\sim\Tvir$ only in subsonic regions, where the CGM is `virialized', the condition for a fully virialized CGM is thus
\begin{equation}\label{e:threshold condition}
 \tcoolsh(\Rcirc) > \ft\tff(\Rcirc) ~,
\end{equation}
where $\ft$ is a factor of order unity and we use the gas circularization radius $\Rcirc$ to define the inner radius of the CGM (see Paper II). The value of $\Rcirc$ can be calculated from 
\begin{equation}\label{e:Rcirc}
 j = \vc(\Rcirc)\Rcirc ~,
\end{equation}
where $j$ is the gas specific angular momentum, $\vc$ is the circular velocity 
\begin{equation}\label{e:vc}
 \vc = \sqrt\frac{G M(<r)}{r}, 
\end{equation}
and $M(<r)$ is the total mass within $r$. 
The value of $\Rcirc$ can be approximated by assuming an isothermal potential, and that $j$ is independent of radius and similar to the average specific angular momentum of the dark matter. This gives $\Rcirc\approx\sqrt{2}\lambda\Rvir\sim 0.05\Rvir$, where the characteristic spin parameter $\lambda\sim0.035$ is independent of halo mass and redshift \citep[e.g.,][]{rodriguezpuebla16}. 

To estimate eqn.~(\ref{e:threshold condition}) we use the definition of the free-fall time:\footnote{We use this definition of $\tff$, which corresponds to the free-fall time under constant gravitational acceleration, since it is commonly used in CGM literature \citep[e.g.][]{Mccourt12}. This definition differs from the definition $\tff=\sqrt{3\pi/(32G\rho)}$ adopted in other contexts by a factor of $\pi/4$.}
\begin{equation}\label{e:tff}
 \tff = \frac{\sqrt{2} r}{\vc} \approx 160 \frac{r}{0.05\Rvir}\fvc^{-1}(1+z)^{-1.4} \Myr ~,
\end{equation}
where 
\begin{equation}\label{e:fvc}
 \fvc(r)\equiv\frac{\vc(r)}{\vc(\Rvir)}
\end{equation}
and the numerical evaluation is based on the virial relation $\tff(\Rvir)=2/(\sqrt{\Delta_{\rm c}}H)$. We define the virial overdensity $\Deltac$ as in \cite{bryan98}, and use the following approximaton of $\sqrt{\Deltac}H$ in our assumed cosmology:
\begin{equation}\label{e:Delta c H}
 \sqrt{\Deltac}H \approx 0.6 (1+z)^{1.4} \Gyr^{-1} ~.
\end{equation}
The cooling time of shocked gas $\tcoolsh$ in condition~(\ref{e:threshold condition}) is defined assuming gas at $\gtrsim \Rcirc$ forms a cooling flow in the pressure-supported limit, i.e.~that it satisfies $v_r\approx -r/\tcool$ and $v_r^2\ll\cs^2$ where $v_r$ and $\cs$ are the radial velocity and sound speed. These conditions yield
\begin{equation}\label{e:tcool0}
 \tcoolsh \equiv \tcool(\Tc,\nHc) = \frac{(3/2)\cdot 2.3 k_{\rm B}\Tc}{\nHc\Lambda},
\end{equation}
where $\Tc$ and $\nHc$ are the temperature and hydrogen number density of a cooling flow in the pressure-supported limit, $\Lambda$ is the cooling function defined such that $\nH^2\Lambda$ is the cooling per unit volume, and we assume 2.3 particles per hydrogen particle. The value of $\Tc$ can be calculated from eqn.~(24) in Paper I:
\begin{equation}\label{e:Tc}
\Tc \equiv \frac{3\mu m_{\rm p}\vc^2}{5Ak_{\rm B}} = 4.5\cdot10^5 A^{-1} v^2_{100}\K,
\end{equation}
where $\mu=0.62$ is the molecular weight, $\vc=100v_{100}\kms$, and the factor $A$ is equal to 
\begin{equation}
 A = \frac{9}{10}\left(1-2\frac{d\log\vc}{d\log r}\right) \approx 1 ~.
\end{equation}
This temperature is comparable to the virial temperature defined as
\begin{equation}
 \Tvir = \frac{\mu m_{\rm p}}{2k_{\rm B}} \frac{G \Mhalo}{\Rvir} 
\end{equation}
and hence
\begin{equation}
\Tc = \frac{6}{5A}\fvc^2\Tvir ~.
\end{equation}
The density of a cooling flow $\nHc$ can be calculated from the assumption of pressure support of the weight of the overlying gas:
\begin{equation}\label{e:pressure support}
 2.3\nHc{k_{\rm B}\Tc} = \int_r {\frac{\rho \vc^2}{r'}{\rm d} r'} 
\end{equation}
where $-\vc^2/r'$ is the gravitational acceleration at radius $r'$. 
Below we use eqn.~(\ref{e:pressure support}) to estimate $\nHc$ in the FIRE simulations. 
To derive a numerical approximation of $\nHc(\Rcirc)$ we assume the gas distribution in the halo follows a power-law $\rho=\rho(\Rvir)(r/\Rvir)^{-a}$ with slope $-a$ and normalization $\rho(\Rvir) = (1-a/3)\fg\cdot (0.158\Deltac\rhocrit)$, where $\rhocrit$ is the critical density and $\fg$ is the ratio of the halo gas mass to the cosmic halo baryon budget $0.158\Mhalo$. We thus get
\begin{equation}\label{e:rhoc}
 \rho(r) \approx 4.9 \fg g^{-1}(a)\left(\frac{r}{0.05\Rvir}\right)^{-a} \Deltac\rhocrit ~,
\end{equation}
where $g$ is defined such that $g(3/2)=1$:
\begin{equation}
 g(a) = 20^{3/2-a}\left(\frac{3}{6-2a}\right) ~.
\end{equation}
Further using $\nH\approx0.7\rho/m_{\rm p}$, $\rhocrit=3H^2/8\pi G$, and eqn.~(\ref{e:Delta c H}), we get
\begin{equation}\label{e:nHc}
 \nHc (r) \approx 2.0\cdot10^{-3} (1+z)^{2.8} \fg  g^{-1}(a)\left(\frac{r}{0.05\Rvir}\right)^{-a} \cm^{-3}  ~.
\end{equation}

While our choice of $\Tc$ and $\nHc$ in the definition of $\tcoolsh$ in eqn.~(\ref{e:tcool0}) assumes the CGM forms a cooling flow in the limit of full thermal pressure support, 
in practice only the pressure support condition is crucial for our results. Any pressure-supported CGM will have a temperature $\sim\Tvir$ and a pressure set by the weight of the overlying gas. Our conclusions are hence not altered if we use e.g.~$\Tvir$ instead of $\Tc$ in eqns.~(\ref{e:tcool0}) and (\ref{e:pressure support}), beyond changing the exact value of the order-unity factor $\ft$.

We emphasize that below we measure $\tcoolsh$ in the FIRE simulations rather than the standard $\tcool$, since the latter depends on the actual gas temperature $T$ in the simulation. 
The reason for this is twofold. First, since $\tcoolsh$ does not depend on $T$, the relation between $\tcoolsh/\tff$ and $T$ in FIRE is non-trivial (see \S\ref{s:virialization} and Fig.~\ref{f:T all}). Second, the cooling function at temperatures $\ll\Tvir$ is immaterial for the question of whether the gas shock-heated to $\sim \Tvir$ can remain pressure-supported or rather cools and free-falls. 
For example, if $\tcoolsh$ is short and the gas does cool, then it could reach the minimum of the cooling curve where $\tcool$ is quite long and hence $\tcool$ would not be a useful indication of the strong cooling. A comparison of $\tcoolsh$ with other ways of evaluating the cooling time in the simulations is presented in Appendix~\ref{a:density}.

Using eqns.~(\ref{e:Tc}) and (\ref{e:nHc}) in eqn.~(\ref{e:tcool0}) we get
\begin{equation}\label{e:tcool}
 \tcoolsh \approx 14\,v_{100}^{3.4} Z_1^{-1} (1+z)^{-2.8} \fg^{-1} g(a) \left(\frac{r}{0.05\Rvir}\right)^{a} \Myr,
\end{equation}
where we used $A=1$ and approximate the cooling function as $\Lambda \approx 1.4\cdot10^{-22}Z_1 T_6^{-0.7}\erg\cm^3\s^{-1}$ where $Z=Z_1 Z_\odot$ is the gas metallicity and $T=10^6 T_6 \K$. This approximation is roughly valid for $Z_1\sim 1$ and $0.1 < T_6 < 10$. 
Using eqn.~(\ref{e:tff}) the ratio of the two timescales is hence 
\begin{equation}\label{e:tratio}
 \frac{\tcoolsh}{\tff} \approx 0.09\, v_{100}^{3.4} \fvc Z_1^{-1} (1+z)^{-1.4} \fg^{-1} g(a) \left(\frac{r}{0.05\Rvir}\right)^{a-1} ~.
\end{equation}
To express this ratio in terms of halo mass we replace $\vc(r)$ with $\fvc\vc(\rvir)$ and use the virial relation $\vc^3(\rvir)=(\sqrt{\Delta_{\rm c}/2}) H G \Mhalo$. With eqn.~(\ref{e:Delta c H}) this gives 
\begin{equation}\label{e:tratio by Mhalo}
 \frac{\tcoolsh}{\tff} \approx 0.17\, M_{12}^{1.1} \fvc^{4.4} Z_1^{-1} (1+z)^{0.2} \fg^{-1} g(a)\left(\frac{r}{0.05\Rvir}\right)^{a-1} ~,
\end{equation}
where $\Mhalo=10^{12}M_{12}\msun$. Note that the ratio $\tcoolsh/\tff$ increases outwards for $a>1$, and that it is almost independent of redshift if other parameters are held fixed. 

The results of Paper I -- II suggest that if $\tcoolsh\gtrsim\tff$ at $\Rcirc$ then we expect the CGM to be virialized down to the galaxy scale. 
If this condition is violated but $\tcoolsh\gtrsim\tff$ in the outer halo (say, at $0.5\Rvir$), then we expect the CGM to be `transonic' with a cool and free-falling inner region and a virialized CGM further out. If $\tcoolsh\lesssim\tff$ also in the outer halo, then we expect the volume-filling phase to be cool and free-falling throughout the halo.

\subsection{The FIRE simulations}

We test the above idealized theory against cosmological `zoom-in' simulations run as part of the Feedback In Realistic Environments project\footnote{\url{https://fire.northwestern.edu/}}, using the second version of these simulations (FIRE-2). The simulation methods are described in detail in \cite{hopkins18}, while the main aspects are summarized here. 

The FIRE-2 simulations use the multi-method gravity and hydrodynamics code GIZMO\footnote{\url{http://www.tapir.caltech.edu/\~phopkins/Site/GIZMO.html}} (\citealt{Hopkins15}) in its meshless finite-mass mode (MFM). 
MFM is a Lagrangian, mesh-free, finite-mass method which combines advantages of traditional smooth particle hydrodynamics (SPH) and grid-based methods.
Gravity is solved using a modified version of the Tree-PM solver similar to GADGET-3 \citep{Springel05} but with adaptive softening for gas resolution elements. 
Radiative heating and cooling rates account for metal line cooling, free-free emission, photoionization and recombination, Compton scattering with the cosmic microwave background, collisional and photoelectric heating by dust grains, and molecular and fine-structure cooling at low temperatures ($<10^4\K$). 
The relevant ionization states are derived from precomputed {\sc cloudy} \citep{Ferland98} tables including the effects of the cosmic UV background from \cite{FaucherGiguere09} and local radiation sources.
Star formation occurs in self-gravitating, self-shielded molecular gas with $\nH>1000\cm^{-3}$ (\citealt{Hopkins13}). The sub-grid implementation of feedback processes from stars includes radiation pressure, heating by photoionization and photoelectric processes, and energy, momentum, mass, and metal deposition from supernovae (core collapse and Ia) and stellar winds. Feedback parameters and their time dependence are based on the stellar evolution models in \cite{Leitherer99} assuming a \cite{Kroupa01} initial mass function.

\subsection{Simulation selection}

\begin{deluxetable}{lcccccc}
\tablecolumns{7}
\tablecaption{FIRE-2 cosmological `zoom' simulations used in this work}
\label{t:sims}
\tablehead{
\colhead{Name} & \colhead{$\mb$} & \colhead{$z_{\rm min}$} & \colhead{$\Mhalo(z_{\rm min})$ } & \colhead{$\Mstar(z_{\rm min})$ } & \colhead{md } & \colhead{ Ref. } \\ 
\colhead{} & \colhead{ $[{\rm M}_\odot]$ } & \colhead{} & \colhead{ $[{\rm M}_\odot]$ } & \colhead{ $[{\rm M}_\odot]$ } & \colhead{} & \colhead{} \\ 
\colhead{(1)} & \colhead{(2)} & \colhead{(3)} & \colhead{(4)} & \colhead{(5)} & \colhead{(6)} & \colhead{(7)} 
}
\startdata
\multicolumn{7}{c}{\textbf{m11's}}\\
\texttt{m11b}  & 2100  & 0 & $0.4\cdot10^{11}$  & $1.2\cdot10^{8}$  & no  & A \\ 
\texttt{m11i}  & 7100  & 0 & $0.7\cdot10^{11}$  & $1.0\cdot10^{9}$  & yes & B \\ 
\texttt{m11e}  & 7100  & 0 & $1.6\cdot10^{11}$  & $1.7\cdot10^{9}$  & yes & B \\ 
\texttt{m11h}  & 7100  & 0 & $1.9\cdot10^{11}$  & $4.0\cdot10^{9}$  & yes & B \\ 
\texttt{m11v}  & 7100  & 0 & $2.6\cdot10^{11}$  & $5.8\cdot10^{9}$  & no  & C \\ 
\texttt{m11d}  & 7100  & 0 & $3.0\cdot10^{11}$  & $5.1\cdot10^{9}$  & yes & B \\
\multicolumn{7}{c}{\textbf{m12's}}\\
\texttt{m12z}  & 4200  & 0 & $0.8\cdot10^{12}$  & $2.5\cdot10^{10}$ & yes & D \\
\texttt{m12i}  & 7100  & 0 & $1.1\cdot10^{12}$  & $7.3\cdot10^{10}$ & yes & E \\
\texttt{m12b}  & 7100  & 0 & $1.3\cdot10^{12}$  & $1.0\cdot10^{11}$ & yes & D \\ 
\texttt{m12m}  & 7100  & 0 & $1.5\cdot10^{12}$  & $1.4\cdot10^{11}$ & no  & C \\ 
\texttt{m12f}  & 7100  & 0 & $1.6\cdot10^{12}$  & $1.0\cdot10^{11}$ & no  & F \\ 
\multicolumn{7}{c}{\textbf{m13's}}\\
\texttt{m13A1} & 33000 & 1 & $0.4\cdot10^{13}$  & $2.8\cdot10^{11}$ & no  & G \\ 
\texttt{m13A4} & 33000 & 1 & $0.5\cdot10^{13}$  & $2.7\cdot10^{11}$ & no  & G \\ 
\texttt{m13A2} & 33000 & 1 & $0.8\cdot10^{13}$  & $5.1\cdot10^{11}$ & no  & G \\ 
\texttt{m13A8} & 33000 & 1 & $1.3\cdot10^{13}$  & $8.0\cdot10^{11}$ & no  & G \\ 
\texttt{z5m13a} & 57000 &4.5& $0.4\cdot10^{13}$ & $2.0\cdot10^{11}$ & no  & H \\ 
\enddata
\tablecomments{
(1) galaxy name; 
(2) initial baryonic particle mass;
(3) final redshift of the simulation; 
(4) central halo mass at the final redshift;
(5) stellar mass of central galaxy at the final redshift; 
(6) whether a prescription for subgrid metal diffusion is included in the simulation;
(7) Reference papers for simulations:
A:~\cite{Chan18},
B:~\cite{elbadry18a},
C:~\cite{hopkins18},
D:~\cite{GarrisonKimmel18},
E:~\cite{Wetzel16}
F:~\cite{GarrisonKimmel17},
G:~\cite{anglesalcazar17b}, 
H:~\cite{ma18}.
}
\end{deluxetable}

\begin{figure*}
\includegraphics{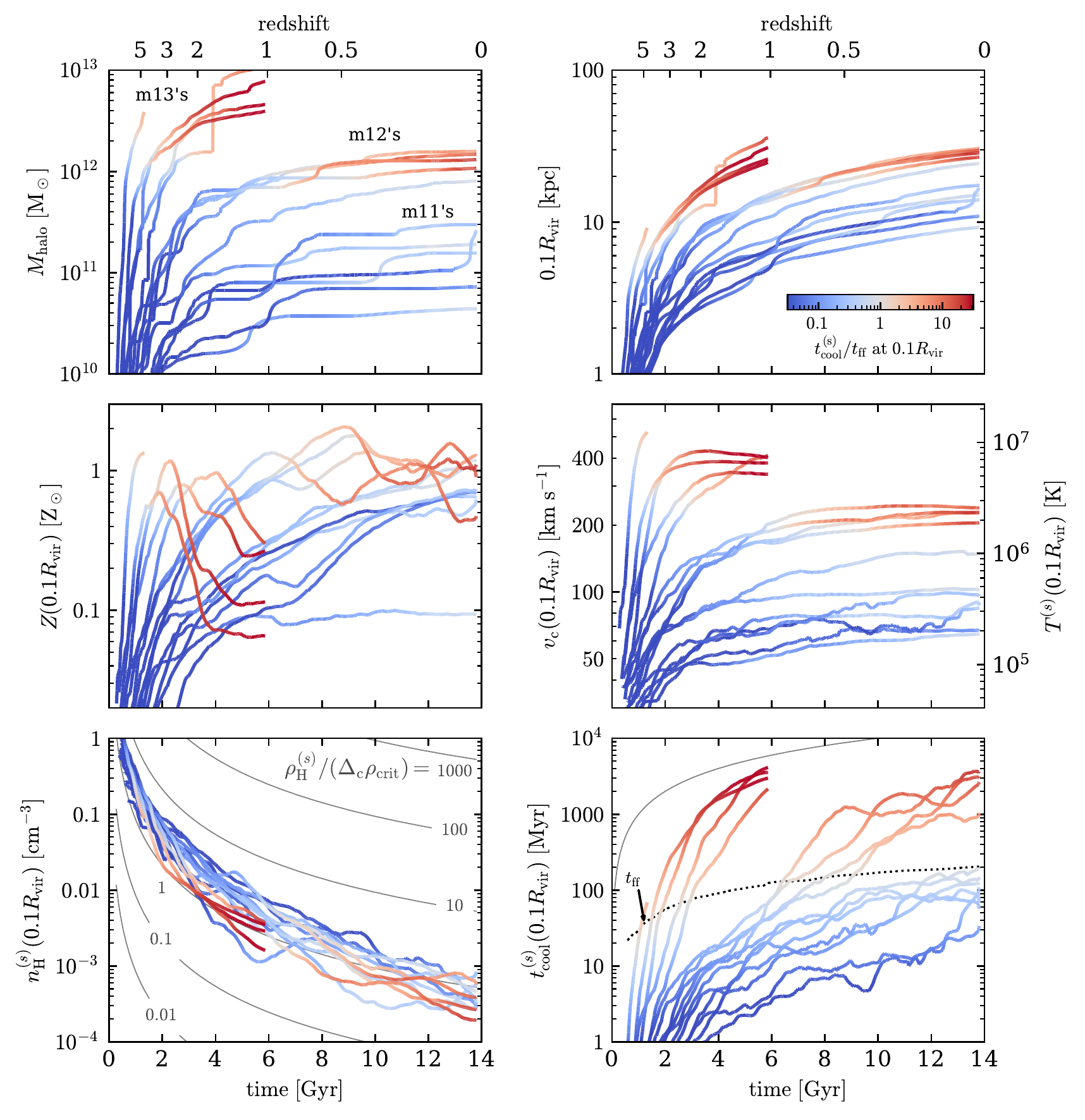}
\caption{
Gas and halo properties versus time in the 16 FIRE zoom simulations used in this work. 
\textbf{(Top-left)} Halo mass. The simulations are divided into three groups of similar mass assembly histories as marked in the panel and in Table~\ref{t:sims}. 
\textbf{(Top-right)} $10\%$ of the virial radius, at which the properties in the four bottom panels are estimated. 
\textbf{(Middle-left)} Mass-weighted gas metallicity. 
\textbf{(Middle-right)} Circular velocity (left axis) and expected temperature if the CGM virializes (right axis, eqn.~\ref{e:Tc} with $A=1$). 
\textbf{(Bottom-left)} 
Expected gas density if the CGM virializes (eqn.~\ref{e:pressure support}). Contours mark the ratio of this density to the mean halo mass density $\Deltac\rhocrit$. 
\textbf{(Bottom-right)} Expected cooling time if the CGM virializes, calculated from the properties shown in the other panels (eqn.~\ref{e:tcool0}). A comparison of $\tcoolsh$ with other ways of evaluate the cooling time is presented in Appendix \ref{a:density}. 
The dotted line marks the median free-fall time, where in individual simulations $\tff$ is within a factor of two of the median. The thin solid line marks the Universe age. 
In all panels line color denotes $\tcoolsh/\tff$ with the colorbar shown in the top-right panel. 
}
\label{f:properties}
\end{figure*}

The subset of FIRE-2 simulations analyzed in this work are listed in Table~\ref{t:sims}. They span a broad range in the mass of the central halo and how it evolves with time, as shown in the top-left panel of Fig.~\ref{f:properties}. The simulations include six ``m11's'' where the main halo at $z=0$ has a mass of $\sim10^{11}\msun$ and hosts a dwarf galaxy, five ``m12's'' with $\Mhalo\sim10^{12}\msun$ and an $\sim$$L^*$ galaxy at $z=0$, and five ``m13's'' in which $\Mhalo$ exceeds $10^{12}\msun$ at high redshift ($2<z<6$)\footnote{The initial conditions for halos m13A1 -- m13A8 \citep{anglesalcazar17b} are identical to that of halos A1 -- A8 simulated previously with the FIRE-1 model \citep{Feldman17}.  The prefix `m13' is added here to distinguish these massive galaxies from the other simulation subgroups.}.
We use this diverse range of halo mass assembly histories to demonstrate that for a given $\tcoolsh/\tff$ the virialization of the inner CGM is independent of redshift. 

In the m11 and m12 simulations the initial mass of a baryonic resolution element is $\mb=7100\msun$ or better. This implies that sub-grid physics are applied at the giant molecular cloud level (or better), while $\sim$$L^\star$ galaxy discs are well-resolved with $\sim10^6$ particles. The more massive m13 simulations have a mass resolution of $\mb=33\,000-57\,000\msun$. Implications of resolution for our results are discussed in appendix~\ref{a:resolution}. 

A subset of the simulations also include a mechanism for subgrid metal diffusion between neighboring resolution elements as described in \cite{Hopkins17} and \cite{Escala18}, though as shown in appendix~\ref{a:resolution} the inclusion of this prescription does not appear to affect our conclusions.

\subsection{Cooling and free-fall times in FIRE}\label{s:tcool in FIRE}

In this section we describe how we estimate $\tcoolsh$ and $\tff$ in the FIRE simulations. The ratio of these two timescales is expected to determine whether the volume-filling phase of the CGM has virialized into a hot and subsonic medium (\S\ref{s:theoretical bkg}). 

For each snapshot in the simulation, we use the Amiga Halo Finder (\citealt{Knollmann09}) to identify the center, virial mass $\Mvir$, and virial radius $\Rvir$ of the main halo, utilizing the virial overdensity definition of \cite{bryan98}. We calculate $\vc(r)$ from $M(<r)$ using eqn.~(\ref{e:vc}), where $M(<r)$ is the sum of all types of resolution elements (gas, stars, dark matter) with centers within $r$. We then use $\vc$ and eqn.~(\ref{e:tff}) to derive $\tff$. 

Calculating $\tcoolsh(r)$ in the simulations requires calculating $\Tc$, $\nHc$, and $\Lambda$ (see eqn.~\ref{e:tcool0}). For $\Tc$ we use eqn.~(\ref{e:Tc}) and the estimated $\vc(r)$. 
For $\nHc$ we use eqn.~(\ref{e:pressure support}), i.e.~we evaluate the weight of the overlying gas to estimate the expected thermal pressure in a virialized CGM, and then divide by $\Tc$ to get $\nHc$. To compute $\rho$ in the integrand in eqn.~(\ref{e:pressure support}) we divide the gas in each snapshot into radial shells with width of $0.05\dex$, and divide the total mass of all gas resolution elements within the shell by the shell volume. The integration is then carried out from the desired $r$ out to $\Rvir$. 
Since the weight of the overlying gas (which enters in the integrand) is typically largest at the smallest radii in the integration range, the exact choice of outer integration limit does not significantly affect the result. 
The values of $\Tc$ and $\nHc$ we derive are comparable to the actual mean temperatures and densities in the simulation after the inner CGM virializes, though they can differ substantially prior to virialization (see \S\ref{s:temperature} and appendix~\ref{a:density}). We derive $\Lambda$ from the \cite{Wiersma09} tables, using $\Tc$, $\nHc$, $z$, and the mass-weighted metallicity of gas in a shell with width $0.05\dex$ around $r$. Our calculation of $\Lambda$ thus assumes that the metals at each radius are uniformly distributed at that radius, as assumed in the spherical steady-state solutions in \S\ref{s:theoretical bkg}.

Estimating $\Rcirc$ (eqn.~\ref{e:Rcirc}) in FIRE is not straightforward. 
In \S\ref{s:implications} below we show that only after the inner CGM virializes there is a well-defined radius where the gas circularizes, with typical values in the range $\Rcirc \sim 0.02-0.07\Rvir$. 
Therefore to estimate condition~(\ref{e:threshold condition}) uniformly at all epochs we instead evaluate $\tcoolsh/\tff$ at $0.1\Rvir$, i.e.\ condition~(\ref{e:threshold condition}) becomes
\begin{equation}\label{e:threshold condition actual}
 \tcoolsh(0.1\Rvir) = \ft'\tff(0.1\Rvir) ~,
\end{equation}
where $\ft'$ is expected to be somewhat larger than $\ft$ (since $0.1\Rvir>\Rcirc$ and $\tcoolsh/\tff$ increases outwards, see eqn.~\ref{e:tratio}). 
Using a constant fraction of $\Rvir$ rather than $\Rcirc$ has the advantages of being conceptually simpler and independent of jitter in the specific angular momentum profile. Also, using a radius which is a factor of $\sim2$ larger than $\Rcirc$ avoids the unwanted increase of our CGM density estimate by gas in the disc.\footnote{In a small number of snapshots the estimated $\nHc$ at $0.1\Rvir$ is more than an order of magnitude larger than at $0.2\Rvir$, due to the gas disc and associated high densities extending beyond $0.1\Rvir$. To avoid this effect in our estimate of the halo gas density, we limit $\nHc(0.1\rvir)$ to no more than $b\cdot\nHc(0.2\rvir)$ with $b=8$. The exact value of $b$ does not affect our results. }
The substitution of $\Rcirc$ with $0.1\Rvir$ however implies that we expect scatter in $\ft'$ due to halo-to-halo variance in $\Rcirc$. 

We experimented also with other choices for the radius at which to estimate $\tcoolsh/\tff$, including the maximum radius where $j$ exceeds some order-unity factor of $\vc r$, and the radius where $\vc r$ equals the average specific angular momentum of all gas at $0.1-1 \Rvir$. Up to order-unity differences in the derived $\ft'$, these alternative choices yielded similar results to estimating $\tcoolsh/\tff$ at $0.1\Rvir$.

\begin{figure}
\includegraphics{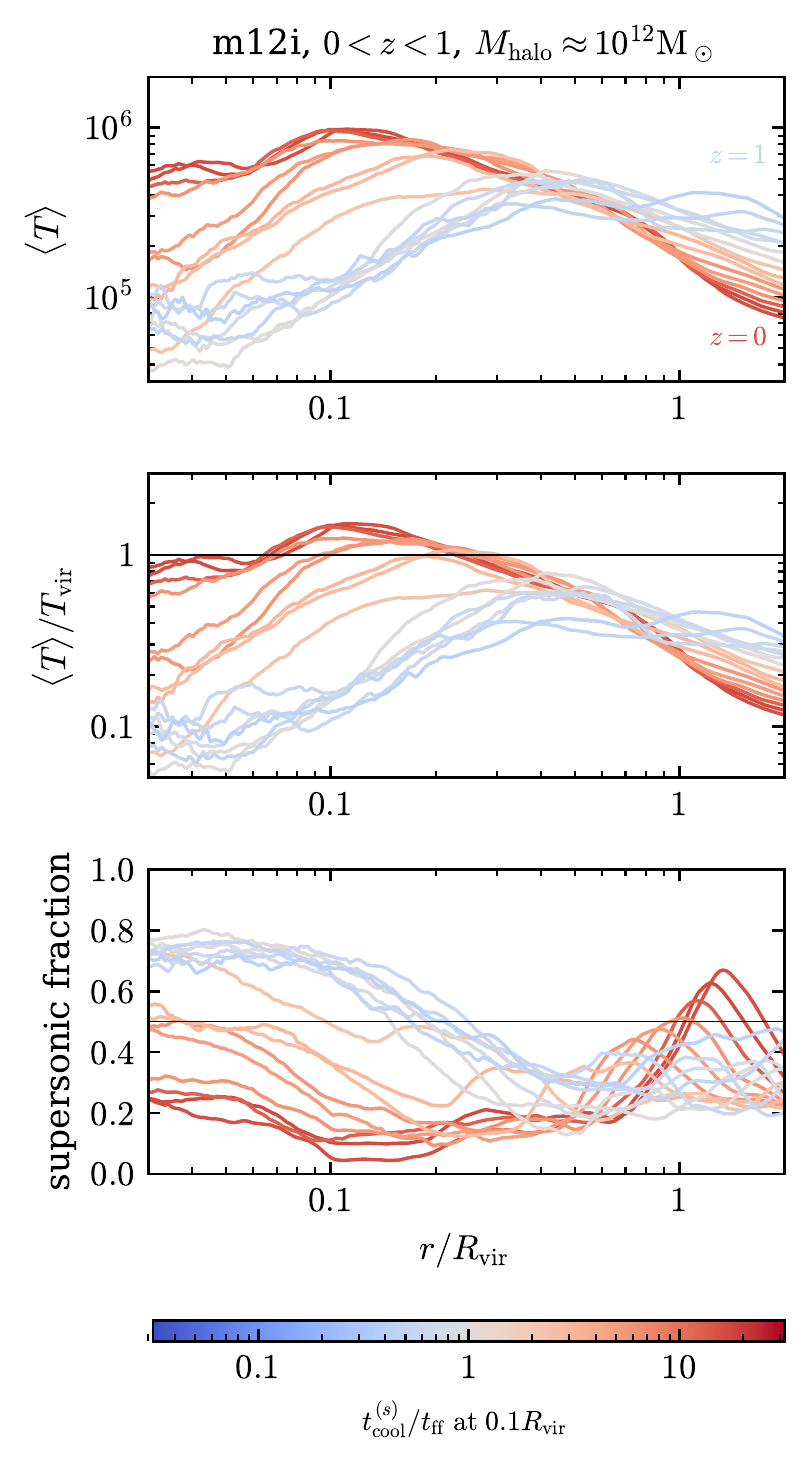}
\caption{\textbf{Top:} Profiles of volume-weighted gas temperature (eqn.~\ref{e:T}) in the main halo of the m12i simulation. Each line corresponds to the median profile of snapshots in a $\Delta t=0.5\Gyr$ window, starting from $z=1$ and down to $z=0$. Line color marks the median $\tcoolsh/\tff$ at $0.1 R_{\rm vir}$, which substantially increases from $0.3$ at $z=1$ to $16$ at $z=0$, in contrast with the mild increase in $\Mhalo$ from $0.9\cdot10^{12}\msun$ to $1.1\cdot10^{12}\msun$. 
\textbf{Middle:} Median temperature profiles normalized by $\Tvir$. 
Note the factor of ten increase in temperature at $0.1\Rvir$ starting when $\tcoolsh$ exceeds $\tff$. In contrast, the change in $\langle T\rangle/\Tvir$ at $\sim 0.5\rvir$ is mild, while at $r\gtrsim\rvir$ the trend is reversed and the temperature decreases with increasing $\tcoolsh/\tff$. 
\textbf{Bottom:} the volume filling fraction of gas with supersonic radial velocities (either inflows or outflows). At low redshifts where $\tcoolsh\gg\tff$ (dark red lines) the volume is dominated by subsonic gas from galaxy scales out to the accretion shock at $\approx\Rvir$. In contrast at higher redshifts at which $\tcoolsh\lesssim\tff$ (blue-gray lines) the gas is mainly supersonic within $\approx0.2\Rvir$ and mainly subsonic outside $\approx0.2\Rvir$.
}
\label{f:Tprofiles m12i}
\end{figure}

\begin{figure}
\includegraphics{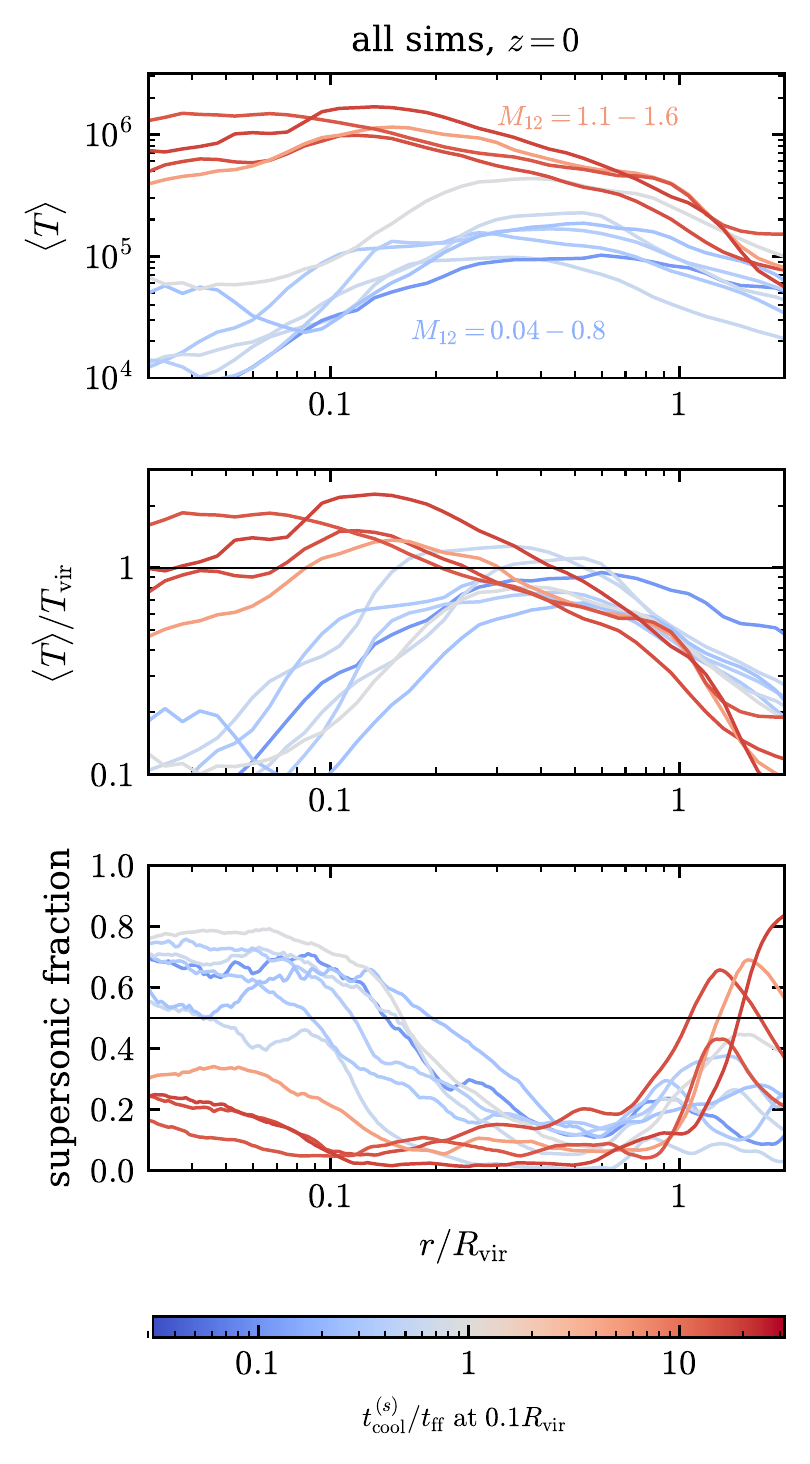}
\caption{\textbf{Top:} Profiles of volume-weighted temperature (eqn.~\ref{e:T}) in different halos at $z=0$. Each line corresponds to the profile of a different simulation. 
Line color marks the median $\tcoolsh/\tff$ at $0.1 R_{\rm vir}$. Noted in the panel is the range of halo mass spanned by halos with $\tcoolsh/\tff\lesssim1$ (blue-gray lines, includes the m11's and m12z simulations) and by halos with $\tcoolsh/\tff\gg1$ (red lines, m12's excluding m12z). 
\textbf{Middle:} Median temperature profiles normalized by $\Tvir$. As in Fig.~\ref{f:Tprofiles m12i}, note the large difference in temperature at $0.1\Rvir$ between halos with $\tcoolsh\gg\tff$ and halos with $\tcoolsh\lesssim\tff$, in contrast with the lack of difference at $0.5\Rvir$. 
\textbf{Bottom:} Volume filling fraction of supersonic gas versus radius. As in Fig.~\ref{f:Tprofiles m12i}, in halos where $\tcoolsh\gg\tff$ the volume is dominated by subsonic gas from galaxy scales out to the accretion shock at $\approx\Rvir$. In contrast, in halos with $\tcoolsh\lesssim\tff$ the gas is predominantly supersonic within $0.1-0.2\Rvir$ and predominantly subsonic at larger radii.
}
\label{f:Tprofiles z 0}
\end{figure}

\begin{figure*}
\includegraphics{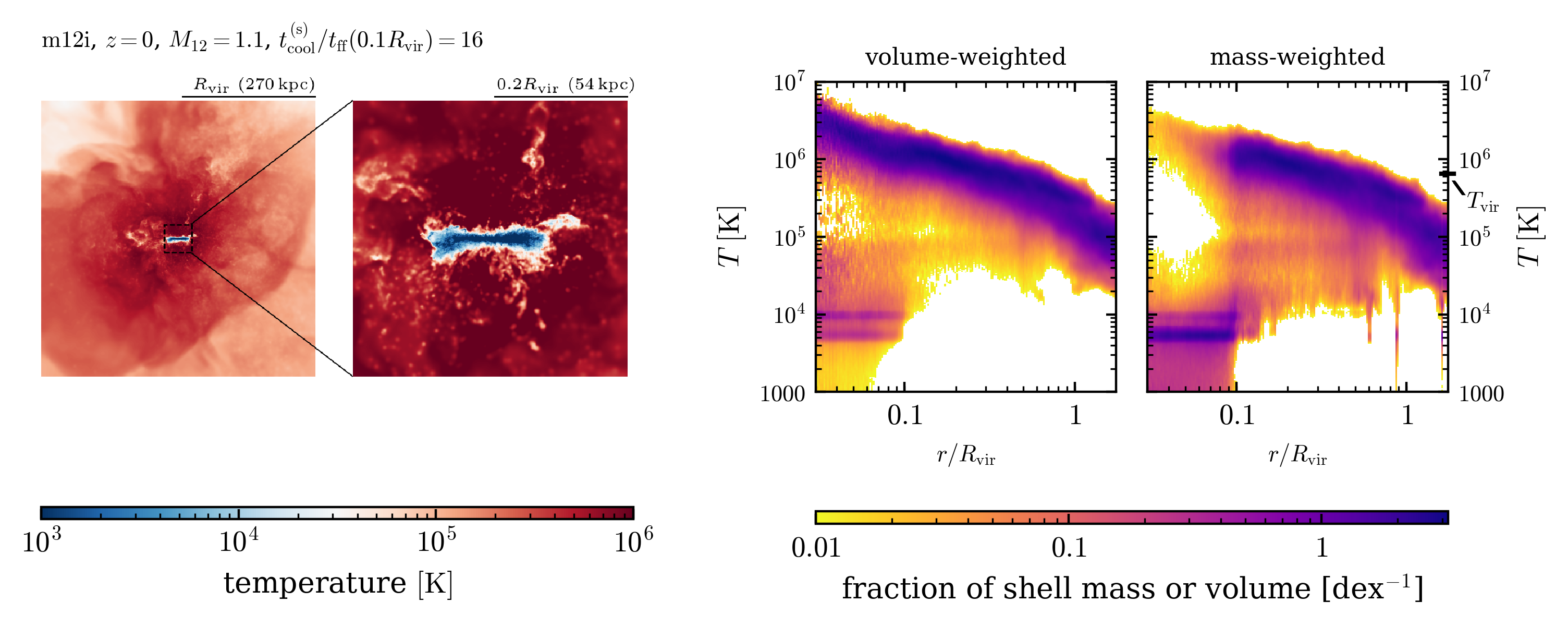}
\caption{Temperature distribution in the main halo of m12i at $z = 0$, at which $\tcoolsh/\tff=16$ at $0.1\rvir$. \textbf{Left panels:} Temperature map of a slice through the snapshot, oriented edge-on to the galaxy disc. The first panel spans the entire halo and the second panel zooms in on the inner $\pm0.2\rvir$. \textbf{Right panels:} Temperature histograms of gas in shells at different radii, weighted by volume (third panel from the left) and weighted by mass (rightmost panel). Note that the hot phase temperature decreases from $T\gtrsim10^6\K$ at galaxy radii to $T\approx10^{5.5}\K$ at $\approx\rvir$, and dominates by volume both on halo and on galaxy scales, and by mass in the halo. The cool phase forms a prominent disc on galaxy scales and is entirely negligible by volume in the halo.
}
\label{f:m12i image}.
\end{figure*}

\begin{figure*}
\includegraphics{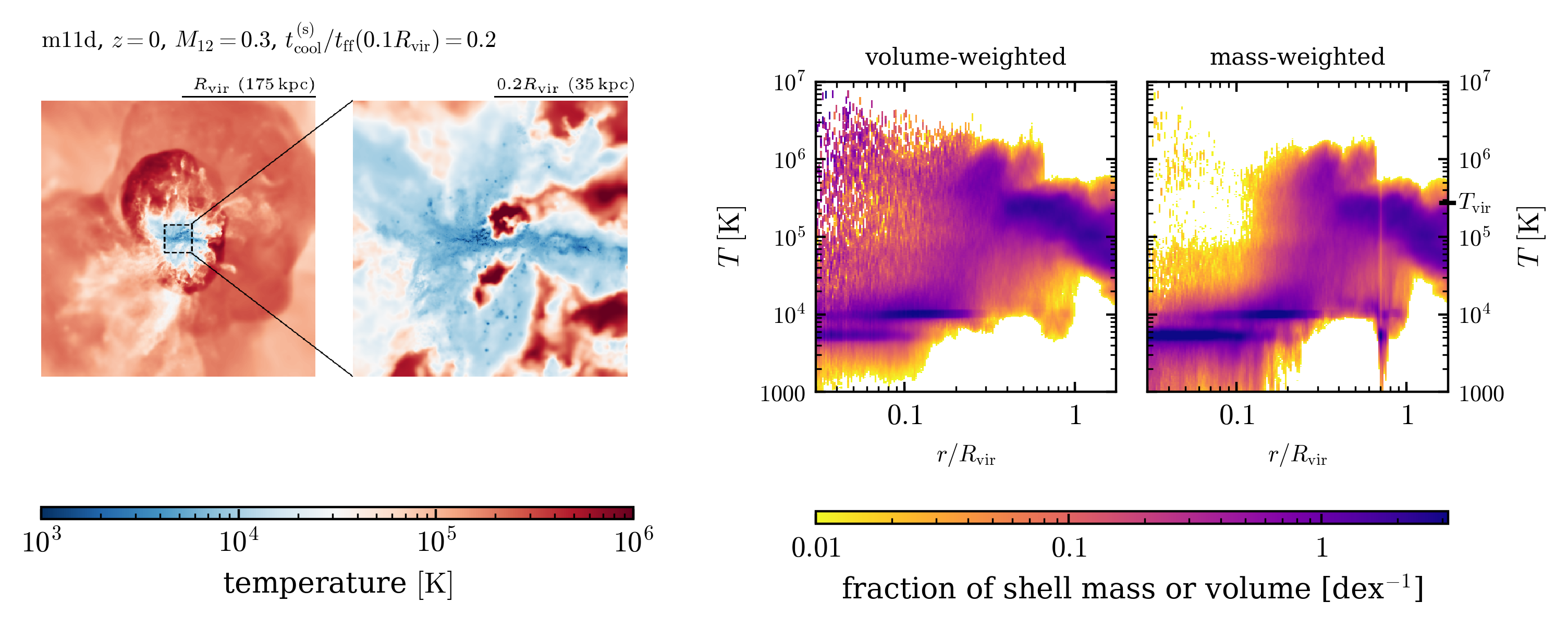}
\caption{Similar to Fig.~\ref{f:m12i image} for m11d at $z = 0$, in which $\tcoolsh/\tff = 0.2$ at $0.1\rvir$. The images on the left are oriented such that the total angular momentum vector is oriented upwards. 
The hot phase dominates the volume in the outer halo as in the m12i simulation shown in Fig.~\ref{f:m12i image}. However, in the inner halo ($\lesssim0.3\rvir$) and on galaxy scales the cool $\lesssim10^4\K$ phase dominates the \emph{volume}, in contrast with Fig.~\ref{f:m12i image}. Note also
the lack of a prominent disc, in contrast with Fig.~\ref{f:m12i image}.
}
\label{f:m11d image}.
\end{figure*}

\begin{figure}
\includegraphics{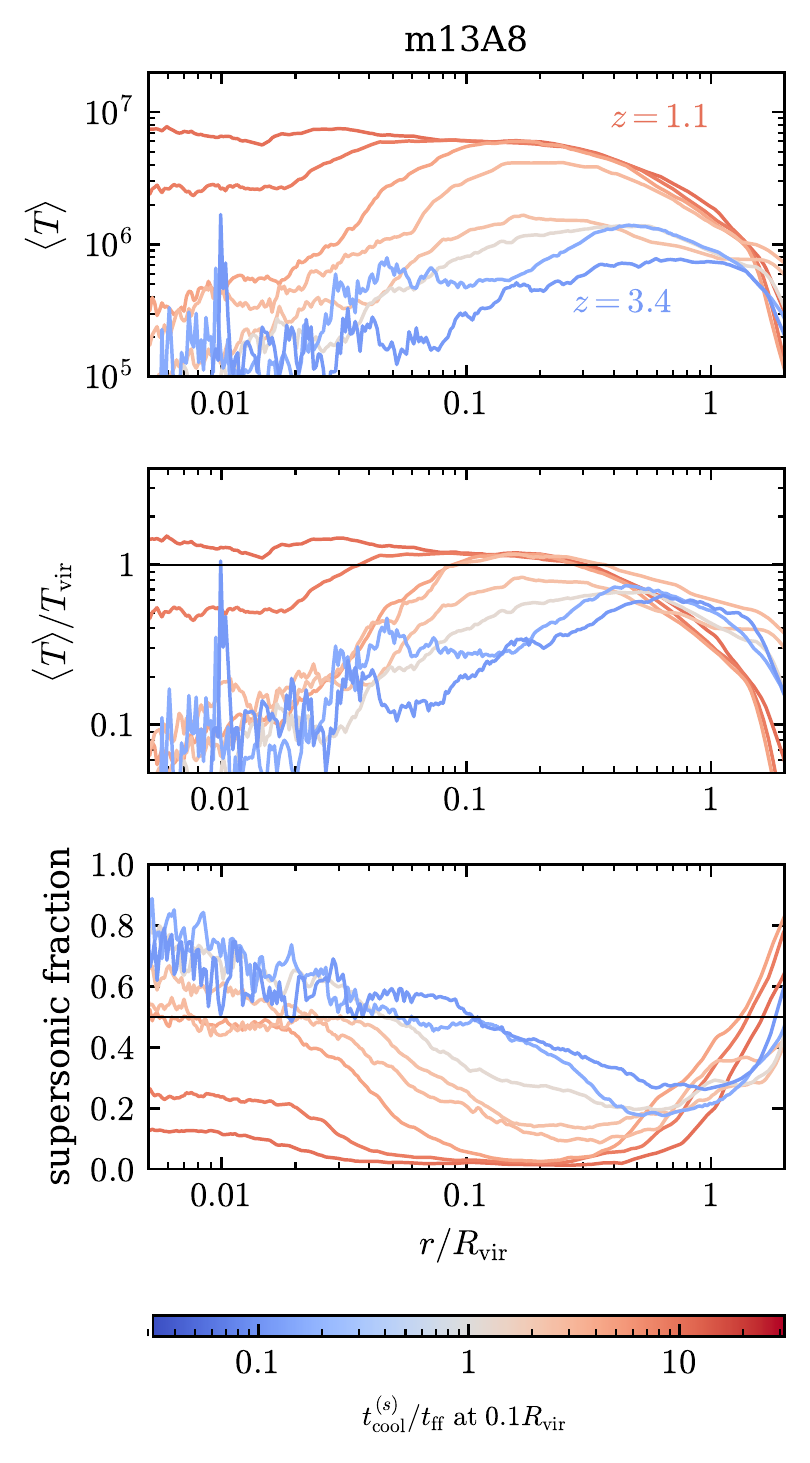}
\caption{Similar to Fig.~\ref{f:Tprofiles m12i} for m13A8, in which $\tcoolsh$ exceeds $\tff$ at high redshift. 
\textbf{Top panels:} 
Profiles of volume-weighted temperature (eqn.~\ref{e:T}). Each line corresponds to the median profile of snapshots in a $\Delta t=0.5\Gyr$ window, where line color marks the median $\tcoolsh/\tff$ at $0.1 R_{\rm vir}$. The shown redshift range spans $3.4$ to $1.1$. Over this period $\tcoolsh/\tff$ increases from $0.1$ to $10$ and the halo mass increases from $0.8\cdot10^{12}\msun$ to $10^{13}\msun$. 
\textbf{Middle panels:} 
Volume-weighted temperature profiles normalized by $\Tvir$. 
Note the large increase with time in $\Tmed/\Tvir$ at small radii compared to the roughly constant $\Tmed/\Tvir$ at $0.5\Rvir$.
\textbf{Bottom:} Volume filling fraction of supersonic gas versus radius. 
When $\tcoolsh\lesssim\tff$ the gas is predominantly subsonic at large radii and predominantly supersonic at small radii. When $\tcoolsh\gg\tff$ the volume is dominated by subsonic gas from galaxy scales out to $\approx\Rvir$. }
\label{f:Tprofiles h2}
\end{figure}

\begin{figure*}
\includegraphics{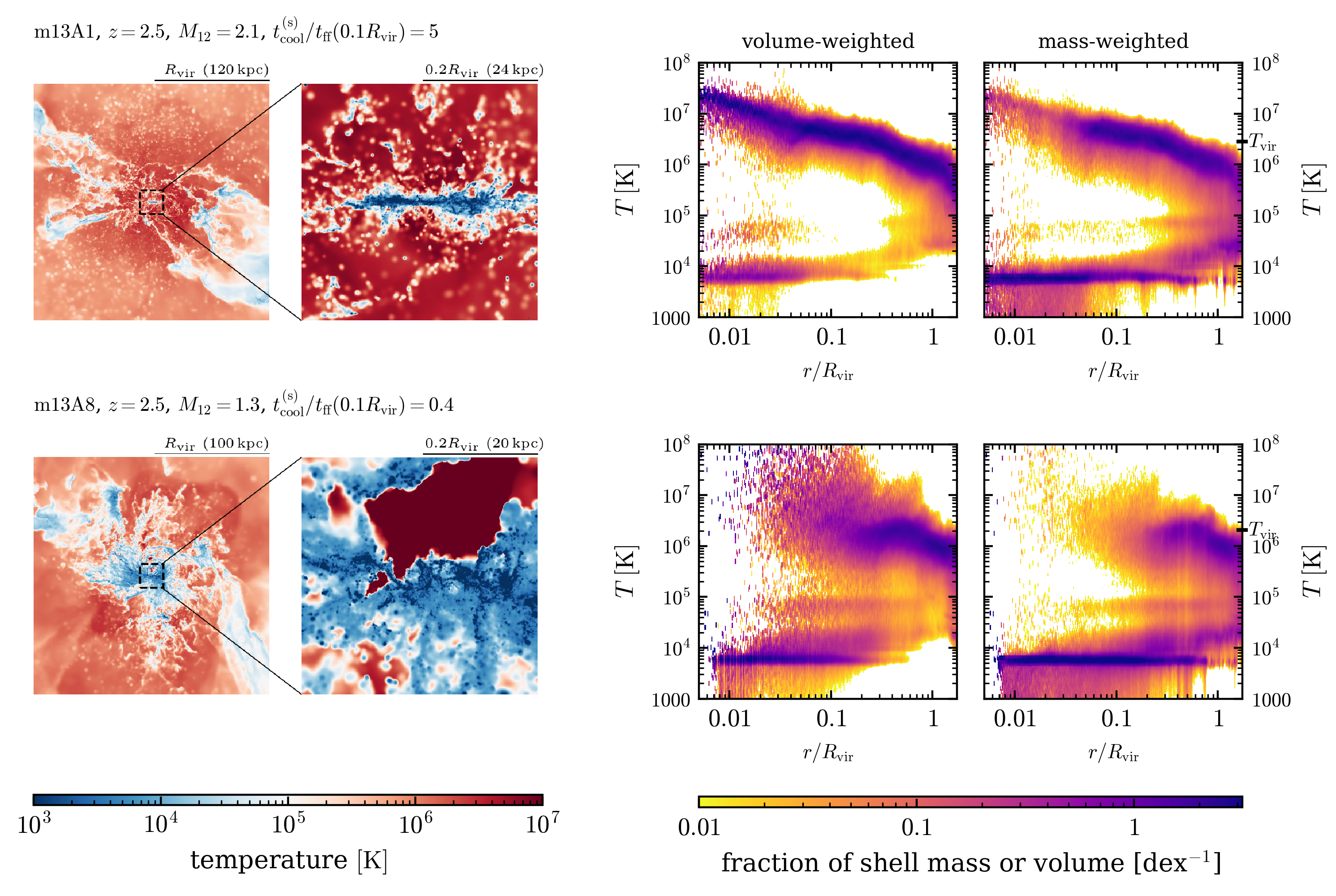}
\caption{Similar to Figs.~\ref{f:m12i image} and \ref{f:m11d image} for $z=2.5$ snapshots of the m13A1 (top row) and m13A8 (bottom row) simulations. In m13A1 the cooling time exceeds the free-fall time at $0.1\Rvir$, while in m13A8 the cooling time is shorter than the free-fall time. 
Images on the left show temperature maps of a slice through the snapshot, oriented so the angular momentum vector of galaxy gas is directed upward. The right panels show temperature histograms of gas in shells at different radii, weighted by volume (third panels from the left) and weighted by mass (rightmost panels). 
In m13A1 (top) gas in the halo is separated into a hot phase and cool filaments. The hot phase dominates the volume both at halo radii and at galaxy radii. 
In m13A8 (bottom) the hot phase dominates the volume in the outer halo, but is sub-dominant in the inner halo and at galaxy radii. Note the prominent disc in m13A1, and its absence in m13A8. 
}
\label{f:h206 h2 image}
\end{figure*}

Figure~\ref{f:properties} plots the results of the above estimates versus time in the 16 FIRE simulations. The top panels show $\Mhalo$ and $0.1\Rvir$, while the bottom four panels show $Z$, $\vc$, $\nHc$, and $\tcoolsh$ measured at $0.1\Rvir$ and smoothed with a ${\rm Gyr}$-wide boxcar to avoid clutter. The value of $\Tc$ is shown in the right-axis of the $\vc$ panel. In the bottom-right panel we also plot the median value of $\tff$, based on the median of all simulations run at the relevant redshift. In individual simulations $\tff$ is within a factor of two of the median so for clarity individual $\tff$ curves are not shown. The curves in Fig.~\ref{f:properties} are colored by the ratio $\tcoolsh/\tff$, as noted in the colorbar in the top-right panel. 

Several trends are apparent in Fig.~\ref{f:properties}. 
The bottom right panel shows that $\tff$ increases rather slowly with time, as expected from eqn.~(\ref{e:tff}). 
In contrast $\tcoolsh$ increases significantly faster, so a typical simulation spans more than three orders of magnitude in $\tcoolsh/\tff$. The increase in $\tcoolsh$ is a result of the decrease in $\nHc$ and increase in $\vc$, while the general increase in $Z$ with time slows the increase in $\tcoolsh$ (see eqn.~\ref{e:tcool}). 
The value of $\nHc$ at $0.1\Rvir$ is naively expected to scale with the mean halo density $\Deltac\rhocrit$ (see eqn.~\ref{e:rhoc}), though in practice they decrease somewhat faster, from typical ratios $\rhoc/\Deltac\rhocrit$ of $1.2-4$ at $z\gtrsim2$, to $0.4-1$ at $z=0$  (see bottom-left panel). Note also that when $\tcoolsh$ exceeds $\tff$ the trend in metallicity reverses and the metallicity starts to decrease with time, suggesting a change in physical conditions in the CGM. We show below that this is likely due to the suppression of outflows once the inner CGM virializes, causing low-metallicity inflowing gas to reduce the overall CGM metallicity. 

The top-left panel of Fig.~\ref{f:properties} shows that for a fixed halo mass $\tcoolsh/\tff$ at $0.1\Rvir$ tends to increase with time.  This is evident in $\sim 10^{12}\msun$ halos where $\tcoolsh\sim \tff$ at high redshift in contrast with $\tcoolsh\gg\tff$ at $z\sim0$, and also in $\sim 10^{11}\msun$ halos where $\tcoolsh\ll\tff$ at high redshift in contrast with $\tcoolsh\sim\tff$ at $z\sim0$. 
The increase in $\tcoolsh/\tff$ with time at fixed halo mass is mainly due to the decrease in $\rhoc/\Deltac\rhocrit$ seen in the bottom-left panel, and due to the increase in $\fvc=\vc(0.1\Rvir)/\vc(\Rvir)$ as a result of the larger concentration of low-redshift halos. We further address this trend in the discussion.

\section{Inner CGM virialization in FIRE}\label{s:virialization}

In this section we demonstrate that the inner CGM virializes in FIRE when $\tcoolsh$ exceeds $\tff$ at $0.1\Rvir$, and that this transition in the inner CGM occurs after the outer CGM has virialized. As discussed in \S\ref{s:theoretical bkg}, we define a `virialized CGM' as one that has a volume-filling phase with temperature $\sim\Tvir$ and subsonic dynamics.

\subsection{Thermal and dynamic CGM properties versus $\tcoolsh/\tff$}\label{s:temperature}

To identify the virialization of the volume-filling CGM phase we measure the volume weighted-temperature in radial shells with centers $r$ and thickness $\Delta \log r=0.05\dex$, calculated via
\begin{equation}\label{e:T}
 \log\ \langle T(r)\rangle \equiv \frac{\sum V_i\log T_i  }{\sum V_i }
\end{equation}
where $T_i$ and $V_i=m_i/\rho_i$ are the temperature and volume of resolution element $i$, with $m_i$ and $\rho_i$ its mass and density. The summations in eqn.~(\ref{e:T}) are over all resolution elements whose centers are within the shell. This weighting by volume deemphasizes the effects of satellites and filaments and focuses on the properties of the volume-filling phase. 
We average the logarithm of the temperature in order to give similar weights to a hot phase with $T\approx\Tvir$ and a cool phase with $T\approx10^4\K$, in contrast with a linear average which would overweight the hot phase relative to the cool phase.

The top panel of Figure~\ref{f:Tprofiles m12i} shows the volume-weighted temperature profiles in the last eight Gyr of the m12i simulation ($0<z<1$). Each curve plots the median temperature profile of all snapshots in a $\Delta t=0.5\Gyr$ window (individual snapshots are separated by $\approx25\Myr$). Taking the median reduces the variability induced by transient CGM heating events due to outflows and allows us to focus on the time-steady effects of virialization \citep[see][]{vandeVoort16}. The distribution of temperatures in individual snapshots is discussed below. 
We color each line in Fig.~\ref{f:Tprofiles m12i} according to the median $\tcoolsh/\tff$ at $0.1 R_{\rm vir}$ during the time window, using the same color scheme as in Fig.~\ref{f:properties}. This ratio increases by a factor of $50$ over the plotted time period, from $0.3$ at $z=1$ to $16$ at $z=0$. For comparison, the halo mass increases only mildly over this time from $0.9\cdot10^{12}\msun$ to $1.1\cdot10^{12}\msun$. 
The middle panel shows the temperature profiles normalized by $\Tvir$.  The figure shows that in the $z=0$ snapshot where $\tcoolsh\gg\tff$ (reddest curve) the volume-weighted temperature profile decreases from $T\approx1.5\Tvir=10^6\K$ at $0.1\Rvir$ to $\approx0.3\Tvir=2\cdot10^5\K$ at $\Rvir$. At $\gtrsim\Rvir$ the temperature profile tends to steepen, while at galaxy radii ($<0.1\Rvir$) the volume weighted temperature is somewhat lower than at $0.1\Rvir$ due to the cool gas disc (see below). 
At higher redshift when $\tcoolsh\lesssim\tff$ (blue-gray curves) the temperature is significantly lower at $0.1\Rvir$, roughly equal to $0.1\Tvir$, indicating that a time-steady volume-filling phase with $T\approx\Tvir$ has not formed in the inner CGM, i.e.~the inner CGM has not virialized. In contrast $\Tmed/\Tvir$ at $\gtrsim0.5\Rvir$ in all shown snapshots is similar to that at $z=0$. If one moves forward in time from blue-gray curves to red curves, the temperature profiles within $\Rvir$ tend to join the $z=0$ profile first at large halo radii and later at small halo radii, indicating that virialization proceeds from the outside-in. 

As another estimate of CGM virialization, we measure the fraction of the volume in each shell with supersonic radial velocities, either inflowing or outflowing (again, taking the median of snapshots in a $\Delta t=0.5\Gyr$ window). As discussed in \S\ref{s:theoretical bkg} and in \cite{Fielding17}, in a virialized CGM we expect most of the volume to have subsonic radial velocities, i.e.~the kinetic energy should be subdominant to the thermal energy. In contrast, prior to virialization we expect a significant fraction of the volume to be supersonic, i.e.~the kinetic energy should dominate. This supersonic fraction profile is shown in the bottom panel of Fig.~\ref{f:Tprofiles m12i}. At low redshifts where $\tcoolsh\gg\tff$ the volume is dominated ($>70\%$) by subsonic gas from galaxy scales out to $\approx0.8\Rvir$. At $\approx\Rvir$ a relatively sharp increase in the supersonic fraction is evidence for an accretion shock. Thus, at $z=0$ the halo gas in m12i is almost entirely subsonic, i.e.~to zeroth-order supported against gravity by thermal pressure. In contrast at higher redshifts at which $\tcoolsh\lesssim\tff$ at $0.1\Rvir$ the gas is predominantly supersonic at small radii and predominantly subsonic at larger radii, i.e.~the halo gas is `transonic'. This panel therefore also supports an outside-in virialization scenario, since the outer CGM is subsonic before the inner CGM becomes subsonic. 

Another interesting result of Fig.~\ref{f:Tprofiles m12i} is the temperature at $r>\Rvir$, i.e.\ outside the halo. At these large radii there appears to be a mild reverse trend relative to the trend at $0.1\Rvir$, with the temperature \emph{decreasing} with time as $\tcoolsh/\tff$ at $0.1\Rvir$ increases. This trend is apparent even if we plot the radius in physical units rather than as a fraction of $\Rvir$. 
The bottom panel shows that this decrease in temperature at $r\gtrsim\Rvir$ is associated with a prominent accretion shock forming at $\sim\Rvir$ after the inner CGM virializes (the causal relationships are not addressed here), similar to a classic virial shock \cite[e.g.,][]{Birnboim03}. Note however that prior to the formation of this shock the kinematics are subsonic at all radii outside $\gtrsim0.2\rvir$ in the FIRE simulations, in contrast with the idealized simulations of \citeauthor{Birnboim03} in which the kinematics are supersonic at all radii prior to shock formation. In the discussion we address the possible origin of this reverse trend outside the halo.

\begin{figure*}
\includegraphics[width=\textwidth]{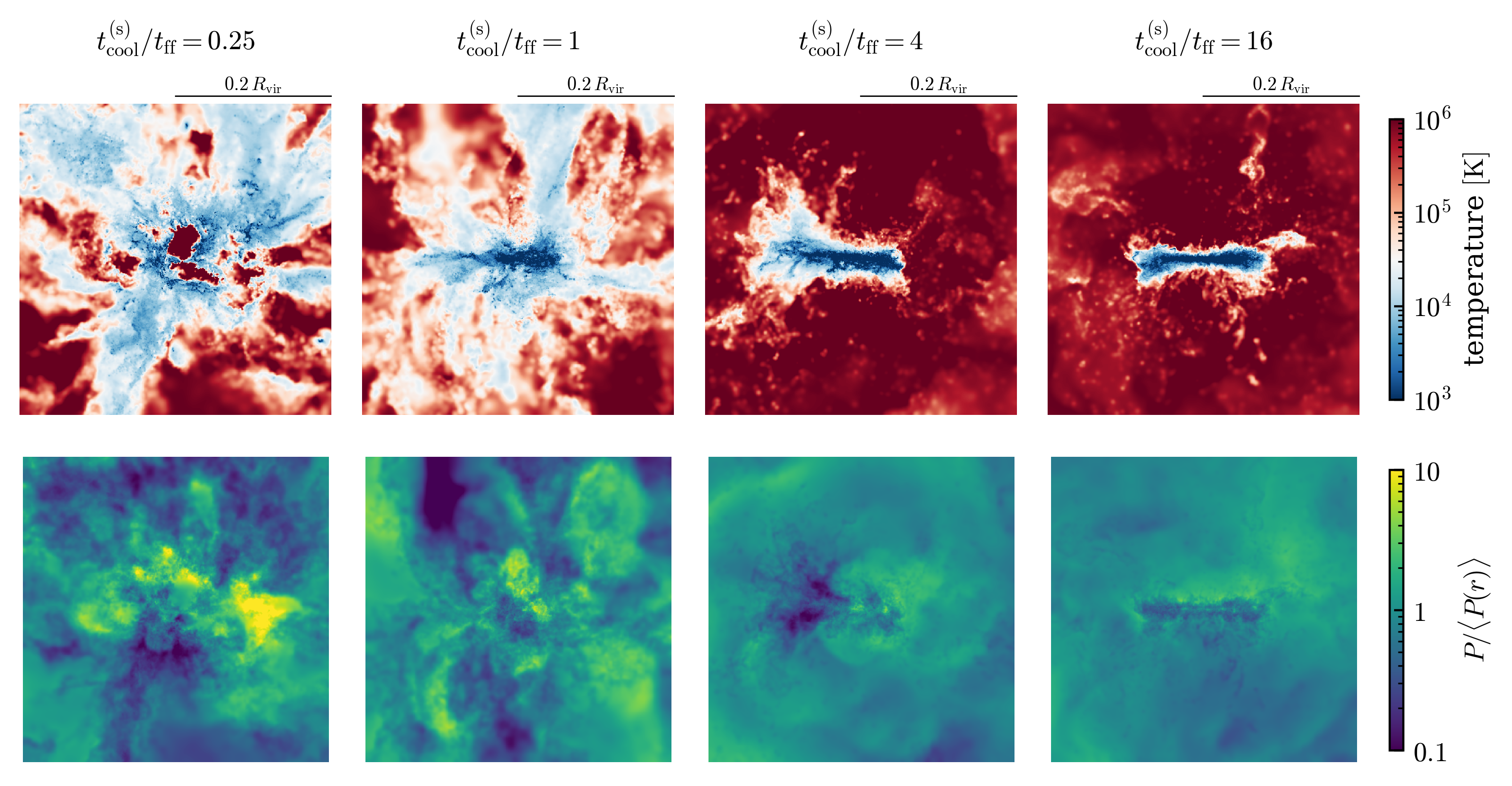}
\caption{Temperature (top) and pressure fluctuations (bottom) in the inner halo versus the cooling time to free-fall time ratio. The panels show snapshots of the m12i simulation at $z=1, 0.4, 0.1$ and $0$ from left to right. The corresponding $\tcoolsh/\tff$ at $0.1\Rvir$ are noted on top. In the bottom row the pressure at each pixel is normalized by the average pressure at the same radius (eqn.~\ref{e:nHT}). The images are oriented such that the angular momentum vector of galaxy gas is directed upward. Note that when $\tcoolsh$ exceeds $\tff$ the inner halo volume becomes uniformly hot, pressure fluctuations decrease, and a prominent disc appears.
}
\label{f:m12i_multiple_images}
\end{figure*}

Figure~\ref{f:Tprofiles z 0} compares the volume-weighted temperature profiles of the $z=0$ snapshots in the m12 and m11 simulations. As in Fig.~\ref{f:Tprofiles m12i} the panels show from top to bottom temperature, normalized temperature and supersonic fraction, while the curves are colored by $\tcoolsh/\tff$ at $0.1\Rvir$. The blue-grayish curves include the m11 subgroup and m12z, with halo masses spanning the range $4\cdot10^{10}-8\cdot10^{11}\msun$, while the red curves include the remaining four m12's with halo masses spanning the range $1.1-1.6\cdot10^{12}\msun$. The trends of temperature and supersonic fraction versus $\tcoolsh/\tff$ are similar to those seen in Fig.~\ref{f:Tprofiles m12i}. The value of $\Tmed/\Tvir$ at $0.1\rvir$ is $1-2\,\Tvir$ in the red group ($\tcoolsh\gg\tff$, higher $\Mhalo$), compared to $0.1-0.5\,\Tvir$ in the blue-gray group ($\tcoolsh\lesssim\tff$, lower $\Mhalo$). In contrast, there is almost no difference between the two groups in $\Tmed/\Tvir$ at $0.5\Rvir$.  A parallel trend is also seen in the supersonic fraction profile shown in the bottom panel. At $0.1\Rvir$ the supersonic fraction is $0.05-0.2$ in the red group, substantially lower than the fraction of $0.4-0.8$ in the blue-gray group. In contrast at $\sim 0.5\rvir$ there is almost no supersonic gas ($<20\%$ of the volume) in either group.
Also evident in this plot is the reverse trend of decreasing $\Tmed/\Tvir$ at $>\Rvir$ with increasing $\tcoolsh/\tff$ at $0.1\Rvir$. 
Figure~\ref{f:Tprofiles z 0} thus supports the conclusion from Fig.~\ref{f:Tprofiles m12i}, that the CGM is fully virialized when $\tcoolsh/\tff\gg1$ at $0.1\Rvir$, while halos with $\tcoolsh/\tff\lesssim1$  at $0.1\Rvir$ are transonic. Furthermore, Fig.~\ref{f:Tprofiles z 0} demonstrates that in all simulations in our sample the CGM at large radii is virialized at $z\sim0$, regardless of their halo mass. Virialization of the CGM in FIRE thus begins at halo masses well-below the classic threshold of $\sim10^{12}\msun$.

Further insight into the differences between snapshots with $\tcoolsh/\tff\lesssim1$ and $\tcoolsh/\tff\gg1$ can be gained by exploring the gas temperature distribution within radial shells. To this end, Figure~\ref{f:m12i image} plots a temperature map and 2D temperature histograms of the $z=0$ snapshot of m12i, in which $\tcoolsh=16\tff$ at $0.1\Rvir$. The images in the two left panels are oriented such that the total angular momentum vector of gas within $0.05\Rvir$ is oriented upwards, i.e.~edge-on to the galaxy disc. The first panel spans $\pm\Rvir$ while the second panel zooms in on the central $\pm 0.2\Rvir$. The images are derived by averaging $\log T$ perpendicular to the image plane over a depth equal to $10\%$ of the image size. 
The Figure shows that in this snapshot hot gas ($>10^5\K$) dominates at practically all locations within the halo except in the galaxy disc. The accretion shock is evident as a temperature drop along a non-spherical contour roughly at a distance $\sim\Rvir$ from the center. 

The two right panels in Fig.~\ref{f:m12i image} show temperature histograms of gas in shells at different radii, weighted by volume (third panel from the left) and weighted by mass (rightmost panel). Color denotes the volume or mass fraction of the shell in each temperature bin. 
As suggested by the images on the left, the hot phase dominates the volume out to $\approx\Rvir$, at which there is a break in the temperature profile due to the accretion shock. In contrast the cool phase ($T\lesssim10^4\K$) is entirely negligible in the halo in terms of volume, and becomes significant (but still subdominant) only at disc radii. In the mass-weighted histogram on the right the hot phase dominates at halo radii ($>0.1\Rvir$) while the cold phase dominates at disc radii ($<0.1\Rvir$). The transition between halo and galaxy radii is sharp (as seen also in the zoomed image in the second panel). We show below that this sharp transition corresponds to the radius $\Rcirc$ where gas in the CGM circularizes (eqn.~\ref{e:Rcirc}).

For comparison, Figure~\ref{f:m11d image} plots temperature maps and 2D temperature histograms for the $z=0$ snapshot of m11d, in which  $\tcoolsh=0.2\tff$ at $0.1\Rvir$. As in Fig.~\ref{f:m12i image}, the orientation of the two left panels is such that the total angular momentum vector of gas within $0.05\Rvir$ is oriented upward. The figure shows that beyond $0.3\Rvir$ the hot phase dominates also in this snapshot, similar to the m12i snapshot shown in Figure~\ref{f:m12i image} albeit with a lower absolute temperature due to the lower $\Tvir$ and a larger dispersion in temperature at a given radius. The gas temperatures of m11d in the inner CGM and at galaxy radii are however completely different from those in m12i. In m11d the cool phase dominates the volume out to $0.3\Rvir$, while in m12i the volume of the cool phase is subdominant at all radii, and negligible beyond $0.1\Rvir$. By mass, the hot phase is completely negligible in m11d out to $\approx0.2\Rvir$, in contrast with m12i where the hot phase mass is significant at $<0.1\Rvir$ and dominant at $>0.1\Rvir$. The zoomed image also shows that m11d lacks the clear sharp disc seen in m12i. 
We conclude that m12i and m11d, which at $z=0$ are respectively above and below the threshold for virialization of the inner CGM, differ mainly in the temperature distribution at inner halo and galaxy radii. This conclusion reinforces the conclusion from Figs.~\ref{f:Tprofiles m12i} -- \ref{f:Tprofiles z 0} based on the 1D temperature profiles.

\begin{figure}
\includegraphics{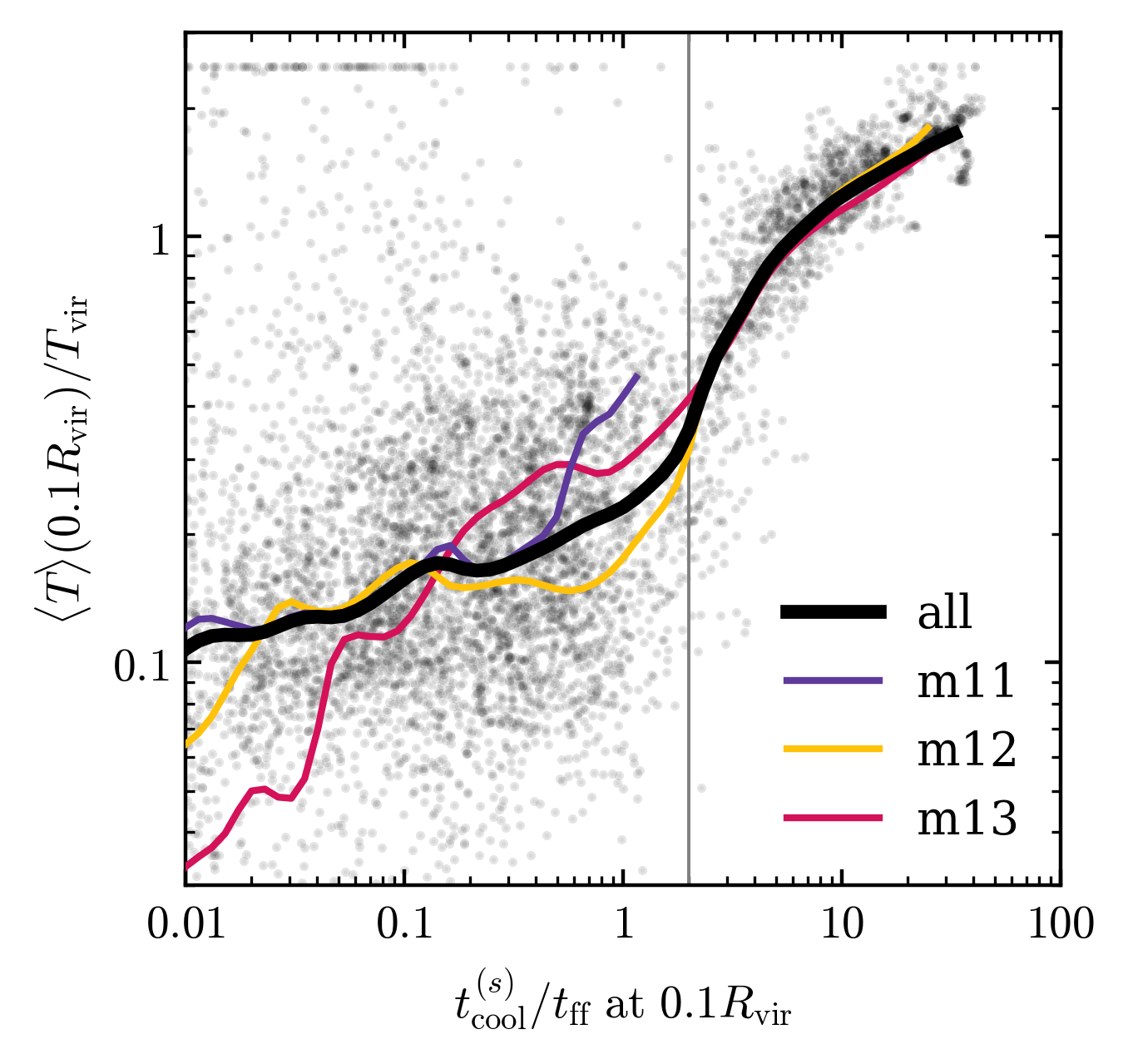}
\caption{Volume-weighted temperature versus $\tcoolsh/\tff$ at $0.1\Rvir$. 
Note that $\tcoolsh$ is calculated independent of the gas temperature in the simulations (eqn.~\ref{e:tcool0}). 
Each gray dot corresponds to a different snapshot, including all snapshots from all 16 FIRE simulations. In individual simulations $\tcoolsh/\tff$ increases with time so each simulation traverses this plot from left to right. To decrease the dynamic range snapshots with $\langle T\rangle>2.5\Tvir$ are plotted at $\langle T\rangle=2.5\Tvir$. 
Thick colored lines plot medians for each of the three simulation subgroups. When $\tcoolsh$ exceeds $\approx2\tff$ the typical temperature increases from $\ll\Tvir$ to $\gtrsim\Tvir$ and the scatter at a given $\tcoolsh/\tff$ decreases. 
}
\label{f:T all}
\end{figure}

Figures~\ref{f:Tprofiles h2} --\ref{f:h206 h2 image} repeat the above analysis for simulations in which $\tcoolsh$ exceeds $\tff$ at high redshift. Fig.~\ref{f:Tprofiles h2} shows the temperature, normalized temperature, and supersonic fraction of m13A8 for different snapshots in the redshift range $1.1<z<3.4$, at which $\tcoolsh/\tff$ increases from $0.1$ to $10$. We plot a wider range of $r/\Rvir$ in this plot than in Figs.~\ref{f:Tprofiles m12i} -- \ref{f:Tprofiles z 0} since the corresponding galaxy sizes are a smaller fraction of $\Rvir$ (see below). 
Fig.~\ref{f:Tprofiles h2} demonstrates that the trends versus $\tcoolsh/\tff$ in m13A8 are similar to the trends in m12i (Fig.~\ref{f:Tprofiles m12i}). The value of $\Tmed/\Tvir$ tends to reach its final value first at large CGM radii and later at small CGM radii and at galaxy radii.  Similarly, when $\tcoolsh\lesssim\tff$ the supersonic fraction is high at the inner CGM and at galaxy radii while it is low at the outer CGM, i.e.~the halo gas is transonic. When $\tcoolsh\gg\tff$ the supersonic fraction is low at all radii within $\Rvir$. These trends again indicate that the outer CGM becomes steadily hot and subsonic prior to the inner CGM, i.e.~the CGM virializes from the outside inwards. 

Figure~\ref{f:h206 h2 image} shows temperature maps and 2D temperature histograms of the $z=2.5$ snapshots of m13A1 (top) and m13A8 (bottom). This redshift is chosen since it is after the inner CGM virializes in m13A1 but before it virializes in m13A8. In m13A1 at radii larger than $0.05\Rvir$ the halo gas is clearly separated into a hot phase with $T\gtrsim10^6\K$ and cool streams with $T\lesssim10^{4.5}\K$, where the hot phase dominates by volume but the two phases are comparable in mass. Within $0.05\Rvir$ the hot phase continues to dominate by volume, but the cool disc evident in the second panel dominates by mass. 
For comparison in m13A8, gas in the outer halo ($>0.3\Rvir$) is similar to gas in the outer halo of m13A1, with a hot phase which dominates by volume and is comparable in mass to the cool streams. At inner halo and galaxy radii however cool gas fills most of the volume and there is no clear disc, similar to m11d and in stark contrast with gas in the inner halos of m13A1 and m12i. 
Figs.~\ref{f:Tprofiles h2} -- \ref{f:h206 h2 image} thus suggest that the behavior of the volume-filling phase with respect to $\tcoolsh/\tff$ at high redshift is similar to its behavior at low redshift (Figs.~\ref{f:Tprofiles m12i} -- \ref{f:m11d image}). This suggests that the process of inner CGM virialization does not strongly depend on redshift or on the existence of cool streams. 

One may wonder how the existence of cool streams does not significantly affect our calculation of $\tcoolsh$, for  which we assume the CGM mass is distributed spherically (eq.~\ref{e:pressure support}). This follows since after virialization the mass in the cool $T<10^5\K$ streams is typically $\lesssim30\%$ of the total CGM mass at $0.1-1\Rvir$. 
The implied change in the expected density of the virialized phase due to the existence of cool streams is thus small. 

\begin{figure}
\includegraphics{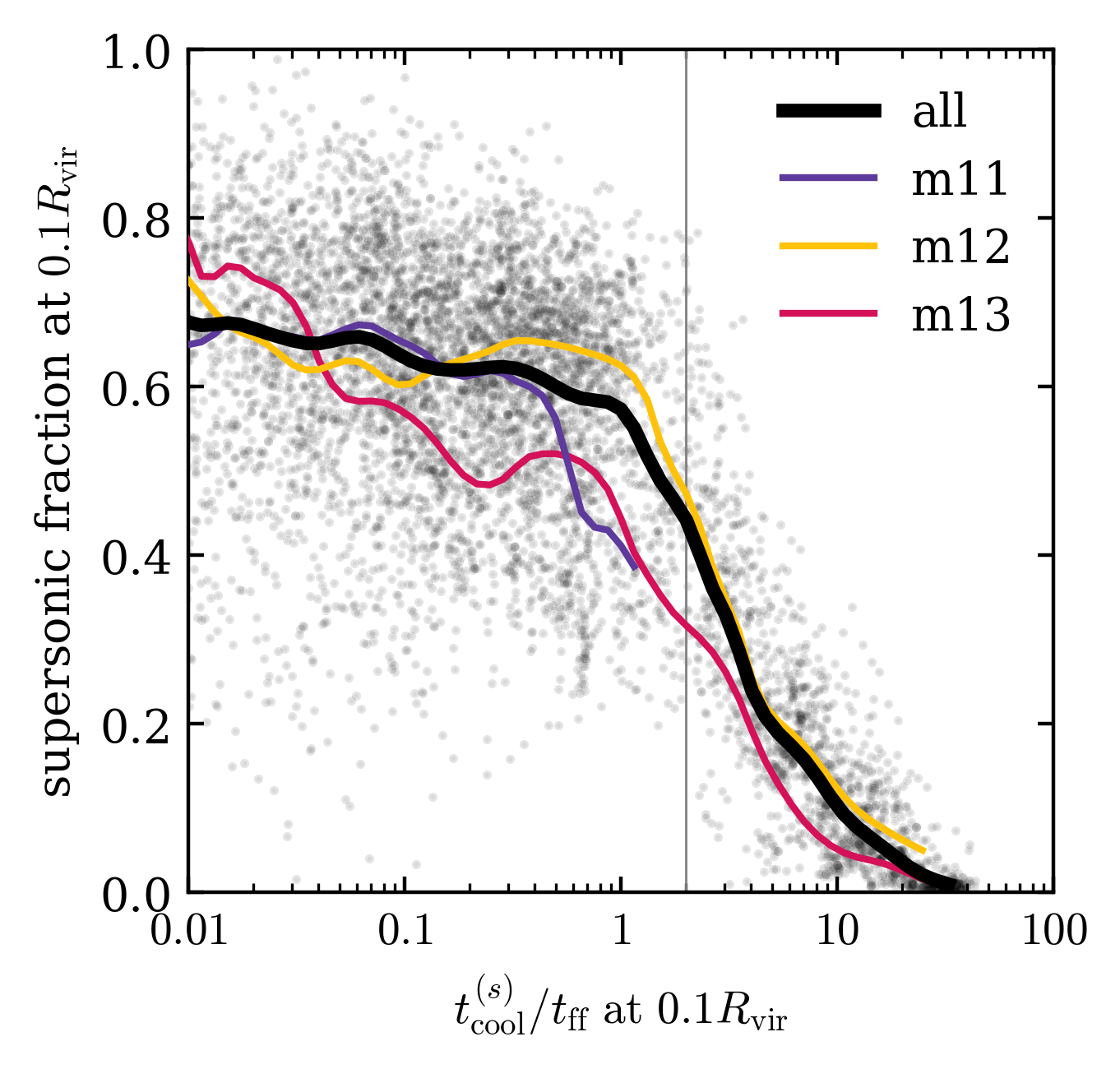}
\caption{Similar to Fig.~\ref{f:T all}, for the volume-weighted supersonic fraction. 
In most snapshots with $\tcoolsh\lesssim\tff$ most of the volume at $0.1\Rvir$ is supersonic, while in snapshots with $\tcoolsh\gg\tff$ the supersonic fraction is close to zero. 
}
\label{f:supersonic all}
\end{figure}

To depict the virialization of the inner CGM in a single simulation, the top row of Figure~\ref{f:m12i_multiple_images} plots temperature maps of the inner $0.2\Rvir$ in different snapshots of m12i, as in the second panel of Fig.~\ref{f:m12i image}. From left to right the redshifts are $z=1, 0.4, 0.1$ and $0$, while $\tcoolsh/\tff$ is $0.25$, $1$, $4$ and $16$. As suggested by the comparison above of different halos at the same redshift, in snapshots with $\tcoolsh\lesssim\tff$ a large fraction of the volume is filled with cool gas, and there is no prominent disc. In contrast in snapshots with $\tcoolsh\gg\tff$ the inner halo volume is filled with hot gas, and a prominent cool disc is apparent. 

The bottom row of Fig.~\ref{f:m12i_multiple_images} plots spatial pressure fluctuations in the same snapshots of m12i shown in the top row. Color indicates the pressure in each pixel relative to the average pressure $\langle P(r)\rangle$ in a shell of the same distance as the pixel. 
The calculation of $\langle P(r)\rangle$  is similar to that of the average temperature in eqn.~(\ref{e:T}):
\begin{equation}\label{e:nHT}
 \log\ \langle P(r)\rangle \equiv \frac{\sum V_i\log P_i  }{\sum V_i }
\end{equation}
where $P_i$ is the thermal pressure of resolution element $i$ and the summation is over all resolution elements whose center is within a shell with thickness $\Delta\log r=0.05\,{\rm dex}$. Normalizing the pressure by the mean removes the radial pressure gradient and allows focusing on the differences between different solid angles.  The figure shows that in the left panels before the inner CGM virializes different directions differ substantially in thermal pressure, with over-pressurized angles having a factor of up to $\sim 100$ higher pressure than under-pressurized angles. In contrast in the right panels after virialization fluctuations are more mild, and the pressure distribution is closer to being spherically symmetric.

\subsection{The condition for inner CGM virialization}\label{s:condition}

Figures~\ref{f:T all} and \ref{f:supersonic all} demonstrate the dependence of inner CGM virialization on $\tcoolsh/\tff$ in all 16 FIRE simulations. Fig.~\ref{f:T all} plots $\Tmed/\Tvir$ at $0.1\Rvir$ against $\tcoolsh/\tff$ at the same radius, while Fig.~\ref{f:supersonic all} plots the supersonic fraction against $\tcoolsh/\tff$ at $0.1\Rvir$. Each grey dot corresponds to a single snapshot, while the plots include all snapshots with $\tcoolsh/\tff>0.01$ from all 16 simulations. Since $\tcoolsh/\tff$ increases with time in individual simulations (see bottom-right panel of Fig.~\ref{f:properties}), the tracks of individual simulations proceed from left to right in these plots. The colored lines show the median value versus $\tcoolsh/\tff$ for each of the three simulation subgroups, derived using a Gaussian kernel density estimator. 

\begin{deluxetable}{lcccccc}
\tablecolumns{7}
\tablecaption{Properties of the galaxy and halo at the redshift where $\tcoolsh=2\tff$ at $0.1\Rvir$ and the inner CGM virializes.}
\label{t:virialization}
\tablehead{
\colhead{Name} & \colhead{$z$} & \colhead{$\Mhalo$} & \colhead{$\vc$ } & \colhead{$\fvc$ } & \colhead{$Z$} & \colhead{ $\rho^{(s)}$} \\ 
\colhead{} & \colhead{ } & \colhead{$[10^{12}\msun]$} & \colhead{ $[{\rm km}\,{\rm s}^{-1}]$  } & \colhead{  } & \colhead{$[\zsun]$ } & \colhead{$[\Deltac \rhocrit]$} \\ 
\colhead{(1)} & \colhead{(2)} & \colhead{(3)} & \colhead{(4)} & \colhead{(5)} & \colhead{(6)} & \colhead{(7)} 
}
\startdata
\multicolumn{7}{c}{\textbf{m12's}} \\
\texttt{m12z}   & --   &  -- & -- & -- & -- & -- \\
\texttt{m12i}   & 0.32 &  $0.9$ & 190 & 1.4 & 1.2 & 0.8 \\
\texttt{m12b}   & 0.70 &  $0.8$ & 210 & 1.4 & 1.0 & 0.7 \\ 
\texttt{m12m}   & 0.44 &  $1.2$ & 240 & 1.6 & 2.3 & 1.1 \\ 
\texttt{m12f}   & 0.26 &  $1.5$ & 220 & 1.4 & 1.2 & 1.5 \\ 
\multicolumn{7}{c}{\textbf{m13's}} \\
\texttt{m13A1}  & 3.6  &  $1.4$ & 320 & 1.2 & 1.0 & 1.2 \\ 
\texttt{m13A4}  & 2.2  &  $2.4$ & 300 & 1.0 & 0.8 & 1.4 \\ 
\texttt{m13A2}  & 2.7  &  $2.4$ & 390 & 1.3 & 1.8 & 1.5 \\ 
\texttt{m13A8}  & 1.7  &  $1.6$ & 310 & 1.3 & 0.9 & 3.4 \\ 
\texttt{z5m13a} & 4.9  &  $2.9$ & 530 & 1.3 & 1.4 & 1.5 \\ 
\enddata
\tablecomments{
(1) galaxy name; 
(2) redshift; 
(3) halo mass;
(4) circular velocity at $0.1\Rvir$;
(5) ratio of circular velocity at $0.1\Rvir$ to virial velocity. 
(6) metallicity at $0.1\Rvir$;
(7) gas density at $0.1\Rvir$ divided by the mean halo density ($\Delta_{\rm c}$ is the \cite{bryan98} virial overdensity and $\rho_{\rm crit}$ is the total critical density of the universe).
}
\end{deluxetable}

Fig.~\ref{f:T all} shows that the typical $\Tmed$ is $\ll\Tvir$ when $\tcoolsh\lesssim\tff$, with occasional snapshots with $\Tmed>\Tvir$. In contrast when $\tcoolsh$ exceeds $\tff$, the median temperature increases to $\gtrsim\Tvir$ and the scatter at a given $\tcoolsh/\tff$ substantially decreases. This transition is paralleled by a sharp transition in the supersonic fraction seen in Fig.~\ref{f:supersonic all}, where the median supersonic fraction is $\approx0.6$ when $\tcoolsh\lesssim\tff$, i.e.\ gas in the inner halo is predominantly supersonic, and drops abruptly when $\tcoolsh\gtrsim\tff$, i.e.\ the inner CGM becomes predominantly subsonic. 
These plots therefore demonstrate that in the majority of snapshots with $\tcoolsh\ll\tff$, most of the \emph{volume} in the inner CGM is occupied by sub-virial gas which is flowing supersonically. 
In contrast, when $\tcoolsh\gg\tff$ the bulk of the inner CGM is constantly occupied by virial-temperature gas which is flowing subsonically. This transition 
is the virialization of the inner CGM. 

Figs.~\ref{f:T all} -- \ref{f:supersonic all} suggest that the inner CGM virializes when $\tcoolsh/\tff$ is in the range $1-4$, indicating that the factor $\ft'$ defined in eqn.~(\ref{e:threshold condition actual}) is larger than unity. We choose $\ft'=2$ as an intermediate value, and in Table~\ref{t:virialization} list for each simulation several properties of the galaxy and halo at the redshift where $\tcoolsh=2\tff$. 

The median $\tcoolsh/\tff$ of the m11 simulations (purple lines in Figs.~\ref{f:T all} -- \ref{f:supersonic all}) does not extend much beyond unity, since this is the maximum ratio reached by these relatively low mass halos (see Fig.~\ref{f:properties}). 
However, the medians of the different subgroups show similar relations of $\Tmed/\Tvir$ and supersonic fraction versus $\tcoolsh/\tff$. This reinforces the conclusion above that the virialization of the inner CGM is mainly a result of the change in $\tcoolsh/\tff$, rather than of other parameters of the system which differ between the simulation subgroups, such as the relation between halo mass and redshift.

\begin{figure}
\includegraphics{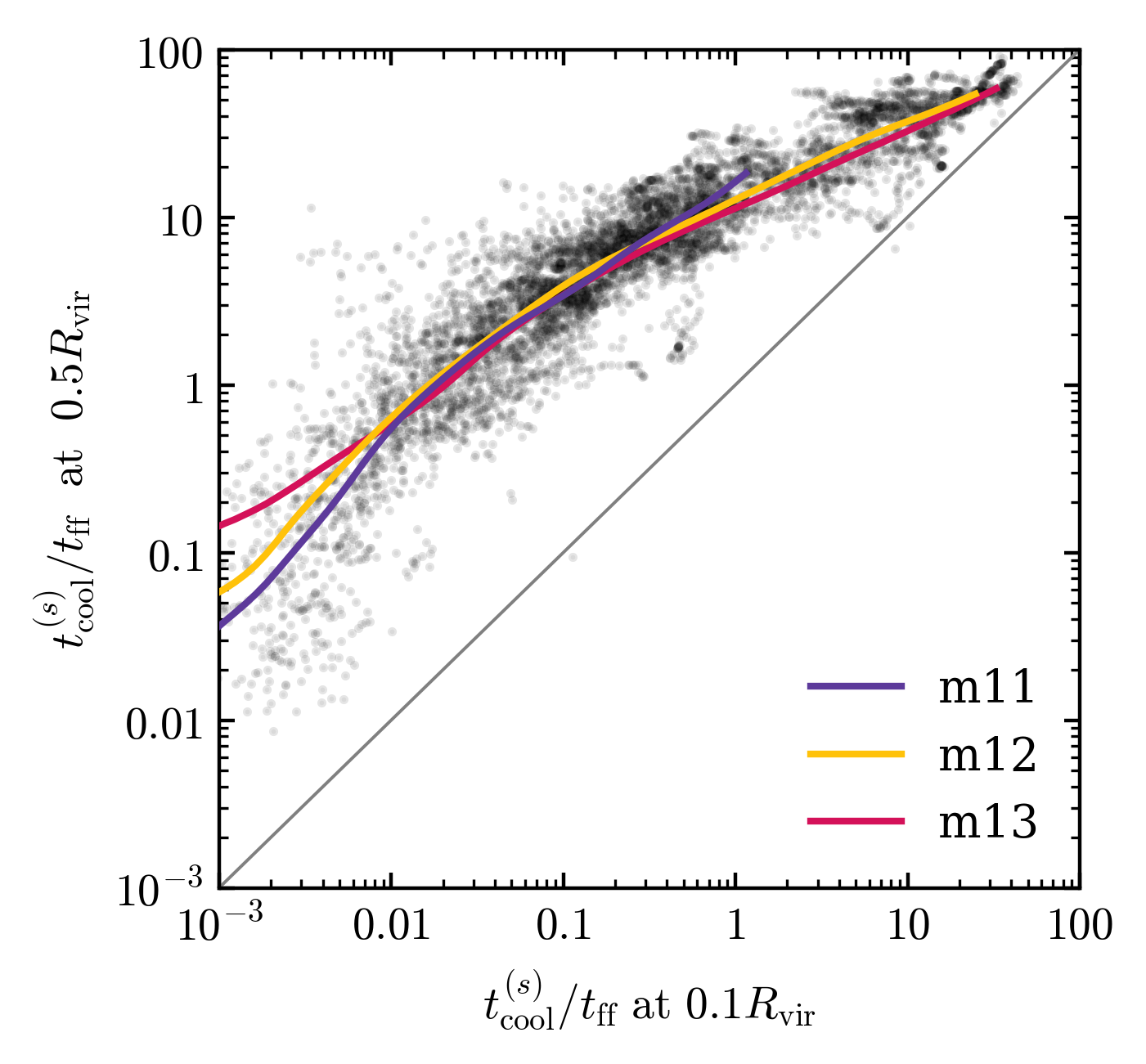}
\caption{A comparison of $\tcoolsh/\tff$ measured at $0.1\Rvir$ with the same ratio measured at $0.5\Rvir$.
Gray markers denote snapshots from all 16 FIRE simulations.
Thick colored lines denote the medians of the three simulation subgroups. In the vast majority of snapshots the ratio $\tcoolsh/\tff$ at $0.5\Rvir$ is larger than at $0.1\Rvir$, indicating $\tcoolsh/\tff$ increases outward. 
}
\label{f:t ratio comparison}
\end{figure}

\subsection{Outside-in virialization in FIRE}\label{s:outer}

Figure~\ref{f:t ratio comparison} compares the ratio $\tcoolsh/\tff$ estimated at $0.1\Rvir$ with the same ratio estimated\footnote{The value of $\tcoolsh$ (eqn.~\ref{e:tcool0}) depends on $\nHc$, whose calculation requires choosing an outer limit for the integral in eqn.~(\ref{e:pressure support}). We integrate out to $\Rvir$, as done above for the calculation of $\nHc(0.1\Rvir)$. A larger outer limit of $2\Rvir$ increases $\nHc(0.5\Rvir)$ typically by $\approx 30\%$, and $\nHc(0.1\Rvir)$ by $<1\%$. Both changes do not affect our conclusions.} at $0.5\Rvir$. 
The figure shows that $\tcoolsh/\tff$ is almost always larger at $0.5\Rvir$ than at $0.1\Rvir$, typically by a factor of $\sim30$ when $\tcoolsh\ll\tff$ at $0.1\Rvir$, and by a factor of $\approx3$ when $\tcoolsh \sim 10\tff$ at $0.1\Rvir$. The median relation between the vertical and horizontal axes is rather similar in the m11, m12, and m13 subgroups. This result is consistent with the expectation of steady-state inflow solutions that $\tcoolsh/\tff$ increases outwards (Papers I and II). Furthermore, since $\tcoolsh/\tff$ generally increases with time this result implies that $\tcoolsh$ exceeds $\tff$ first in the outer halo and then in the inner halo.

\begin{figure}
\includegraphics{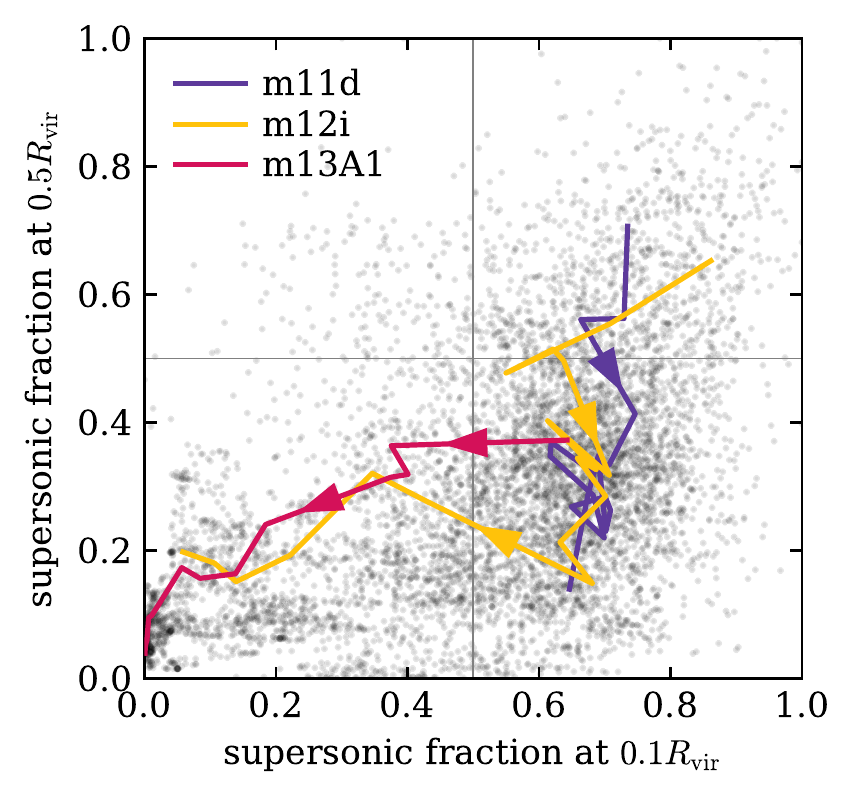}
\caption{Volume fraction of supersonic gas in the inner and outer halo. Gray dots denote snapshots from all 16 simulations. Colored lines and arrows show the tracks of three individual simulations using the median values in windows of $1\Gyr$ (m11d and m12i) and windows of $500\Myr$ (m13A8). 
The tendency of the tracks to go through the lower-right quadrant of the plot, rather than through the upper left quadrant, indicates that the quasi-static virialized CGM forms from the outside-in rather than from the inside-out. 
}
\label{f:supersonic comparison}
\end{figure}

In Figure~\ref{f:supersonic comparison} we compare the supersonic fractions at $0.1\Rvir$ and $0.5\Rvir$. Gray dots denote individual snapshots in all 16 simulations, while the lines and arrows plot the tracks of three simulations, one from each simulation subgroup. The tracks are calculated using median values in $1\Gyr$ time windows for m11d and m12i and using $500\Myr$ windows for m13A1. 
In m12i the supersonic fraction decreases first in the outer CGM and then in the inner CGM. 
In m11d the supersonic fraction decreases in the outer CGM, but remains high ($\approx65\%$) down to $z=0$ in the inner CGM. In m13A1 the supersonic fraction is always below $50\%$ in the outer CGM and decreases with time in the inner CGM. 
All three tracks, and the small number of snapshots in the upper-left quadrant, suggest that the volume-filling phase becomes predominantly subsonic first in the outer CGM and then in the inner CGM, again indicating that the CGM virializes from the outside inward.

\section{Implications for the central galaxy and feedback}\label{s:implications}

\subsection{Inner CGM virialization coincides with disc formation}
\label{s:disc_formation}

To explore the formation of the gaseous disc we calculate the specific angular momentum profile of gas at galaxy and CGM radii. As above, we divide the gas in each snapshot into radial shells and calculate:
\begin{equation}\label{e:jvec}
 \langle \vec{j}\rangle= \frac{\sum m_i \left(\vec{r_i} \times \vec{v_i}\right)  }{\sum m_i } ~,
\end{equation}
where $m_i$, $\vec{r_i}$ and $\vec{v_i}$ are the mass, position and velocity of resolution element $i$. We then project the total angular momentum onto the axis of rotation $z$
\begin{equation}\label{e:jz}
 \langle j_z\rangle= \langle \vec{j}\rangle\cdot \hat{z}
\end{equation}
where $\hat{z}$ is defined as the direction of the total angular momentum vector of all gas resolution elements within $0.05\Rvir$ in the snapshot. 
The rotational velocity is hence
\begin{equation}\label{e:Vrot}
 \langle\Vrot(r)\rangle= \frac{\sum m_i \left(\vec{v_i}\cdot \hat{\phi}\right)}{\sum m_i }  ~.
\end{equation}
where $\hat{\phi}$ is the corresponding azimuthal coordinate. 
The summations in eqns.~(\ref{e:jvec}) and (\ref{e:Vrot}) are over all resolution elements within a shell centered at $r$ and with thickness $\Delta\log r=0.05\dex$. 

\begin{figure}
\includegraphics{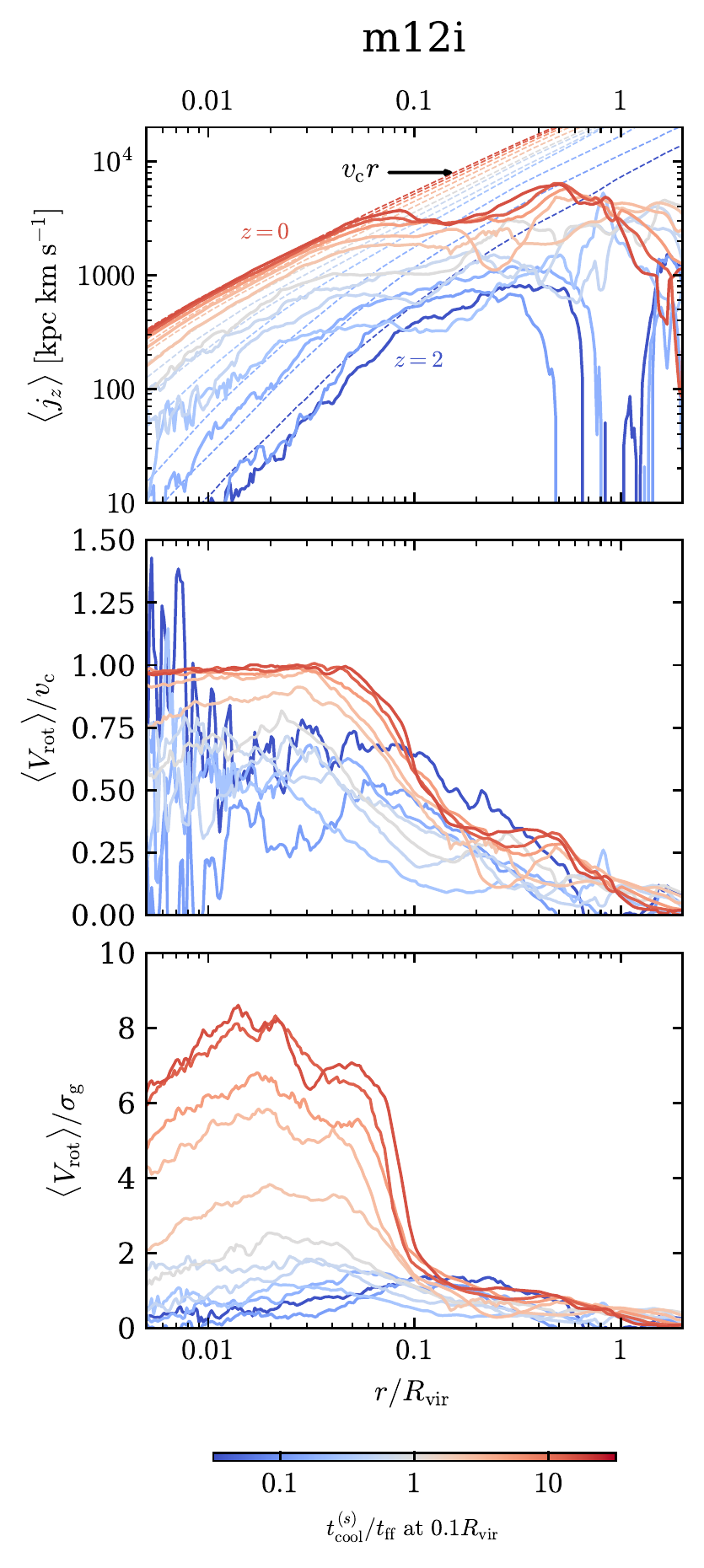}
\caption{Formation of rotation-dominated galactic discs versus the ratio of the cooling time to free-fall time.
\textbf{Top:} 
Median specific angular momentum profiles (see eqn.~\ref{e:jz}) in $\Delta t=0.5\Gyr$ windows in the m12i simulation. Line color marks $\tcoolsh/\tff$ at $0.1 R_{\rm vir}$, while dashed lines mark the specific angular momentum of circular orbits corresponding to each time window. Redshifts of the first and last plotted time windows are noted in the panel. 
Note that when $\tcoolsh\gg\tff$ (red lines) the angular momentum profile roughly equals $\vc r$ at small radii and flattens at large radii. 
\textbf{Middle:}
The ratio of the rotation to circular velocities in the same time windows as in the top row.
\textbf{Bottom:} The ratio of the gas rotation velocity to its velocity dispersion (eqn.~\ref{e:sigma}). 
A rotation-dominated disc forms at $\lesssim0.07\Rvir$ when $\tcoolsh$ exceeds $\tff$. Similar plots for the other 15 simulations are available in the online journal.
}
\label{f:jProfiles by sim}
\end{figure}

The top panel of Figure~\ref{f:jProfiles by sim} shows the specific angular momentum profiles $\langle j_z\rangle$ in the m12i simulation. 
As above we plot the median profiles of snapshots in $\Delta t=0.5\Gyr$ windows, and use line color to denote the median $\tcoolsh/\tff$ at $0.1 R_{\rm vir}$. 
For each time window we also mark the specific angular momentum corresponding to circular orbits $\vc r$ using thin dashed lines with the same color. The middle panel shows $\langle\Vrot\rangle/\vc$, while the lines span the redshift range at which $-1.5<\log\tcoolsh/\tff<1.5$.  Similar plots for the other simulations are available online.

The top two panels in Fig.~\ref{f:jProfiles by sim} demonstrate a clear trend with increasing $\tcoolsh/\tff$. In snapshots with $\tcoolsh\lesssim\tff$ (before the inner CGM virializes, blue-gray lines) there is no range of radii where $\Vrot$ follows $\vc$. In contrast in snapshots with $\tcoolsh\gg\tff$ there is an extended region where $\Vrot\approx\vc$ and the gas is on circular orbits. 
The range of radii where $\Vrot\approx\vc$ corresponds to the range of radii where the gas is predominantly cool by mass (see right panel of Fig.~\ref{f:m12i image}). 
These panels thus indicate that a rotating disc forms once $\tcoolsh$ exceeds $\tff$ and the inner CGM virializes. This association of a rotating thin disc with virialization is also suggested by the images shown in Figures~\ref{f:m12i image}, \ref{f:m11d image}, \ref{f:h206 h2 image} and \ref{f:m12i_multiple_images}, and is apparent also in the other simulations (see online figures). 

The top panels in Fig.~\ref{f:jProfiles by sim} also show that in snapshots after a disc forms the gas transitions from circular orbits at $r<0.06\Rvir$ to a roughly flat angular momentum profile at $0.06\rvir < r < \rvir$. This radius corresponds to $\Rcirc$ defined in eqn.~(\ref{e:Rcirc}), the radius where halo gas can be supported by angular momentum. In snapshots after the inner CGM virialized we find 
a trend of decreasing $\Rcirc/\Rvir$ and disc-to-halo size ratio with increasing redshift, with $\Rcirc\approx 0.05\Rvir$ for the m12 subgroup and $\Rcirc\approx0.02\Rvir$ for the m13 subgroup. We defer exploring the origin of this trend to future work.

The bottom panel of Fig.~\ref{f:jProfiles by sim} plots the profile of $\langle\Vrot\rangle/\sigmag$ in the same time windows as in the top panels. The ratio $\langle\Vrot\rangle/\sigmag$ has been highlighted by recent observations as a measure of `disc settling' (e.g., \citealt{kassin12,simons17}). We calculate  $\sigmag$ following \cite{elbadry18a}:
\begin{equation}\label{e:sigma}
 \sigmag = \sqrt{\langle\Vrot^2\rangle - \langle\Vrot\rangle^2} ~,
\end{equation}
i.e.~$\sigmag$ equals the dispersion in the rotational velocity. 
\citeauthor{elbadry18a}\ demonstrated that $\sigmag$ defined in this way is similar to the velocity dispersion measured in a mock slit aligned along the major axis of the simulated galaxy (see their figure~2). 
Fig.~\ref{f:jProfiles by sim} shows that $\langle\Vrot\rangle/\sigmag$ increases rapidly at galaxy radii ($r\lesssim0.1\rvir$) as $\tcoolsh$ exceeds $\tff$, from a value of $\approx1$ when $\tcoolsh\ll\tff$ to a value of up to $\approx8$ when $\tcoolsh\gg\tff$. 
This result also indicates that inner CGM virialization coincides with the formation of a rotation-dominated galactic disc.

\begin{figure*}
\includegraphics{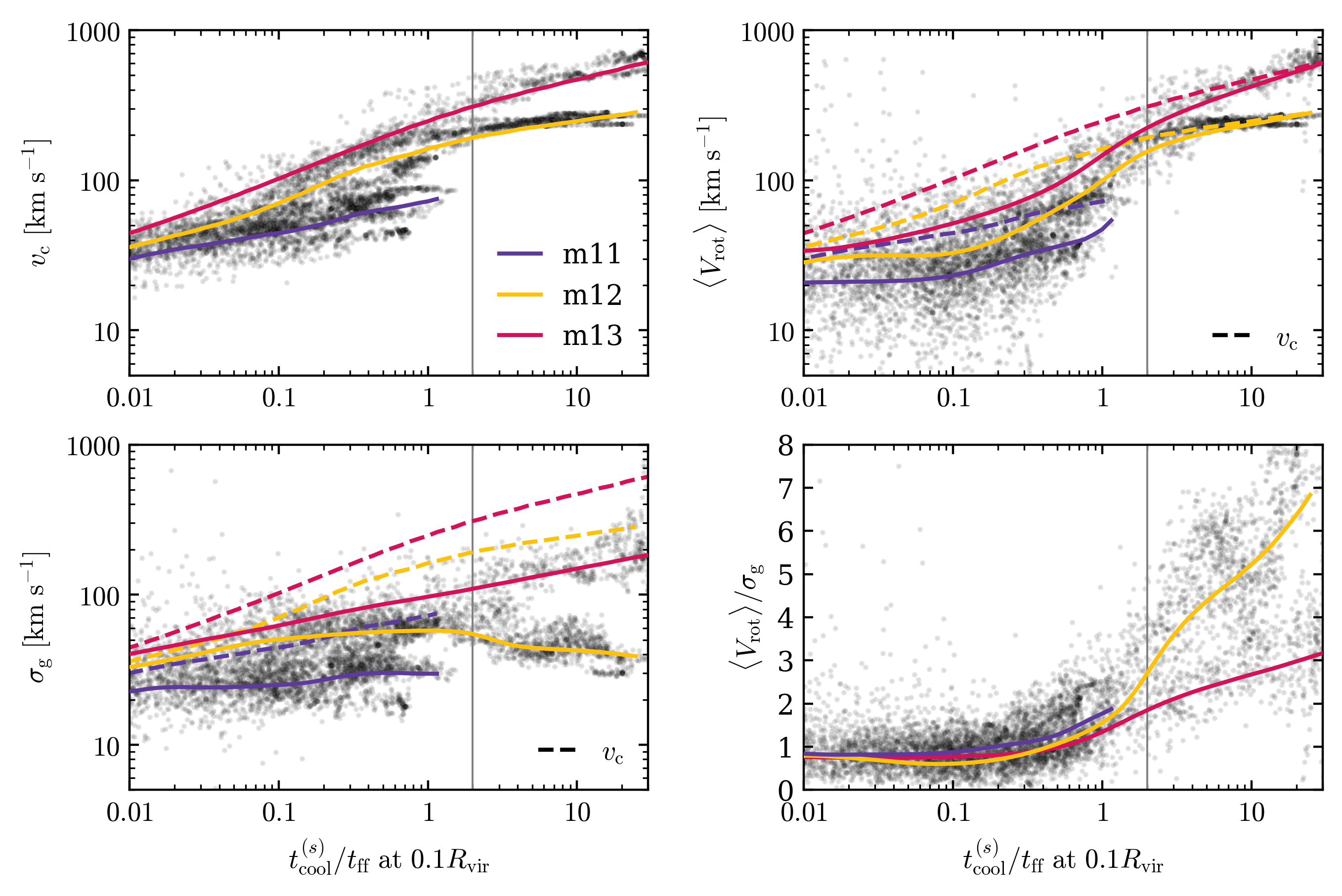}
\caption{Kinematics of gas in the central galaxy versus the ratio of cooling time to free-fall time. Panels show the circular velocity (top-left), rotational velocity (top-right), gas dispersion (bottom-left), and the ratio of rotation to dispersion (bottom-right). In each snapshot the properties are averaged over gas in the central $0.05\Rvir$ and weighted by its \hi-mass. 
Gray dots denote snapshots in all 16 simulations in our sample, while colored solid lines show the medians for each of the three simulation subgroups. Dashed lines in the $\langle\Vrot\rangle$ and $\sigmag$ panels plot the median $\vc$ for comparison. 
Note that $\langle\Vrot\rangle\approx\sigmag$ when $\tcoolsh\ll\tff$, while $\langle\Vrot\rangle/\sigmag\sim 3-6$ when $\tcoolsh\gg\tff$. 
}
\label{f:Vrot to sigma}
\end{figure*}

The calculations of $\langle\Vrot\rangle$ and $\sigmag$ shown in Fig.~\ref{f:jProfiles by sim} are weighted by gas mass, regardless of whether it is ionized or neutral. This allows us to understand the properties of the angular momentum profile independent of phase changes in the gas. 
To also explore the gas kinematics integrated over the cool star-forming disc, we recalculate $\hat{z}$, $\langle\Vrot\rangle$ and $\sigmag$ using eqns.~(\ref{e:jvec}) -- (\ref{e:sigma}), but weighting the resolution elements by their \hi\ mass or by their SFR rather than by the total gas mass.  Figure~\ref{f:Vrot to sigma} plots \hi-weighted quantities versus $\tcoolsh/\tff$, while SFR-weighted quantities (not shown) exhibit similar trends. 
The panels in Fig.~\ref{f:Vrot to sigma} show \hi-mass weighted $\langle\vc\rangle$ (top-left), $\langle\Vrot\rangle$ (top-right),  $\sigmag$ (bottom-left) and $\langle\Vrot\rangle/\sigmag$ (bottom-right). 
As above, gray dots represent all individual snapshots in the 16 FIRE simulations, while colored lines show the medians for the three simulation subgroups derived using a Gaussian kernel density estimator. For comparison, we also plot the median $\vc$ of each subgroup in the $\langle\Vrot\rangle$ and $\sigmag$ panels.

\begin{figure*}
\includegraphics{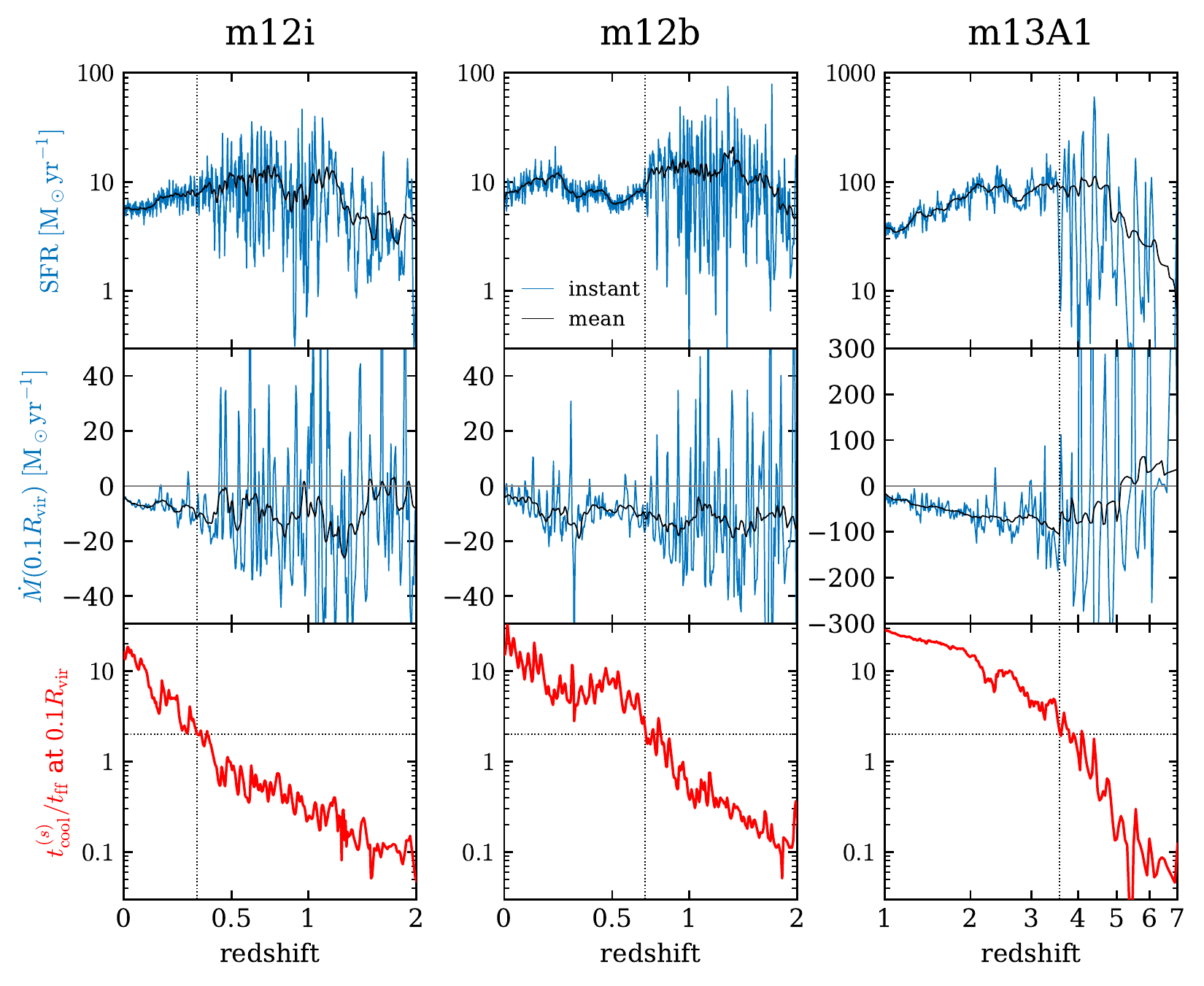}
\caption{Implications of inner CGM virialization for SFR and outflow properties. 
\textbf{Top row:} Instantaneous SFR (blue) and average SFR in $300\Myr$ windows (black) versus redshift in the central galaxy of three FIRE simulations.
\textbf{Middle row:} 
Instantaneous and average radial mass flow rate at $0.1\Rvir$.
\textbf{Bottom row:} Cooling time to free-fall time ratio at $0.1\Rvir$. Vertical lines mark where $\tcoolsh$ crosses $2\tff$, which roughly indicates when the inner CGM virializes. 
Note how fluctuations in SFR and $\Mdot$ are large prior to virialization and small afterwards. 
Similar plots for the other 13 simulations are available in the online journal.
}
\label{f:SFR_Mdot}
\end{figure*}

\begin{figure}
\includegraphics{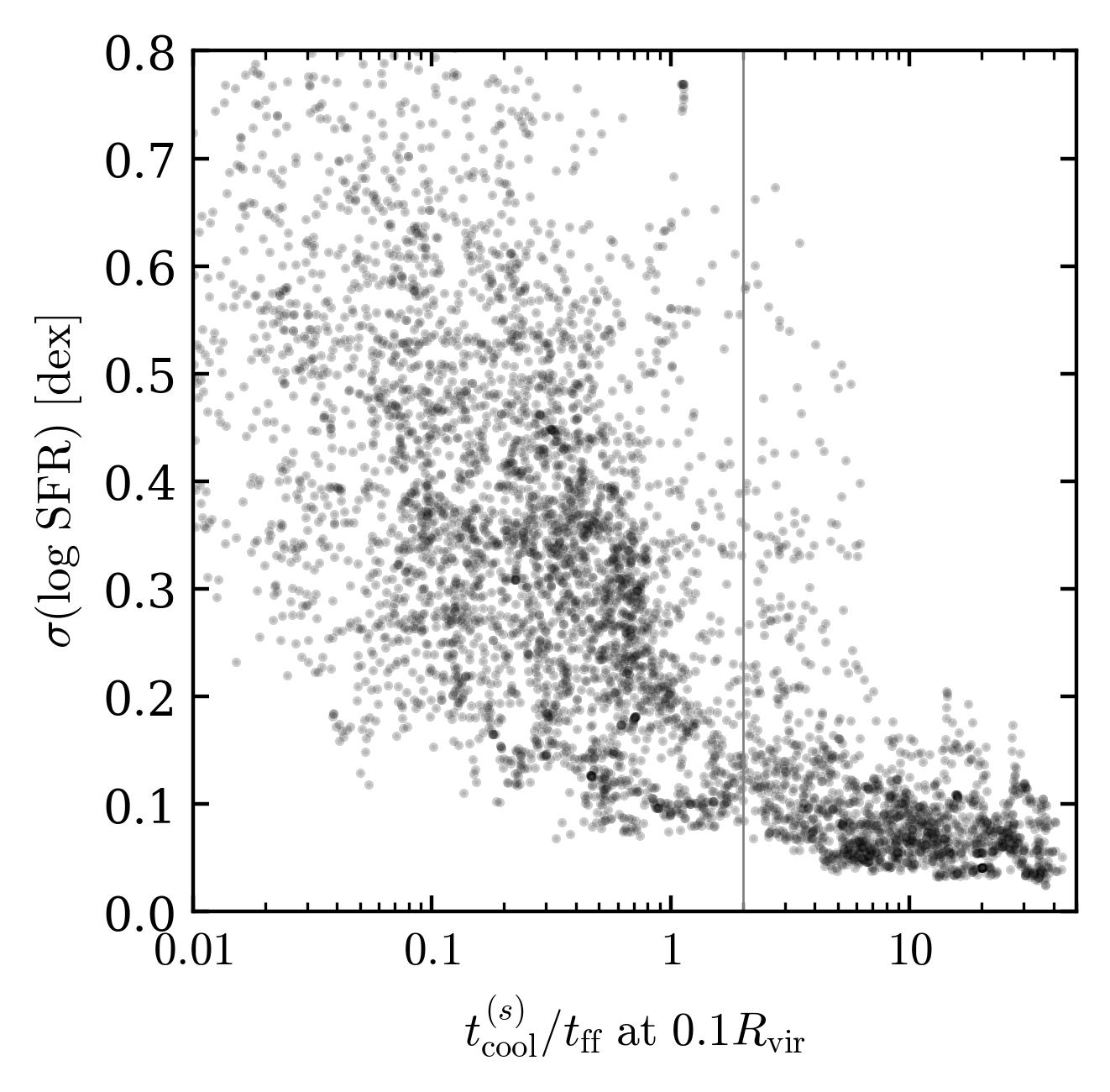}
\caption{
The SFR dispersion in $300\Myr$ windows versus the ratio of cooling time to free-fall time in the inner CGM. Gray dots denote snapshots in all 16 simulations in our sample. 
The dispersion in SFR drops when $\tcoolsh$ exceeds $\tff$ and the inner CGM virializes.
}
\label{f:SFR_all}
\end{figure}

The top-right panel in Fig.~\ref{f:Vrot to sigma} shows how the median $\Vrot$ transitions from $\approx0.5\vc$ when $\tcoolsh\ll\tff$ to $\approx\vc$ when $\tcoolsh\gg\tff$, similar to the transition seen in the mass-weighted $\langle\Vrot\rangle$ at small radii (middle panel of Fig.~\ref{f:jProfiles by sim}). 
Also, the bottom-left panel shows that $\sigmag$ decreases relative to $\vc$ when $\tcoolsh$ exceeds $\tff$, especially in the m12 subgroup. These two trends combine to a sharp transition in $\langle\Vrot\rangle/\sigmag$ when $\tcoolsh$ exceeds $\tff$. At small $\tcoolsh/\tff$ the ratio $\langle\Vrot\rangle/\sigmag$ is close to unity with a scatter of $\approx0.5$ in all subgroups, indicating dispersion-dominated kinematics. In contrast at $\tcoolsh\sim\tff$ the ratio $\langle\Vrot\rangle/\sigmag$ starts to increase, reaching $\approx6$ in the m12 subgroup and $\approx3$ in the m13 subgroup when $\tcoolsh/\tff\approx10$, indicating rotation-dominated kinematics.
The m11s do not reach values of $\tcoolsh$ significantly larger than $\tff$ (see Fig.~\ref{f:properties}). This plot thus again shows that the formation of a rotation-dominated disc indicated by $\langle\Vrot\rangle/\sigmag$ larger than unity is strongly linked to the virialization of the inner CGM indicated by $\tcoolsh/\tff$ larger than unity at $0.1\Rvir$.

\subsection{Inner CGM virialization is associated with transition to steady star formation}

In this subsection we discuss how the characteristics of the SFR change with the virialization of the inner CGM.
To focus on star formation within the central galaxy, we include only stars formed at $r<0.1\Rvir$. 
We refer to the average SFR in $10\Myr$ windows as the `instant' SFR, and to the average SFR in $300\Myr$ windows as the `mean.' These two quantities are plotted in the top row of Figure~\ref{f:SFR_Mdot} for the m12i, m12b and m13A1 simulations. 
All three simulations show a transition from large fluctuations in the instant SFR at early times to small fluctuation at late times. This transition has been previously identified for the m12 subgroup in \citet[see also \citealt{Sparre17, FaucherGiguere18}]{muratov15} and for the m13 subgroup in \cite{anglesalcazar17b}. The transition though occurs at substantially different redshifts in the different simulations, at $z<1$ in m12i and m12b compared to at $z\approx3.5$ in m13A1. 
For comparison, red lines in the bottom row of Fig.~\ref{f:SFR_Mdot} plot $\tcoolsh/\tff$ at $0.1\Rvir$. The transitions from `bursty' to `steady' SFR roughly coincide with where $\tcoolsh$ exceeds $\approx 2\tff$. 

Figure~\ref{f:SFR_all} plots the dispersion in $\log$ SFR in $300\Myr$ windows against $\tcoolsh/\tff$ in all 16 simulations in the sample. Each gray dot corresponds to a single snapshot, with $\tcoolsh/\tff$ measured at this snapshot and the $300\Myr$ window centered on the snapshot time. The figure shows that the dispersion in $\log$ SFR tends to be large when $\tcoolsh\lesssim\tff$, typically $0.2-0.6\dex$ (a factor of $1.5-4$), while it tends to be small when $\tcoolsh\gg\tff$, typically $\lesssim0.1\dex$ ($\lesssim15\%$). This result also suggests a physical connection between the burstiness of the SFR and the virialization of the inner CGM.  

\subsection{Inner CGM virialization coincides with suppression of star formation-driven galactic winds}
\label{s:wind_suppression}

The middle row of Figure~\ref{f:SFR_Mdot} plots the net mass flow rate $\Mdot$ versus redshift in the m12i, m12b and m13A1 simulations. 
The value of $\Mdot$ is calculated via:
\begin{equation}\label{e:Mdot}
 \Mdot(r) = \frac{\sum m_i v_{r, i}}{\Delta r}~,
\end{equation}
where $v_{r, i}$ is radial velocity of resolution element $i$ and the summation is over all resolution elements whose center is within a shell of thickness $\Delta\log r=0.05\dex$ and center at $r=0.1\Rvir$. 
All three simulations show two distinct phases, as previously identified for the m12 subgroup in 
\cite{muratov15} and for the m13 subgroup in \cite{anglesalcazar17b}. At early times the flow shows episodes with strong outflow bursts which exceed the inflow rate (positive net $\Mdot$). 
At late times the flow is rather steady and almost always has a net inflow. Note that the qualitative behavior across the transition is independent of the redshift at which the transition occurs. 
The bottom row demonstrates that the disappearance of outflow bursts roughly coincides with when $\tcoolsh$ exceeds $2\tff$ and the inner CGM virializes. This suggests a physical connection between the properties of stellar-driven galactic outflows and the virialization of the inner CGM.

\section{Discussion}\label{s:discussion}

In this paper we revisit the long standing questions of how the CGM virializes and how CGM virialization affects galaxy evolution, questions discussed since the advent of modern galaxy formation theory (e.g.~\citealt{White78}). 
We utilize the FIRE cosmological simulations to both extend the idealized analysis of the CGM in Papers I and II to the more realistic conditions implemented in FIRE, and to investigate the relation between CGM virialization and the simulated central galaxy. 
We demonstrate that in FIRE gas in the inner CGM goes through a transition when the local cooling time of shocked gas $\tcoolsh$ exceeds the local free-fall time $\tff$. When $\tcoolsh\ll\tff$ the inner CGM has sub-virial temperatures, supersonic velocities and large pressure fluctuations, while when $\tcoolsh\gg\tff$ the gas has virial temperatures, subsonic velocities and relatively small pressure fluctuations (Figs.~\ref{f:Tprofiles m12i} -- \ref{f:supersonic all}). The physical changes associated with this transition do not depend strongly on redshift, despite occurring over a large range of redshifts $0\lesssim z \lesssim 5$ in our simulations (Table~\ref{t:virialization}). 
We further showed that this transition in the inner CGM occurs when the outer CGM is already subsonic and has virial temperatures, indicating that the CGM virializes from the outside inward (Fig.~\ref{f:supersonic comparison}). 

Previous studies have suggested that CGM virialization facilitates the transition between blue star-forming and red-and-dead galaxies, due to the 
increase in halo gas susceptibility to black hole feedback \citep[e.g.,][]{Keres05, Croton06,Dekel06, Bower06,  Cattaneo06}. 
Our results suggest an alternative -- and in some ways orthogonal -- role for CGM virialization in galaxy evolution, in which it initially facilitates the transition between thick irregular galaxies and thin rotation-dominated discs, i.e.\ `disc settling' \citep{kassin12}. 
This connection between CGM virialization and thin discs is evident in FIRE from the relation between $\tcoolsh/\tff$ and several properties of the simulated galaxy, including the ratio of gas rotational velocity to velocity dispersion $\Vrot/\sigmag$ (Figs.~\ref{f:jProfiles by sim} -- \ref{f:Vrot to sigma}), the transition from `bursty' to `steady' star formation (Figs.~\ref{f:SFR_Mdot} -- \ref{f:SFR_all}), and the suppression of outflow bursts driven by star-formation (Fig.~\ref{f:SFR_Mdot}). 
In this section we discuss several aspects of our results.

\subsection{Why does the CGM virialize from the outside in?}\label{s:outside in formation}

We find that the inner CGM virializes after the outer CGM is virialized, i.e.~with time the radius where $\tcoolsh\approx\tff$ moves inward with respect to $\rvir$.\footnote{In physical units the radius where $\tcoolsh\approx\tff$ either decreases or increases with time. We emphasize that we refer here to the `sonic radius' where $\tcoolsh\approx\tff$, rather than to the `cooling radius' where $\tcoolsh=\thubble$ (see \S\ref{s:theoretical bkg}).} This outside-in virialization scenario is opposite to the direction of virialization in the 1D simulations of \cite{Birnboim03} and \cite{Dekel06}, which account for radiative cooling and angular momentum but neglect feedback from the galaxy. 
In their simulations IGM inflows are initially cool, free-falling and supersonic down to the disc radius $\Rcirc$. When the halo mass exceeds a threshold of $\sim10^{11.5}\msun$ a shock forms at $\Rcirc$ and moves outwards, i.e.~the postshock subsonic phase forms first in the inner halo and then expands into the outer halo. 
In contrast in FIRE the CGM becomes predominantly subsonic near the disc radius after it is already subsonic at larger radii (Figs.~\ref{f:Tprofiles m12i}, \ref{f:Tprofiles h2}, and \ref{f:supersonic comparison}).

An outside-in CGM virialization scenario is consistent with expectations from spherical steady-state CGM solutions without ongoing heating by feedback (Papers I and II). 
In such solutions $\tcoolsh/\tff$ increases with radius and with halo mass. At radii where $\tcoolsh>\tff$ the gas is hot and subsonic since radiative cooling is balanced by compressional heating of the inflow, while at radii where $\tcoolsh<\tff$ radiative cooling is rapid so the gas is cool and supersonic. These solutions thus indicate that the hot phase can be long-lived (i.e.~reach steady-state) only at radii where $\tcoolsh>\tff$. If we apply these conclusions to a time-dependent scenario where the halo mass is growing and shocks which seed the hot phase are prevalent, then we expect a long-lived hot phase to form first at large radii and later at small radii. These expectations based on steady-state solutions appear to hold in FIRE: a long-lived hot CGM phase exists only when $\tcoolsh$ exceeds $\tff$ (Figs.~\ref{f:T all} -- \ref{f:supersonic all}), $\tcoolsh/\tff$ increases outwards (Fig.~\ref{f:t ratio comparison}), and the fraction of gas moving supersonically drops first in the outer CGM and then in the inner CGM (Fig.~\ref{f:supersonic comparison}). 

We emphasize that hot subsonic gas often exists in the inner CGM also prior to virialization, in contrast with the steady-state solutions. However the volume-filling fraction of this hot gas fluctuates rapidly, with typical values less than 50\% (Figs.~\ref{f:m11d image} and \ref{f:T all}--\ref{f:supersonic all}). The rapid fluctuations prior to virialization are likely due to cycles of rapid cooling followed by heating by outflow bursts \citep[see][]{vandeVoort16}. Hot gas powered by outflows will also propagate outwards, in contrast with the inward direction in which the CGM virializes. Thus, the FIRE simulations appear to be consistent with the steady-state solutions in the sense that when $\tcoolsh>\tff$ a time-steady hot phase forms with volume filling fraction approaching unity (i.e., the CGM `virializes'). The FIRE simulations, though, exhibit also a transient hot phase prior to virialization which is absent from the steady-state solutions.

One may wonder why the outside-in scenario suggested by cooling flow solutions is not realized in the simulations in \citeauthor{Birnboim03}. We emphasize that while steady-state solutions imply a subsonic hot phase would be long-lived at radii where $\tcoolsh>\tff$, an initial shock is still required to seed a subsonic hot phase at these radii. At large radii such initial shocks could be a result of the interaction of inflows with outflows as seen in the idealized simulations of \cite{Fielding17}, or as a result of merger events \citep[e.g.,][]{Shi20}. As these mechanisms are absent from the 1D simulations of \citeauthor{Birnboim03}, a subsonic hot phase may not form even if the conditions for it to be long-lived are satisfied.

We note that while our results indicate that the CGM virializes from the outside inward, we do see evidence for some, more subtle changes in the properties of the outer CGM  \emph{following} the virialization of the inner CGM. This includes the decrease in temperature beyond $\Rvir$ when $\tcoolsh$ exceeds $\tff$ at $0.1\Rvir$, and the associated appearance of an accretion shock (Figs.~\ref{f:Tprofiles m12i} -- \ref{f:Tprofiles z 0}). Also, the volume-filling fraction of supersonic gas in the outer halo appears to decrease from $\approx0.3$ to $\approx0.1$ when the inner CGM virializes (Figs.~\ref{f:Tprofiles m12i} and \ref{f:supersonic comparison}). Potentially, these effects are due to the stifling of outflows when the inner CGM virializes (Fig.~\ref{f:SFR_Mdot}), which affects the physical conditions also in the outer halo. We leave exploring this effect to a future study. 

Outside-in virialization has several implications for the X-ray and SZ (\citealt{sunyaevzeldovich70}) signals from the halo. As mentioned in the introduction, \cite{vandeVoort16} identified in FIRE a transition between highly variable X-ray emission at halo masses below $\Mthres\approx10^{12}\msun$ to time-steady X-ray emission above $\Mthres$, a transition which they associated with CGM virialization. Our results add to their conclusions by demonstrating that it is the \emph{inner} CGM which virializes at this mass scale in FIRE. 
Indeed, the X-ray emission is dominated by the densest gas and is thus most sensitive to the physical conditions in the inner CGM.
In contrast, the SZ signal is roughly weighted by mass and hence dominated by gas in the outer halo, so a drop in the SZ signal is expected only at lower halo masses $\lesssim 10^{11}\msun$ where the outer halo has not yet formed (the median halo mass at which $\tcoolsh$ exceeds $\tff$ at $0.5\Rvir$ in our simulations is $0.8\cdot10^{11}\msun$). 
We note that additional effects such as gas depletion in low mass halos (\citealt{vandeVoort16,oppenheimer20}) may also affect the SZ and X-ray signals and thus complicate the interpretation. We postpone a more quantitative analysis of the implications of our results for these observational signatures to future work. 

\subsection{The threshold halo mass for inner CGM virialization}\label{s:Mthres}

Table~\ref{t:virialization} lists for each simulation the halo mass and other physical properties in the snapshot where $\tcoolsh$ equals $2\tff$ and the inner CGM virializes. 
Defining the halo mass when $\tcoolsh=2\tff$ as $\Mthres$, we find $\Mthres\approx0.8 - 1.5\cdot 10^{12}\msun$ in the m12 simulations which virialize at redshift $z<1$ and a somewhat higher $\Mthres\approx1.4-2.9\cdot10^{12}\msun$ in the m13 simulations, which virialize at higher redshift. The somewhat lower $\Mthres$ for halos which virialize at low redshift are mainly driven by the lower gas density relative to the cosmic mean and the higher $\fvc$ factor (see eq. \ref{e:fvc}), where the latter is a result of the higher concentration of low redshift halos. This can be seen by comparing the halo and gas properties of the m12 and m13 simulations at virialization (Table~\ref{t:virialization}) with the analytic calculation of $\tcoolsh/\tff$ (eqn.~\ref{e:tratio by Mhalo}; note $\tcoolsh/\tff \propto \fvc^{4.4}$). 

\begin{figure}
\includegraphics{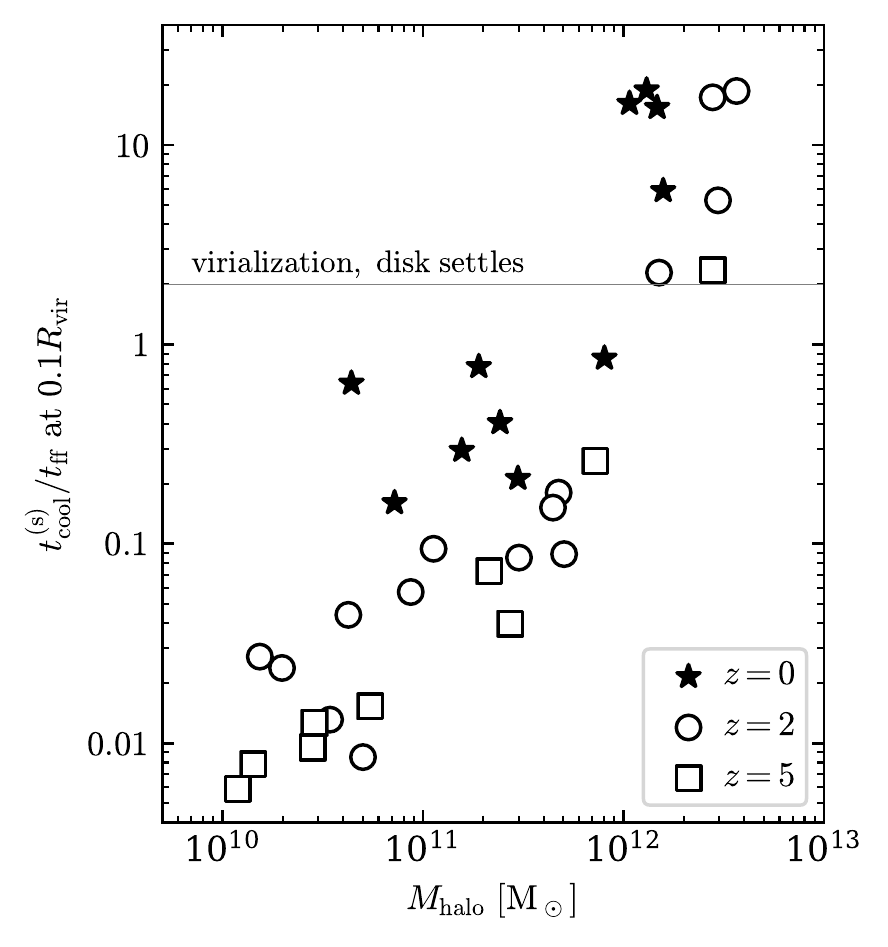}
\caption{The relation between halo mass and cooling to free-fall time ratio in the inner CGM at different redshifts in the FIRE simulations. The horizontal line denotes the approximate threshold in $\tcoolsh/\tff$ for inner CGM virialization and the formation of rotation-dominated discs. Note that in $10^{10.5}-10^{11.5}\msun$ halos at $z=0$ the ratio $\tcoolsh/\tff$ can approach the threshold, suggesting that virialized inner CGM and settled discs would form also in these halos if their $\tcoolsh/\tff$ is somewhat underpredicted in the FIRE simulations analyzed here (see \S\ref{s:Mthres}). 
}
\label{f:mvir vs tratio}
\end{figure}

Although we deduce a weak dependence of $\Mthres$ on redshift, we find that in halo masses of $10^{10.5}-10^{11.5}\msun$ at $z\sim0$ the ratio $\tcoolsh/\tff$ can be comparable to unity, in contrast with $\tcoolsh\ll\tff$ at the same halo masses at high redshift. This trend is shown in Figure~\ref{f:mvir vs tratio} in which we plot $\tcoolsh/\tff$ as a function of halo mass for different redshifts.  This difference is again due to the lower gas mass and higher concentration of low-redshift halos compared to their high-redshift counterparts.  
The result that $\tcoolsh/\tff$ can be close to unity in $\ll10^{12}\msun$ halos at $z\sim0$ implies that the properties of their CGM and central galaxies strongly depend on the mass and metallicity of the CGM, which in turn depend on the integrated enrichment and depletion of the CGM by outflows over cosmic time. 
Halo-to-halo variance in these quantities, or if these quantities are somewhat overpredicted by FIRE, could drive $\tcoolsh/\tff$ above unity in $\sim 10^{11}\msun$ halos at $z\sim0$, allowing them to develop virialized inner CGM and rotation-dominated disks. Our results thus allow for the possibility that $\Mthres$ decreases to $\sim10^{11}\msun$ or less at late cosmic times. 

The possibility that $\Mthres(z\sim0)\sim10^{11}\msun$ could explain the tension between FIRE and observations of low redshift SF galaxies discussed in \cite{elbadry18a,elbadry18b}. They showed that in FIRE only above a stellar mass of $\sim10^{10}\msun$ do galaxies predominantly form rotation-dominated discs, in contrast with observations, which find rotation dominated discs above $\sim10^9\msun$. If as suggested above the ratio $\tcoolsh/\tff$ is somewhat above unity at $\Mhalo\sim10^{11}\msun$ rather than comparable to unity as in FIRE, then we would expect also $\Vrot\gg\sigmag$  (Fig.~\ref{f:Vrot to sigma}) at the corresponding stellar mass of $\sim10^9\msun$, which would be more consistent with observations. Similarly, $\tcoolsh>\tff$ at $\Mhalo\sim10^{11}\msun$ would also suggest a steady rather than bursty SFR at stellar masses $\gtrsim10^9\msun$ (Fig.~\ref{f:SFR_all}), which could explain the apparent overprediction of SFR burstiness in FIRE at this mass scale \citep{Sparre17,Emami19}.

It is worth noting also that when $\tcoolsh$ exceeds $\tff$ and the inner CGM virializes the metallicity at $0.1\Rvir$ stops increasing and in some cases starts to drop with time (middle-left panel of Fig.~\ref{f:properties}). This is likely due to the suppression of outflows when the inner CGM virializes (Fig.~\ref{f:SFR_Mdot}), so the CGM becomes dominated by fresh accretion and previously ejected winds, both of which tend to have a lower metallicity than new winds (see \citealt{hafen19} for the metallicities of different components). 
This causes the CGM and ISM metallicies to diverge, a phenomenon previously identified by \cite{muratov17}, and our results suggest that this divergence is associated with the virialization of the inner CGM. 
Also, the drop in CGM metallicity causes $\tcoolsh/\tff$ to further increase once it crosses the threshold for virialization. 

\subsection{How can inner CGM virialization enable disc settling?}\label{s:hot halo disc connection}

The virialization of the inner CGM is expected to both confine galaxy outflows and change the nature of inflows onto the galaxy.
In this subsection we discuss these transformations and how they might cause the associated transitions in galaxy properties shown in Figs.~\ref{f:jProfiles by sim} -- \ref{f:SFR_all}. 

\newcommand{\Teq}{T_{\rm eq}}

The virialized inner CGM extends down to a few density scale heights above the galaxy midplane (see Fig.~\ref{f:m12i image} and top row of Fig.~\ref{f:h206 h2 image}), and provides a homogeneous confining medium for the gaseous galactic disc. The homogeneity of a virialized CGM is evident from the relatively small pressure fluctuations shown in the bottom-right panel of Fig.~\ref{f:m12i_multiple_images}, in contrast with the large fluctuations prior to virialization when the flow is supersonic and dynamic pressure dominates (bottom-left panels in Fig.~\ref{f:m12i_multiple_images}). A homogeneous medium is more effective in confining outflows since the outflows cannot expand through paths of least resistance. 
For example, in the $z=1$ snapshot of m12i at which $\tcoolsh=0.25\tff$ (see Fig.~\ref{f:m12i_multiple_images}), half the volume at $\sim0.1\Rvir$ is filled with gas with a pressure equal to $15\%$ or less of the mean (volume-averaged) pressure, while a tenth of the volume is filled with gas with a pressure of $2\%$ or less than the mean pressure. 
Similar volume-filling fractions are found for regions with a low density compared to the mean value. 
The required outflow ram pressure to expand through such under-pressurized and underdense regions is lower, by similar factors, than the required ram pressure to expand in a homogeneous medium with the same mean pressure. 

How does confinement affect the physical conditions in the galaxy? 
Hydrodynamic simulations and analytic considerations suggest that without confinement, a `superbubble' produced by supernovae (SNe) of a single stellar cluster can breakout of the disc and drive a strong outflow \citep[e.g.,][]{kim17,fielding18}, which significantly perturbs the disc vertical structure \citep{martizzi20}. 
The disappearance of outflow bursts when the inner CGM virializes (Fig.~\ref{f:SFR_Mdot}) may suggest that the homogeneous virialized CGM confines such outflows. This can be understood by comparing the CGM thermal pressure with the expected pressure in SNe-powered superbubbles.
Assume a CGM characteristic of the Milky Way with pressure $P_{\rm h}/k\sim1000\cm^{-3}\K$ (estimated from the pressure of high-velocity clouds, \citealt{Dedes10}) at the disc scale of $R_{\rm d}\approx\Rcirc\approx 10\kpc$. The work required to lift the virialized CGM beyond the scale of the disc is $\sim (4\pi/3) R_{\rm d}^3 P_{\rm h} \sim 2 \cdot 10^{55}\erg$. For comparison, the formation of a stellar cluster with mass $M_{\rm cl}=10^5 M_5\msun$ causes $\sim 1000 M_5$ core-collapse SNe to explode and inject an energy of $E\sim 10^{54}M_5\erg$ into their surroundings. 
A virialized CGM similar to that of the Milky Way is thus capable of confining a superbubble produced by a stellar cluster as massive as $\sim 2\cdot 10^6\msun$, or more if cooling losses in the bubble are substantial. 
Since $P_{\rm h}R_{\rm d}^3$ is roughly a constant fraction of the halo gas gravitational binding energy (assuming $R_{\rm d}\propto \Rvir$), which scales as $\Mhalo^2/\Rvir$, confinement is even more effective at higher redshift where $\Rvir$ is smaller. 
A similar conclusion that the CGM of Milky-Way mass galaxies can confine stellar-driven galactic outflows was reached by \cite{Hopkins20_winds}, who implicitly assumed the CGM was virialized (see section 3.3.1 there).

If superbubbles produced by individual stellar clusters are confined, then multiple superbubbles produced by different stellar clusters would build up pressure in the ISM until it equals that of the inner CGM. Any additional injected feedback energy not lost to radiation would then be channeled to an outflow, e.g.~via buoyant bubbles.\footnote{Note that buoyancy arguments which assume pressure equilibrium between the outflow and ambient medium (e.g.~\citealt{Bower17}) are unlikely to be applicable prior to virialization since the ambient medium does not have a uniform pressure. }
We thus suspect that inner CGM virialization fundamentally changes how feedback energy is distributed. Prior to virialization the energy injected by individual superbubbles vents relatively easily through the porous inner CGM, causing large spatial and temporal pressure fluctuations in the galaxy disc and inner halo. 
In contrast after virialization the feedback energy is confined, and will regulate the pressure in the ISM. That is, we suspect that the classic ansatz that SNe keep the characteristic ISM pressures uniform \citep{mckeeostriker77} applies only after the inner CGM virializes. 
This transition could explain the sharp increase in $\Vrot/\sigmag$ when the inner CGM virializes (Fig.~\ref{f:Vrot to sigma}) since the large pressure gradients prior to virialization would induce non-rotational motions and thus enhance $\sigmag$, while after virialization these pressure gradients would drop and $\sigmag$ would decrease. 
Also, the transition to pressure balance can enable galaxies to realize a Kennicutt-Schmidt-type star formation relation as in equilibrium ISM models \citep[see \citealt{Gurvich20} for a test of these models in FIRE]{Thompson05,OstrikerShetty11, FaucherGiguere13}, which could explain the transition to a steady SFR (Fig.~\ref{f:SFR_all}).

A similar argument for confinement of outflows applies to the outer CGM, during the `transonic' phase of halo gas in which the outer halo is smooth and subsonic while the inner halo is clumpy and supersonic. 
During this phase outflows would potentially halt at the `sonic radius' which separates between the two layers. 
It would thus be interesting to compare the sonic radius in simulations (roughly the radius where $\tcoolsh(r)=\tff(r)$) to the typical maximum radius reached by outflows, known as the `recycling radius' (\citealt{Oppenheimer10, Ford14, anglesalcazar17a, muratov17, hafen19}). 

We note that a prerequisite for outflows to be confined by the inner CGM is that they manage to break out of the disc, which depends on stellar cluster mass \cite[$M_{\rm cl}\gtrsim10^5\msun$ for the fiducial parameters in][]{fielding18}. Thus an alternative possibility for the suppression of outflows following inner CGM virialization is that $M_{\rm cl}$ decreases below the threshold for breakout. This could be induced by the drop in scale height $h$ when the disc settles, if $M_{\rm cl}$ scales with the Toomre mass which in turn scales as $h^2$.

\subsection{Comparison to the distinction between `hot mode' and `cold mode' accretion}


Inner CGM virialization initiates the `hot accretion mode', in which the inflow is subsonic (i.e., pressure supported) from the accretion shock down to the galaxy. 
Prior to inner CGM virialization, the accretion flow resembles the previously discussed `cold accretion mode' \citep[e.g.,][]{Keres05, Keres09, dekelnature09}, in the sense that the gas accretes supersonically onto the galaxy. However, our results suggest that prior to inner CGM virialization the thermal history of accreted gas can differ from what has often been envisioned for cold mode accretion. In intermediate-mass ($\sim10^{11}-10^{12}\msun$) halos in FIRE, the outer CGM is virialized, corresponding to a hot inflow at large radii followed by a cool flow at small CGM radii. 
In contrast, it is often assumed that in cold mode accretion the gas remains cool throughout the CGM. This difference suggests that the maximum temperature reached by gas prior to accretion, which was the focus of many previous studies of the cold mode versus the hot mode in cosmological simulations \citep[e.g.,][]{Nelson13}, is not in general an accurate indicator of gas properties upon accretion onto the galaxy. 

The accretion of cold gas onto galaxies can persist after inner CGM virialization if cool streams penetrate the hot phase, as expected at high redshift when cosmic web filaments are narrow relative to $\gtrsim10^{12}\msun$ halos (e.g.~\citealt{Keres05,Dekel06}), and as seen in high-redshift snapshots in FIRE (Fig.~\ref{f:h206 h2 image}). The relation between inner CGM virialization and the distribution of physical states of the gas as it accretes onto galaxies is thus redshift-dependent. 
Cold accreting gas is expected to stir turbulence in the gaseous disc when the supersonic accretion flow shocks against the ISM \citep{dekelsariceverino09}, in contrast with hot mode accretion where the subsonic flow is expected to smoothly connect to the disc velocity field (see figures~2 and 7 in Paper II, and also \citealt{Cowie80}). 
This difference may explain the decrease in $\sigmag$ when $\tcoolsh$ exceeds $\tff$ at low redshift and the cold mode disappears, which is not seen when $\tcoolsh$ exceeds $\tff$ at high redshift but accretion via cold streams persists (compare the `m12' curve with the `m13' curve in the bottom panels of Fig.~\ref{f:Vrot to sigma}). 
However, the transitions in galaxy properties highlighted in \S\S \ref{s:disc_formation}--\ref{s:wind_suppression} do not appear to depend on redshift for a fixed $\tcoolsh/\tff$, and thus appear to be independent of the existence of cold filaments. 
This suggests that these transitions are more directly related to inner CGM virialization and the associated confinement of outflows discussed in \S\ref{s:hot halo disc connection}, rather than to the relative fractions of cold and hot gas accreting onto the central galaxy.

\subsection{Additional physics: black hole feedback and cosmic rays}\label{s:AGN}

Feedback from the central black hole (BH) is believed to have a central role in the evolution of galaxies, and forms the leading paradigm for quenching star formation in red-and-dead galaxies \citep[e.g.,][]{SomervilleDave15,naabostriker17}. 
The FIRE simulations used in this work passively followed BH accretion assuming it proceeds via gravitational torques in the circumnuclear material \citep[see][]{anglesalcazar17b} but neglected AGN feedback, which in principle could limit the applicability of our conclusions regarding CGM virialization. 
We argue here that, for a wide range of supermassive black hole growth and feedback models, AGN feedback can likely be assumed to have only modest or negligible effects on our main results regarding virialization. This follow from the result of \cite{anglesalcazar17b} that BH growth experiences a transition coincident with the other galactic transitions discussed above.
At early times when the SFR is bursty BH accretion is weak and irregular, and the BH mass remains $\lesssim 10^5\msun$. At later times when the SFR becomes steady BH accretion becomes significantly stronger and more steady, and the BH mass increases to $\sim0.1\%$ of the galaxy mass. Byrne et al.~(in prep.) demonstrates that this transition to significant BH accretion also coincides with when $\tcoolsh$ exceeds $\tff$ at $0.1\rvir$. 
If, as suggested by the results just described, substantial BH growth commences only after the inner CGM virializes, then the energy released by BH feedback is likely small before and up to the virialization of the inner CGM. 
We thus do not expect BH feedback to significantly alter the conditions in the CGM at this early phase, and hence not to alter our conclusions regarding inner CGM virialization. 

The properties of halo gas may also be modified by interaction with cosmic rays (CRs), which are not accounted for in the simulations used here but which some studies have suggested could be an important source of feedback \citep[e.g.,][]{Booth13,salem16,Pfrommer17}. Explicit injection of CRs by supernovae, CR transport and CR-gas interactions have been implemented in a separate suite of FIRE simulations (\citealt{chan19,Ji20,hopkins20_whatabout}). 
\cite{hopkins20_whatabout} demonstrated that in the m12 simulations non-thermal pressure gradients in the CR fluid can potentially support the halo gas against gravity at low redshift, thus effectively increasing the free-fall time and preventing the formation of a pressure-supported virialized CGM. This conclusion is however sensitive to the assumed CR transport model, which is uncertain especially in the CGM \citep{Hopkins20_differentCRmodels}. Thus, it is not yet clear whether and how CRs affect our conclusions concerning CGM virialization. 

\subsection{Implications for SAMs}

Our results differ from the prescriptions typically employed by semi-analytic models (SAMs) of galaxy formation in three main ways. First, SAMs usually determine whether the inner CGM virializes by comparing the cooling time to the dynamical time at the virial radius \citep[or equivalently they compare the cooling radius with $\rvir$, e.g.,][]{Somerville08_SAM,Lu11,Croton16}. Our results in contrast suggest that CGM virialization is complete and hence the nature of accretion onto the galaxy and confinement by the CGM change only when $\tcoolsh$ exceeds $\tff$ at the galaxy scale $\approx 0.1\Rvir$ (Figs.~\ref{f:T all}--\ref{f:supersonic all}), as expected from the steady-state solutions in Paper II. 
Second, SAMs typically assume central galaxies form thin discs with a specific angular momentum comparable to that of their parent halo \citep[e.g.,][]{Somerville08_SAM}. As far as we are aware the thermal properties of halo gas do not play a role in shaping discs in SAMs, and so these models may be missing a critical ingredient which enables the formation of thin rotationally-supported disks. It would thus be interesting to incorporate into SAMs the connection between the virialization of the inner CGM and the formation of discs found above (Figs.~\ref{f:jProfiles by sim} -- \ref{f:Vrot to sigma}). The predictions of such SAMs could then be compared to the observed demographics of blue irregulars with $\Vrot\sim\sigmag$ and blue thin disks with $\Vrot\gg\sigmag$ \citep[e.g.,][]{kassin12,Karachentsev13_catalog,simons17}. 
Third, many SAMs assume there is a direct connection between CGM virialization and SF quenching, which is implemented by `turning on' radio mode feedback from the black hole when the condition for CGM virialization is satisfied, thus offsetting cooling and shutting the gas supply for further SF \citep{Croton06, Bower06, Bower12, Cattaneo06, Cattaneo08, Somerville08_SAM, Lu11}.
Our results indicate that inner CGM virialization is instead more directly associated with disc settling, suggesting that the connection between virialization and quenching is not so direct. Potentially, if disc settling enables accelerated BH growth as discussed in the previous section, CGM virialization would remain associated with galaxy quenching by BH feedback, but there may be a non-negligible delay between inner CGM virialization and the time needed for sufficient BH accretion energy to be released.

\section{Summary}\label{s:summary}

In this work we use the FIRE-2 cosmological zoom-in simulations to study how the CGM virializes, and how CGM virialization is connected to the evolution of the central galaxy. We utilize a suite of simulations spanning a large range of halo mass assembly histories, including halos which reach a mass of $10^{12}\msun$ over a wide redshift range of $0 \lesssim z \lesssim 5$, and halos which at $z=0$ have a mass of $10^{10.5}-10^{12}\msun$. Our results can be summarized as follows:

\begin{enumerate}
 \item At times when the cooling time of shocked gas $\tcoolsh$ (eqn.~\ref{e:tcool}) is shorter than the free-fall time $\tff$, the volume-weighted temperature of the inner CGM ($\lesssim0.1\Rvir$) is typically $\ll\Tvir$, with occasional snapshots where $T>\Tvir$. In contrast when $\tcoolsh\gg\tff$ the temperature of the inner CGM is consistently $\sim\Tvir$ (Fig.~\ref{f:T all}). This transition in temperature is accompanied by a drop in the volume fraction of gas with supersonic radial velocities (Fig.~\ref{f:supersonic all}) and a transition from large spatial pressure fluctuations to a more spherically-symmetric pressure distribution (Figs.~\ref{f:m12i_multiple_images}).  We identify this transition as the \emph{virialization of the inner CGM}. 
 
 \item The inner CGM virializes when the outer CGM ($\sim0.5\Rvir$) is already predominantly subsonic and has a temperature $\sim\Tvir$, indicating an outside-in virialization scenario (Figs.~\ref{f:Tprofiles m12i} -- \ref{f:h206 h2 image} and \ref{f:supersonic comparison}). This scenario is consistent with expectations based on steady-state solutions where $\tcoolsh/\tff$ increases with halo radius (Papers I and II), and in contrast with the inside-out scenario indicated, e.g., by the 1D virial shock simulations of \cite{Birnboim03}.

 \item In our simulations $\tcoolsh$ exceeds $\tff$ and the inner CGM virializes when the halo mass surpasses $\sim10^{12}\msun$, roughly independent of the redshift at which this mass is reached (Table~\ref{t:virialization}). The outer CGM virializes above a lower halo mass of $\sim10^{11}\msun$ (e.g.~Fig.~\ref{f:Tprofiles z 0}). However, in halo masses $\sim10^{11}\msun$ at $z\sim0$, $\tcoolsh$ can approach $\tff$ also in the inner CGM (Fig.~\ref{f:mvir vs tratio}). 
This suggests that if actual CGM densities or metallicities are a factor of $2-3$ lower than predicted by FIRE, such that $\tcoolsh>\tff$, then at low redshift the inner CGM may virialize even in $\ll10^{12}\msun$ halos.

 \end{enumerate}

The virialization of the inner CGM coincides with several transitions in the properties of the central galaxy and of SF-driven outflows, which collectively have significant implications for interpreting observations of star formation and galactic winds in galaxies and how they correlate with the CGM: 

 \begin{enumerate}
 \setcounter{enumi}{3}
 \item We find that Milky Way-mass and more massive galaxies in FIRE-2 experience a transition from a disordered velocity field to a rotation-dominated disc, over a wide range of redshifts $0\lesssim z \lesssim 5$. This is consistent with previous results using the FIRE-1 simulations \citep{ma18, GarrisonKimmel18}. We demonstrate that the transition to rotation-dominated discs (`disc settling') coincides with the virialization of the inner CGM (Figs.~\ref{f:jProfiles by sim} -- \ref{f:Vrot to sigma}), which differs from the common assumption in SAMs that CGM virialization is associated with SF quenching \citep[e.g.,][]{ Croton06}. 
 Disc settling is also suggested by observations of star-forming galaxies \citep{kassin12, simons17}, albeit above a somewhat lower mass threshold in the local Universe than in FIRE \citep{elbadry18a,elbadry18b}. This may indicate that densities/metallicities in the halos of some local dwarfs is lower than in FIRE, thus promoting CGM virialization and the formation of rotation-dominated discs.
 
 \item We find that star formation in FIRE evolves in time, from occurring in bursts at higher $z$ to being distributed more uniformly in time at lower $z$, consistent with previous results \citep{muratov15, Sparre17, anglesalcazar17b,FaucherGiguere18}. We find that this transition coincides with the virialization of the inner CGM, and is independent of the redshift at which virialization occurs (Figs.~\ref{f:SFR_Mdot} -- \ref{f:SFR_all}).
 
 \item We show that the transition in SFR coincides with suppression of stellar-driven galactic outflows, consistent with \cite{muratov15} and \cite{anglesalcazar17a,anglesalcazar17b}. Suppression of outflows thus also coincides with the virialization of the inner CGM (Fig.~\ref{f:SFR_Mdot}).
 \end{enumerate}

We hypothesized that a virialized inner CGM enables the formation of stable discs because its uniform pressure confines superbubbles powered by clustered SNe, allowing them to enforce pressure balance in the ISM. In contrast, superbubbles can escape the ISM more easily through paths of least resistance in the clumpy CGM prior to virialization. SNe-driven outflows may also be inherently less powerful following virialization if there is an associated drop in the characteristic mass of stellar clusters. 
The virialization of the inner CGM also enables the hot accretion mode in which rotating hot gas smoothly accretes onto the disc's outskirts, which may also be conducive to the formation of stable discs as suggested by \cite{Sales12}. 
In a follow-up study we will further explore the implications of CGM virialization for the physical conditions in the ISM. 
 
To conclude, our analysis suggests that the inner CGM of blue galaxies with an irregular `thick disc' morphology is predominantly cool ($T\ll\Tvir$), supersonic and clumpy, compared to a hot ($T\sim\Tvir$), subsonic and smooth inner CGM around thin disc galaxies. This prediction can be tested  observationally by comparing the morphology of blue galaxies with the emission and absorption signatures of gas in the inner CGM.

\acknowledgements

We thank Andrey Kravtsov and Clarke Esmerian for useful discussions, and the anonymous referee for a highly insightful and thorough report. 
JS is supported by the CIERA Postdoctoral Fellowship Program.
CAFG was supported by NSF through grants AST-1517491, AST-1715216, and CAREER award AST-1652522, by NASA through grant 17-ATP17-0067, by STScI through grants HST-GO-14681.011, HST-GO-14268.022-A, and HST-AR-14293.001-A, and by a Cottrell Scholar Award and a Scialog Award from the Research Corporation for Science Advancement. 
DF is supported by the Flatiron Institute, which is supported by the Simons Foundation.
EQ was supported in part by a Simons Investigator Award from the Simons Foundation and by NSF grant AST-1715070.
AG was supported by the National Science Foundation and as a Blue Waters graduate fellow. 
TKC is supported by Science and Technology Facilities Council astronomy consolidated grant ST/T000244/1. 
RF acknowledges financial support from the Swiss National Science Foundation (grant no 157591).
DK was supported by NSF grant AST-1715101 and the Cottrell Scholar Award from the Research Corporation for Science Advancement. 
AW received support from NASA through ATP grant 80NSSC18K1097 and HST grants GO-14734, AR-15057, AR-15809, and GO-15902 from STScI; the Heising-Simons Foundation; and a Hellman Fellowship.
Some of the simulations were run using XSEDE (TG-AST160048), supported by NSF grant ACI-1053575, and Northwestern University's computer cluster `Quest'.


\bibliography{fire}{}
\bibliographystyle{aasjournal}

\appendix

\section{Cooling time and density in the inner CGM}\label{a:density}

\begin{figure*}
\begin{center}
 \includegraphics{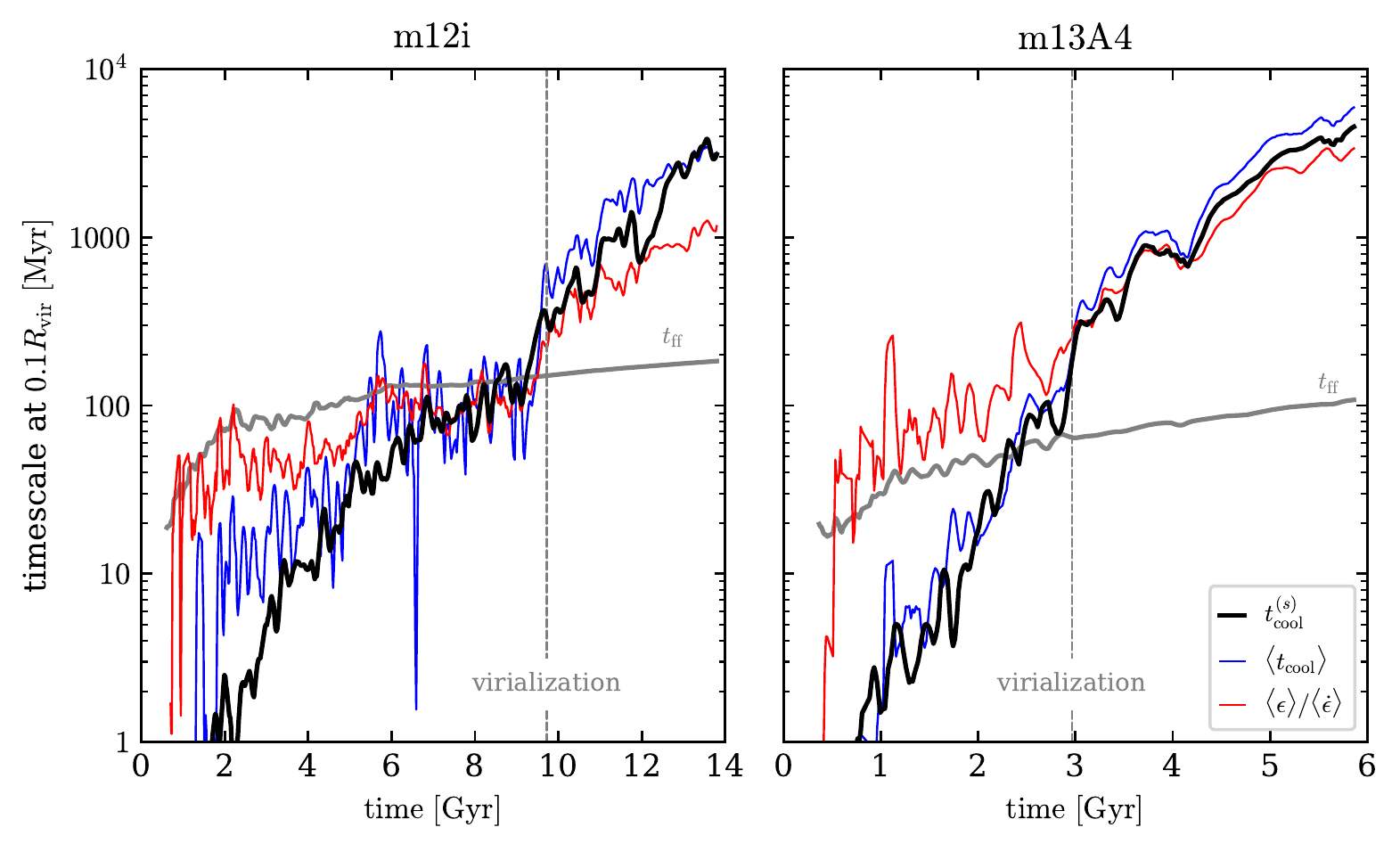}
\end{center}
 \caption{
 The cooling time of shocked gas used in this work (black, eqn.~\ref{e:tcool0}) versus other cooling time averages in FIRE. 
 Left and right panels show the m12i and m13A4 simulations in which the inner CGM virializes at $t\approx10\Gyr$ ($z=0.36$) and $t\approx3\Gyr$ ($z=2.2$), respectively. The mass-weighted average $\tcool$ in a shell at $0.1\Rvir$ is plotted in blue, while the ratio of the shell energy to its luminosity is plotted in red. After virialization the different cooling times are comparable, though they can differ substantially prior to virialization. Similar behaviour is seen in other simulations in our sample. This result supports our assumption that $\tcoolsh$ is an estimate of the gas cooling time if it were virialized. 
 }
 \label{f:tcool comparison}
\end{figure*}

In this section we compare the expected cooling time in a virialized CGM $\tcoolsh$ (eqn.~\ref{e:tcool0}) with other averages of the CGM cooling time. To this end we measure the cooling time of individual particles (see eqn.~\ref{e:tcool0})
\begin{equation}\label{e:tcool actual}
 \tcool = \frac{(3/2)\cdot 2.3 k_{\rm B}T}{\nH\Lambda(\nH,T,Z,z)},
\end{equation}
using the temperature, density and metallicity of each particle. We then take the mass-weighted average of all particles in a shell centered at $0.1\Rvir$ with width of $0.05\dex$. This average is plotted versus time in Figure~\ref{f:tcool comparison} for two FIRE simulations (blue curves, marked as $\langle\tcool\rangle$).
We also plot the ratio of the total energy of particles in the shell with the total luminosity of particles in the shell (red, marked as $\langle\epsilon\rangle/\langle\dot{\epsilon}\rangle$), which gives an estimate of the shell cooling time in the limit that the energy of individual particles is efficiently exchanged via hydrodynamics interactions. Black curves plot $\tcoolsh$ versus time.
Fig.~\ref{f:tcool comparison} shows that after virialization of the inner CGM (marked by a vertical line) the different cooling times estimates are comparable to a factor of $2-3$.  Prior to virialization $\tcoolsh\sim\langle\tcool\rangle$ except at early times in m12i, while $\langle\epsilon\rangle/\langle\dot{\epsilon}\rangle$ is significantly larger than $\tcoolsh$ and comparable to $\tff$. Similar behavoir with respect to the epoch of virialization is seen in the other 14 simulations of our sample. This result that $\tcoolsh$ is comparable to other cooling time averages after virialization supports our assumption that $\tcoolsh$ is an estimate of the gas cooling time if it were virialized. The result that $\langle\epsilon\rangle/\langle\dot{\epsilon}\rangle\sim\tff$ prior to virialization suggests that in the bursty phase heating and cooling occur on a dynamical timescale.

\begin{figure*}
\begin{center}
 \includegraphics{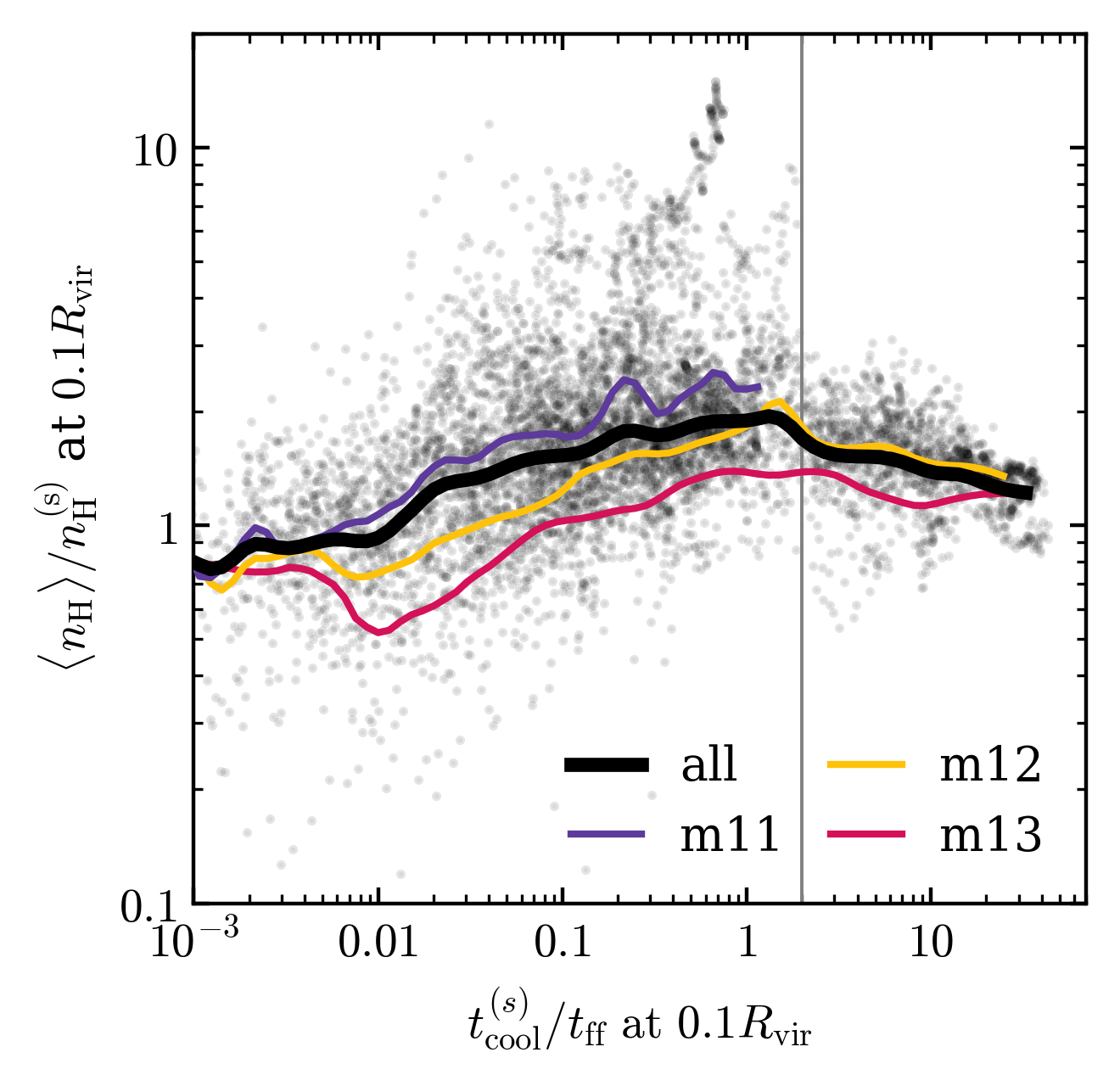}
\end{center}
 \caption{The ratio of the density in the inner CGM (eqn.~\ref{e:nH}) to the expected density if the inner CGM is virialized (eqn.~\ref{e:pressure support}), versus $\tcoolsh/\tff$. Note that the scatter decreases when $\tcoolsh\gg\tff$ and the inner CGM virializes.   }
 \label{f:nH comparison}
\end{figure*}

In Figure~\ref{f:nH comparison} we plot the ratio of the expected density in a virialized CGM $\nHc$ (eqn.~\ref{e:pressure support}) to the shell-averaged CGM density. The average density is measured via
\begin{equation}
 \langle n_{\rm H}(r)\rangle = \frac{\sum X_i m_i/m_{\rm p}}{\sum m_i/\rho_i} \approx \frac{X M_{\rm shell}/m_{\rm p}}{4\pi r^2 \Delta r} ~,
\label{e:nH}
\end{equation}
where $m_i$, $X_i$, and $\rho_i$ are the mass, hydrogen mass fraction, and density of resolution element $i$, the summations  are over all resolution elements within a shell centered at $r$ with thickness $\Delta \log r=0.05\dex$, and $X M_{\rm shell}/m_{\rm p}$ is the total number of hydrogen particles in the shell. Gray dots denote individual snapshots from all 16 simulations, while lines denote medians of the entire sample and the three simulation subgroups. 
The median density ratio is comparable to unity for all values of $\tcoolsh/\tff$, but the scatter between individual snapshots decreases once $\tcoolsh$ exceeds $\tff$ and the inner CGM virializes. Fig.~\ref{f:nH comparison} thus indicates that $\nHc$ is a reasonable approximation of the post-virialization density.

\section{The effect of subgrid metal diffusion and resolution}\label{a:resolution}

\begin{figure*}
\begin{center}
 \includegraphics{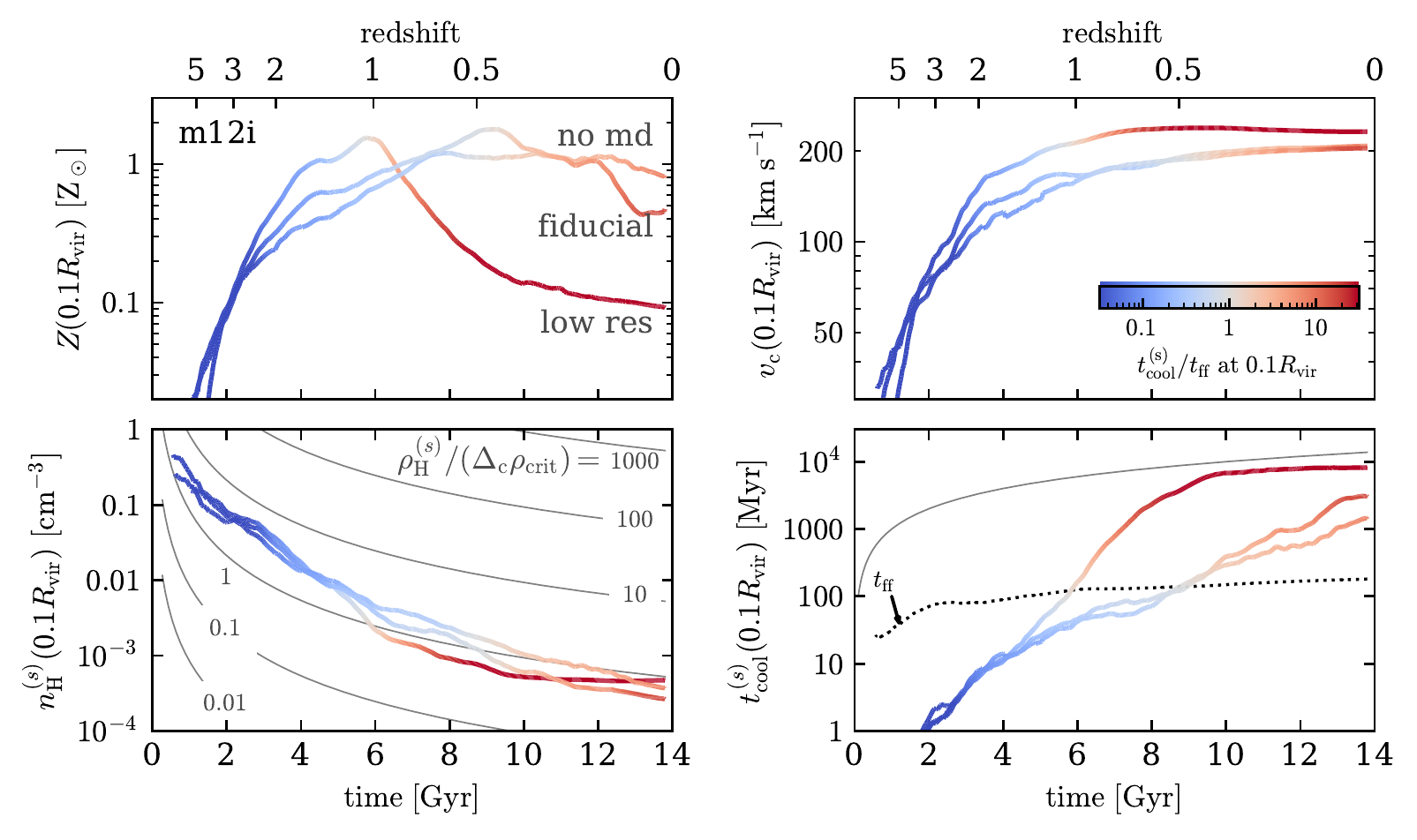}
\end{center}
 \caption{Similar to Figure~\ref{f:properties}, for three simulations of the m12i halo: the fiducial run used in the main text, a run without subgrid metal diffusion, and a lower resolution simulation with an initial baryon mass resolution of $\mb=57000\msun$, in contrast with $\mb=7100\msun$ in the fiducial run. 
 The lower resolution simulation has somewhat higher $\vc$ at $z>1$ (top-right panel), causing $\tcoolsh$ to exceed $\tff$ about $3\Gyr$ sooner than in the higher resolution simulations. 
}
 \label{f:params appendix}
\end{figure*}

\begin{figure*}
\begin{center}
 \includegraphics{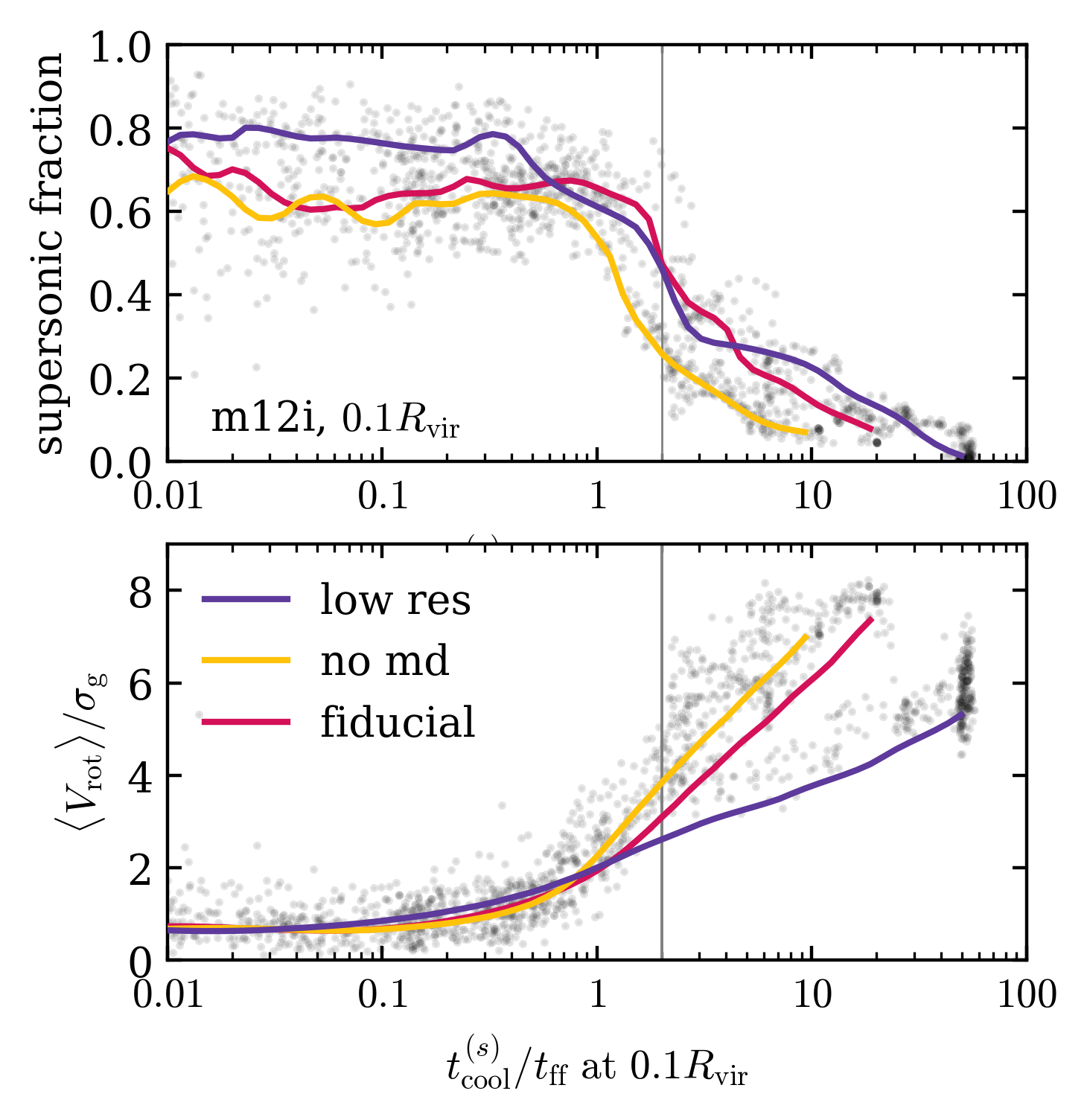}
\end{center}
 \caption{Similar to Figs.~\ref{f:supersonic all} and \ref{f:Vrot to sigma}, for three simulations of the m12i halo. A drop in supersonic fraction and increase in $\langle\Vrot\rangle/\sigma_{\rm g}$ when $\tcoolsh$ exceeds $\tff$ is apparent in all simulations.}
 \label{f:supersonic fraction appendix}
\end{figure*}

In this section we explore the implications of resolution and subgrid metal diffusion on our results. Figures~\ref{f:params appendix} and \ref{f:supersonic fraction appendix} repeat the analysis in Figures~\ref{f:properties}, \ref{f:supersonic all}, and \ref{f:Vrot to sigma} above, for three different runs of the m12i simulation. The `fiducial' simulation is the simulation used in the main text, the `no md' simulation is run without the prescription for subgrid metal diffusion \citep{Hopkins17,Escala18}, and the `low-res' simulation is run with an initial baryon mass resolution of $\mb=57000\msun$, eight times lower than the fiducial $\mb=7100\msun$. 

The top-right panel of Fig.~\ref{f:params appendix} shows that the `low-res' simulation has a somewhat higher $\vc$ at $z>1$ than the `fiducial' and `no md' simulations. The bottom-right panel shows that this difference causes $\tcoolsh$ to exceed $\tff$ about $3\Gyr$ sooner than in the higher resolution simulations. When $\tcoolsh$ exceeds $\tff$ and the inner CGM virializes the `low-res' simulation the metallicity at $0.1\Rvir$ drops, which in turn causes $\tcoolsh/\tff$ to further increase above those in the high resolution simulations. This drop in metallicity upon virialization is also seen in the main simulation sample (see Fig.~\ref{f:properties} and \S\ref{s:Mthres}). 

Fig.~\ref{f:supersonic fraction appendix} shows that all three simulations show a drop in supersonic fraction and increase in $\langle\Vrot\rangle/\sigma_{\rm g}$ when $\tcoolsh$ exceeds $\tff$. Thus, Figs.~\ref{f:params appendix} -- \ref{f:supersonic fraction appendix} suggest that while resolution can affect galaxy/CGM properties and hence the epoch of virialization, our general conclusions 
in any given simulation do not heavily depend on resolution or on the inclusion of subgrid metal-diffusion. We note also that other simulations show a weaker dependence on resolution than m12i \citep[see][]{hopkins18}.

\end{document}